\documentclass[aps,onecolumn,prd,showpacs,showkeys,preprintnumbers,superscriptaddress,nobibnotes,floatfix,longbibliography,notitlepage,nofootinbib]{revtex4-1}

\pdfoutput=1
\usepackage{amsmath}
\usepackage{amsfonts}
\usepackage{amssymb}
\usepackage{mathrsfs}
\usepackage{graphicx}
\usepackage{xcolor}
\usepackage{xspace}
\usepackage{placeins}
\usepackage{fontawesome}

\usepackage{array}
\usepackage{bigints}
\usepackage{tabularx}

\usepackage[colorlinks]{hyperref}
\hypersetup{
    colorlinks,
    linkcolor={red!50!black},
    citecolor={blue!50!black},
    urlcolor={blue!80!black}
}
\usepackage{multirow}
\usepackage{upgreek}
\usepackage[capitalise]{cleveref}
\usepackage{soul}

\newcommand{\eq}{Eq.}

\newcommand{\rhobh}{\rho_{\bullet}}
\newcommand{\fbh}{f_{\bullet}}
\newcommand{\fbhToday}{f_{\bullet,0}}
\newcommand{\Mbh}{M_{\bullet}}
\renewcommand{\Mbh}{M}

\newcommand{\githubmaster}{\href{https://github.com/songningqiang/CosmoLED}{
\faGithub}\xspace}

\begin{document}

\title{Primordial Black Hole Dark Matter in the Context of Extra Dimensions}

\author{Avi Friedlander}
\email{avi.friedlander@queensu.ca}
\affiliation{Department of Physics, Engineering Physics and Astronomy, Queen's University, Kingston ON K7L 3N6, Canada}
\affiliation{Arthur B. McDonald Canadian Astroparticle Physics Research Institute, Kingston ON K7L 3N6, Canada}

\author{Katherine J. Mack}
\email{kmack@ncsu.edu}
\affiliation{North Carolina State University, Department of Physics, Raleigh, NC 27695-8202, USA}
\affiliation{Perimeter Institute for Theoretical Physics, Waterloo ON N2L 2Y5, Canada}

\author{Sarah Schon}
\email{sqs7027@psu.edu}
\affiliation{Department of Physics, Pennsylvania State University, State College, PA 16801, USA}
\affiliation{Department of Physics, Engineering Physics and Astronomy, Queen's University, Kingston ON K7L 3N6, Canada}

\author{Ningqiang Song}
\email{ningqiang.song@liverpool.ac.uk}
\affiliation{Department of Mathematical Sciences, University of Liverpool, \\ Liverpool, L69 7ZL, United Kingdom}
\affiliation{Department of Physics, Engineering Physics and Astronomy, Queen's University, Kingston ON K7L 3N6, Canada}

\author{Aaron C. Vincent}
\email{aaron.vincent@queensu.ca}
\affiliation{Department of Physics, Engineering Physics and Astronomy, Queen's University, Kingston ON K7L 3N6, Canada}
\affiliation{Arthur B. McDonald Canadian Astroparticle Physics Research Institute, Kingston ON K7L 3N6, Canada}
\affiliation{Perimeter Institute for Theoretical Physics, Waterloo ON N2L 2Y5, Canada}

%\date{\today}

\begin{abstract}
Theories of large extra dimensions (LEDs) such as the Arkani-Hamed, Dimopoulos \& Dvali scenario predict a ``true'' Planck scale $M_\star$ near the TeV scale, while the observed $M_{pl}$ is due to the geometric effect of compact extra dimensions. These theories allow for the creation of primordial black holes (PBHs) in the early Universe, from the collisional formation and subsequent accretion of black holes in the high-temperature plasma, leading to a novel cold dark matter (sub)component. Because of their existence in a higher-dimensional space, the usual relationship between mass, radius and temperature is modified, leading to distinct behaviour with respect to their 4-dimensional counterparts. Here, we derive the cosmological creation and evolution of such PBH candidates, including the \textit{greybody factors} describing their evaporation, and obtain limits on LED PBHs from direct observation of evaporation products, effects on big bang nucleosynthesis, and the cosmic microwave background angular power spectrum. Our limits cover scenarios of 2 to 6 extra dimensions, and PBH masses ranging from 10 to $10^{21}$ g. We find that for two extra dimensions, LED PBHs represent a viable dark matter candidate with a range of possible black hole masses between $10^{17}$ and $10^{23}$~g depending on the Planck scale and reheating temperature. For $M_\star = 10$~TeV, this corresponds to PBH dark matter with a mass of $M \simeq 10^{21}$~g, unconstrained by current observations. We further refine and update constraints on ``ordinary'' four-dimension black holes. \githubmaster
\end{abstract}

\maketitle

\tableofcontents
\section{Introduction}
\label{sec:intro}
It has long been appreciated that black holes (BHs) could constitute an ideal dark matter (DM) candidate. Cosmological data tells us that 85\% of the matter content of the Universe must behave as a cold, pressureless fluid, and that it must have been present in the early Universe \cite{Planck:2018nkj}. When neither evaporating nor accreting, black holes exhibit this behavior, with the obvious caveat that a new primordial creation mechanism must be postulated, as stellar remnant black holes are a product of the late Universe. Typical creation scenarios invoke large inhomogeneities at small scales created during inflation, leading to BH creation during subsequent matter or radiation domination \cite{Hawking:1971ei,Carr:1974nx}. As black holes evaporate via Hawking radiation, a minimum BH mass of $\sim 10^{15}$g ($10^{-18}M_\odot$) is required for them to survive until today \cite{Hawking:1974rv}. At present, there exist strong constraints on the fraction of DM that could be in the form of these primordial (P)BHs over masses ranging from this lifetime threshold all the way up to the ``incredulity limit'' $\gg 10^{10} M_\odot$, the requirement that at least one PBH exist per dynamical object. These constraints stem from a variety of physical processes including milli-/micro-/femto-/pico-lensing; disruption of binaries, globular clusters and galaxies; heating of stars; (non) observation of accretion X-rays; and the distortion of the cosmic microwave background (CMB). The presence of lighter PBHs at earlier epochs is constrained down to $10^{10}$\,g by the imprint of their Hawking evaporation on big bang nucleosynthesis (BBN), the CMB, extragalactic background light, and antimatter in the Milky Way. We point the reader to Refs.~\cite{Carr:2020gox,Green:2020jor} for reviews of current constraints.

Most searches thus far have relied on PBHs behaving as semiclassical, 4D BHs, as described by Hawking \cite{Hawking:1975vcx}. However, another tantalizing scenario exists, which does not rely on the details of an earlier inflationary epoch. In the presence of large extra dimensions (LEDs), as described e.g. by \cite{arkanihamed:1998rs, Antoniadis:1998ig}, the ``true'' Planck scale $M_\star$ is lowered to the $\sim$ TeV scale. This has the effect of vastly increasing the horizon radius of BHs that are smaller than the scale of these extra dimensions, such that collisions of high-energy particles can produce microscopic black holes. In the late Universe, these are short-lived, evaporating nearly immediately with a large Hawking temperature $T_H \gg$ GeV. Bounds on the length scale (or equivalently, $M_\star$) of LEDs mainly come from collider searches for energetic, high-multiplicity events, typical of isotropic black hole evaporation to standard model products~\cite{Dimopoulos:2001hw,Giddings:2001bu,CMS:2018ucw,CMS:2018ozv}, which indicates $M_\star$ must be greater than a few TeV, depending on the number of extra dimensions.

\begin{figure}
    \centering
    \includegraphics[width=1\textwidth]{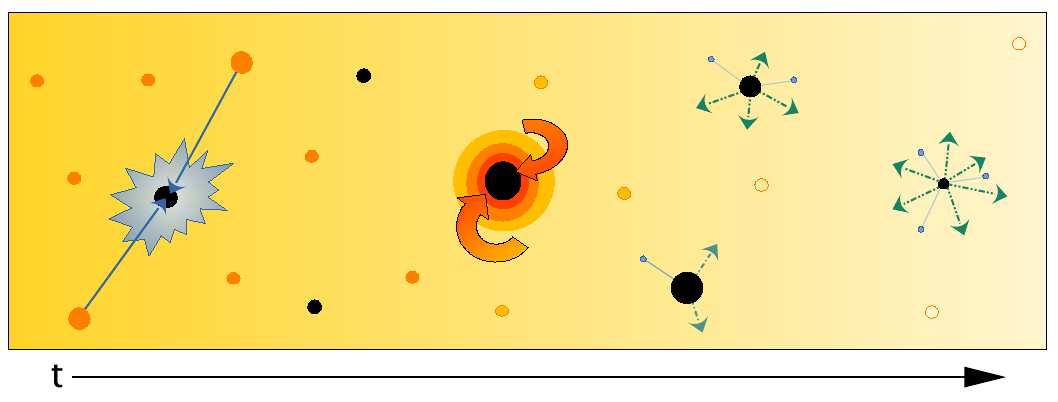}
    \caption{Evolution of large extra dimension black holes in the early Universe. Planck scale black holes are created in particle collisions, which then accrete the surrounding plasma and grow into massive black holes, whilst evaporating to standard model products and gravitons due to Hawking radiation.}
    \label{fig:BHsketch}
\end{figure}

If LED black holes are produced in the high-temperature plasma of the early Universe, their behaviour can be markedly different. As depicted in Fig.~\ref{fig:BHsketch}, because the horizon radii---and thus surface areas---of these collision-initiated BHs are much larger than in the 4D case for a given BH mass, they are able to much more efficiently accrete plasma in the radiation-dominated Universe, and can grow to macroscopic masses. Depending on the number of extra dimensions $n$, the Planck scale $M_\star$, and the reheating temperature $T_{\rm RH}$, this process can occur very rapidly, leading to a population of primordial black holes that can survive until today \cite{Conley:2006jg}. 
In this sense, LED PBHs not only offer an alternate production mechanism to 4-dimensional PBHs, but also present very different phenomenology, and are therefore subject to different constraints, as well as presenting intriguing new possibilities for a role in the late Universe. 

There has been some ambiguity about the mass function expected of primordial black holes in standard scenarios. Indeed, if the mass function of PBHs is not monochromatic, constraints must be recomputed and reinterpreted \cite{Kannike:2017bxn,Carr:2017jsz,Bellomo:2017zsr}. Because PBHs from LEDs are produced in high temperature collisions and rapidly accrete in a predictable way, we find that such scenarios actually predict a relic abundance of BHs with nearly single mass that is set by  $n$,  $M_\star$, and $T_{\rm RH}$, leading to much more straightforward interpretation of results.

We limit our discussion here to the implications of primordial black hole formation in the context of the LED model proposed in \citet{arkanihamed:1998rs} that allows for two or more LEDs, which we will refer to as the ADD model. The Randall-Sundrum model \cite{Randall:1999ee,Randall:1999vf} can also result in the formation of microscopic black holes; their phenomenological implications have been discussed in other works \cite{Guedens:2002sd,Majumdar:2002mra,Sendouda:2003dc,Sendouda:2004hz,Tikhomirov:2005bt}. Like the PBHs produced in the ADD model, those produced in a 5D Randall-Sundrum Type II model can accrete at early times, during the high-energy regime of the braneworld cosmology. This allows the PBHs to have longer lifetimes than 4D PBHs produced at the same era and to produce evaporation radiation that can be constrained by observations at late times. However, the amount of growth is dependent on the accretion efficiency. For concreteness and simplicity we neglect these models here.

In this work, we therefore revisit the full cosmology of primordial black holes in the presence of extra dimensions, with three important results 1) we will find a full set of constraints on LED PBHs based on recent astrophysical data, 2) we will identify the region of parameter space in which LED black holes from particle collisions in the Universe could constitute a viable dark matter candidate, and 3) we will update constraints on low-mass ($\lesssim 10^{17}$~g) ``ordinary'' four-dimensional primordial black holes. 

Black holes from LEDs are constrained by two important effects: first, if they are overproduced in the early Universe, they may lead to rapid absorption and loss of the primordial plasma, leading to a matter-dominated Universe incompatible with $\Lambda$CDM. BHs that do survive into observable cosmological epochs will be constrained by their evaporation products. We will compute the so-called \textit{greybody factors} that describe evaporation of these BHs, along with the spectra of secondary particles, and use these to place limits on LED PBHs from their effects on BBN, the CMB, galactic and extragalactic gamma rays. The new greybody factors and constraints are packed in the \texttt{CosmoLED} code, which will soon be made publicly available. In all cases, the BHs produced in LED collisions are light enough that lensing and dynamical constraints do not apply.

We will find that, in the case of $n = 2$ extra dimensions only, PBHs can be produced which survive until today and reproduce the observed cold dark matter abundance. These dark matter candidates require a specific combination of the Planck scale $M_\star$ and reheating temperature. For $M_\star = 10$ TeV, this leads to a population of PBH dark matter with a monochromatic mass $M \simeq 10^{22}$ g, which lie in the open window between evaporation and lensing constraints. 

Finally, we will provide updated constraints in the low mass range on the evaporation of ordinary 4D primordial black holes. Our inclusion of secondary particles and angular information in the 511 keV flux from positron annihilation will lead to some of the strongest constraints yet from galactic gamma rays. Our updated BBN and CMB constraints also include more precise greybody and secondary particle production than prior work, leading to similar, but modified parameter space constraints. 

This article is structured as follows. In Sec.~\ref{sec:theory}, we describe the formation of PBHs in the LED scenario and model their accretion and evaporation, including the greybody factors appropriate to 4+n-dimensional BHs, and the hadronization and decay products from primary particles. In Sec.~\ref{sec:constraints}, we present the observational constraints we have derived from PBH evaporation's impact on: high-energy Galactic radiation (\ref{sec:galactic}), isotropic photon backgrounds (\ref{sec:extragalactic}), the rescattering of CMB photons (\ref{sec:CMB}), and the relic abundances of primordial elements from Big Bang nucleosynthesis (\ref{sec:BBN}). In Sec.~\ref{sec:fullconstraints}, we combine the above constraints---our full results are summarized in Fig.~\ref{fig:combinedConstraints}. We present our conclusions and a discussion of future prospects in Sec.~\ref{sec:conclusions}.

Throughout the text, we use units in which $c = \hbar = k_\textrm{B} = 1$ and Planck 2018 cosmological parameters of $H_0~=~67.36$~km/s/Mpc, $\Omega_m = 0.3153$, $\Omega_\Lambda = 0.6847$, and $\Omega_b = 0.0493$  \cite{Planck:2018nkj}.

\section{Theory}
\label{sec:theory}
In this section, we examine the production of microscopic black holes in the early Universe and their subsequent evolution. The initial number density will be set by a brief period of BH production from high-energy collisions in the plasma, which will rapidly shut off as the Universe cools. At subsequent times, the density of black holes will be determined by two competing effects: accretion of radiation in the plasma and Hawking evaporation. The cosmology of LED BHs was explored in Ref.~\cite{Conley:2006jg}. Here, we improve on that treatment by simultaneously solving the Friedmann equations governing the evolution of the Universe, deriving and applying exact greybody factors to account for the full Standard Model particle content, and providing more exact numerical solutions to the BH evolution equations. In Sec.~\ref{sec:BHbasics} we summarize the properties of BHs in LEDs, and derive the greybody spectra for the emission of Standard Model (SM) particles on the brane, and gravitons in the bulk. In Sec.~\ref{sec:BHformation} we compute the production rate of LED BHs in the primordial plasma. Following that, we describe the accretion and decay of BHs in Sec.~\ref{sec:BHevo} along with their mass spectrum. Finally, in Sec.~\ref{sec:evapproducts}, we obtain the full spectra of BH evaporation products after hadronization and decay, relevant for cosmological observations. 

\subsection{Black holes in large extra dimensions}
\label{sec:BHbasics}

In the ADD model, gravity acts on a $4+n$-dimensional spacetime where the additional $n$ spatial dimensions are compactified to a submillimeter characteristic length, $R$. While, gravity can propagate through the \textit{bulk} consisting of all $3+n$ spatial dimensions, all Standard Model contents are confined to a 3-dimensional \textit{brane}. Despite the fundamental \textit{bulk} energy scale of quantum gravity $M_\star$ being comparable to the electroweak scale, gravity on the \textit{brane} feels much higher Planck scale $M_{pl}$---and thus a much weaker gravitational coupling $G = 1/M_{pl}^2$.
The fundamental Planck scale in the \textit{bulk} including extra dimensions $M_\star$ is related to the Planck scale on the 3-dimensional \textit{brane} by
\begin{equation}
    M_{pl}^2\sim M_\star^{2+n}R^n\,.
\end{equation}  
 For $n\geq2$ and $M_\star \gtrsim 1$ TeV, this implies the LED are of sub millimeter size. However, for $n=1$, for any $M_\star$ sufficiently small to produce PBHs, the size of the LED would be on the scale of the solar system and therefore not viable. \cite{arkanihamed:1998rs}.
 
The exact relation between $M_{pl}$ and $M_\star$ depends on the compactification scheme, but has been studied with different conventions. Setting $M_{pl}^2 = M_\star^{2+n}(2\pi R)^n$ and the bulk gravitational constant $G_\star=1/M_\star^2$ while matching the Schwarzschild solution in higher dimensional general relativity~\cite{Myers:1986un}, yields the horizon radius of a bulk black hole~\cite{Argyres:1998qn} in the\textit{ Dimopoulos convention}:
\begin{equation}
    r_h=\dfrac{a_n}{M_\star}\left(\dfrac{M}{M_\star}\right)^{1/(n+1)}\,,
    \label{eq:rh}
\end{equation}
where
\begin{equation}
    a_n = \left[8\pi^{-(n+1)/2}\frac{\Gamma((n+3)/2)}{n+2}\right]^{1/(n+1)}\,.
    \label{eq:an}
\end{equation}
One could instead set $M_{pl}^2 = 8\pi M_\star^{2+n}R^n$ where $M_\star$ is understood as the reduced Planck mass in the bulk, which leads to the same relation as in Eq.~\eqref{eq:rh}, but replacing $a_n$ with $k_n$ as defined in the \textit{collider convention }~\cite{Giudice:1998ck,Abe:2001swa,Dai:2007ki} 
\begin{equation}
    k_n = \left[2^n\pi^{(n-3)/2}\frac{\Gamma((n+3)/2)}{n+2}\right]^{1/(n+1)}\,.
    \label{eq:kn}
\end{equation}
The bulk Planck scales in the two conventions are related by
\begin{equation}
    M_\star^\mathrm{Dimopoulos} = \left(\dfrac{8\pi}{(2\pi)^n}\right)^{\frac{1}{(n+2)}}M_\star^\mathrm{collider}\,.
\end{equation}
In this article, we will exclusively use the Dimopoulos convention since the horizon radius in Eq.~\eqref{eq:rh}  reduces to the Schwarzschild radius of a 3+1 dimensional black hole when $n=0$ and $M_\star=M_{pl}$.

BHs remain spherically symmetric in all spatial dimensions when the horizon radius is much smaller than the size of extra dimensions, i.e. $r_h\ll R$. As BH mass increases, the horizon approaches the boundary of extra dimensions. Larger BHs saturate the bulk and the majority horizon area will lie in the brane. For $r_h \gg R$, LED BHs will behave identically to classical 4D BHs, i.e., will share the same Hawking temperature, greybody spectra and lifetime, feeling the weak 4D gravitational constant rather than the true fundamental scale $M_\star$. The exact mass above which BHs behave like ordinary 4D BHs depends on the compactification scheme. We estimate it with the mass of a 4D BH whose Schwarzschild radius matches the size of extra dimensions,
\begin{equation}
    M_{\rm 4D}=\dfrac{1}{4\pi}\dfrac{M_{pl}^2}{M_\star}\left(\dfrac{M_{pl}}{M_\star}\right)^{2/n}\,.
    \label{eq:m4d}
\end{equation}
As displayed in Table~\ref{tab:Mmax}, at $M_\star=10$~TeV, the maximum LED BH mass ranges from about $10^{24}$~g to $10^{14}$~g as the number of extra dimensions increase from $n=2$ to $n=6$.
\begin{table}[!htb]
\centering
\setlength\extrarowheight{3pt}
\begin{tabular}{l | l | l | l | l | l}
\hline\hline
	$n$ & 2 & 3 & 4 & 5 & 6\\ \hline
    $M_{\rm 4D}$~[g] & $2.57\times 10^{24}$ & $2.41\times 10^{19}$ & $7.36\times 10^{16}$ & $2.28\times 10^{15}$ & $2.25\times 10^{14}$ \\
    $M_{\rm survive}$~[g] & $2.43\times 10^{7}$ & $5.50\times 10^{10}$ & $1.79\times 10^{13}$ & $1.46\times 10^{15}$ & $4.96\times 10^{14}$*  \\
    
\hline \hline
\end{tabular}
\caption{$M_{\rm 4D}$: Mass above which BHs saturate the size of the extra dimensions, causing them to behave like classical four-dimensional black holes (Eq. \eqref{eq:m4d}). $M_{\rm survive}$: Mass above which BHs do not fully evaporate before today if created in the early Universe. The asterisk for $n=6$ indicates that $M_{\rm survive} > M_{\rm 4D}$ therefore all BHs with $n=6$ that survive until today act like 4D BHs. All values in this table assume the fundamental Planck scale is $M_\star=10$~TeV.}
\label{tab:Mmax}
\end{table}

\begin{figure}
    \centering
    \includegraphics[width=0.45\textwidth]{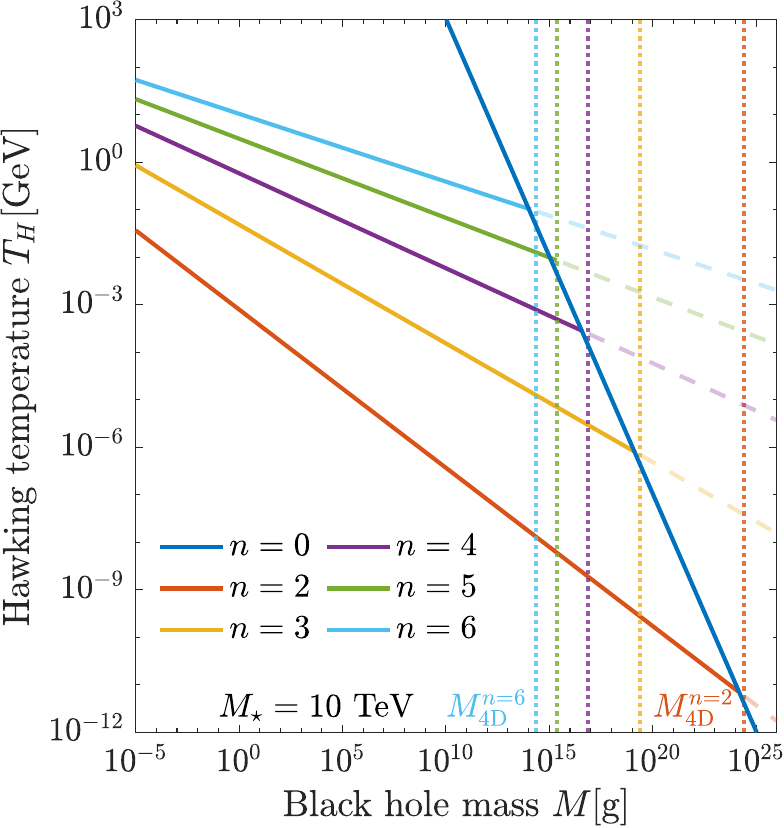}\hspace{0.4cm}
    \includegraphics[width=0.45\textwidth]{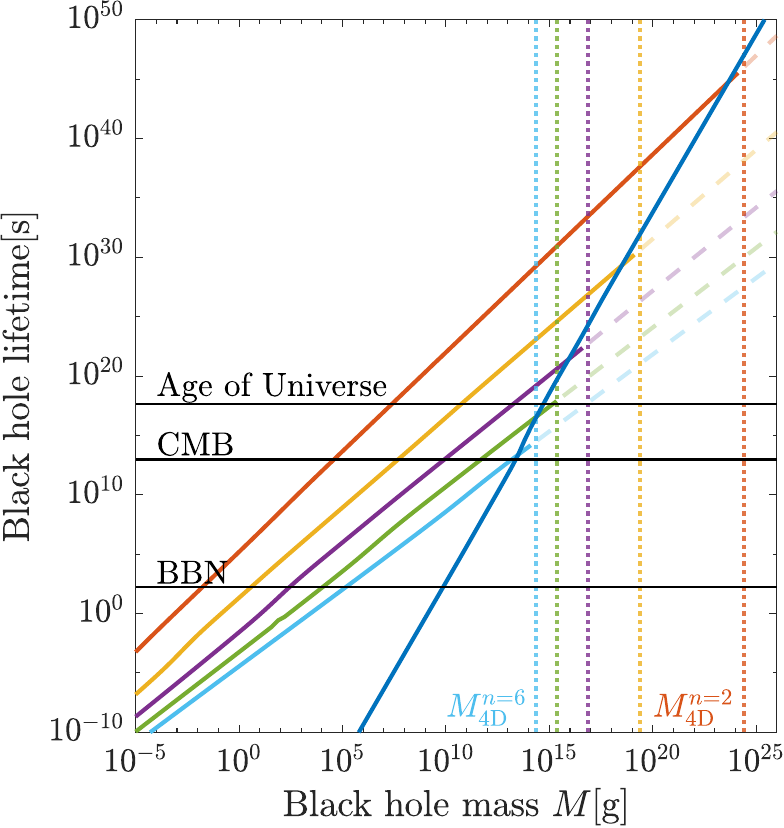}       
    \caption{\textit{Left}: Hawking temperature as a function of black hole mass. \textit{Right}: Lifetime of black holes as a function of their mass. In both panels, vertical dotted lines indicate $M_{\max}$, the mass at which the horizon radius $r_h$ approximately saturates the size of the extra dimensions. Above this mass, LED BHs behave the same as ordinary 4D BHs. The fundamental Planck scale $M_\star=10$~TeV is assumed for $n>0$ LED BHs. }
    \label{fig:TBH}
\end{figure}

As with ordinary four-dimensional BHs, LED BHs also lose mass via Hawking evaporation. However, since Hawking evaporation is geometric and the horizon area of a black hole depends on the number of extra dimensions, LED black holes will have a modified Hawking temperature~\cite{Argyres:1998qn}
\begin{equation}
T_H =    \frac{n+1}{4\pi r_h}.
\label{eq:TH}
\end{equation}
The Hawking temperature of BHs in different dimensions is depicted in the left panel of Fig.~\ref{fig:TBH}. LED BHs in fewer extra dimensions typically radiate particles at a lower temperature than high-$n$ BHs. They also remain considerably colder than 4D BHs, benefiting from the low bulk Planck scale. It is also worth noting that an LED BH with mass $M_{\rm 4D}$ may not share precisely the same Hawking temperature with 4D BHs of the same mass, i.e. some discontinuity might be observed during extra dimension-to-4D transition. This is expected for two reasons: 1) The radius of $M_{\rm 4D}$ LED BHs is not identical to the size of extra dimensions due to the different mass-radius relations for $n > 0$ and $n = 0$. 2) The LED Hawking temperature given in Eq.~\eqref{eq:TH} explicitly contains $n$. This discontinuity is not very large: it can be seen in Fig.~\ref{fig:TBH} by observing that the solid $n \neq 0$ lines do not end exactly on the blue $n = 0$ (4D) line.

\begin{figure}
    \centering
    \includegraphics[width=0.48\textwidth]{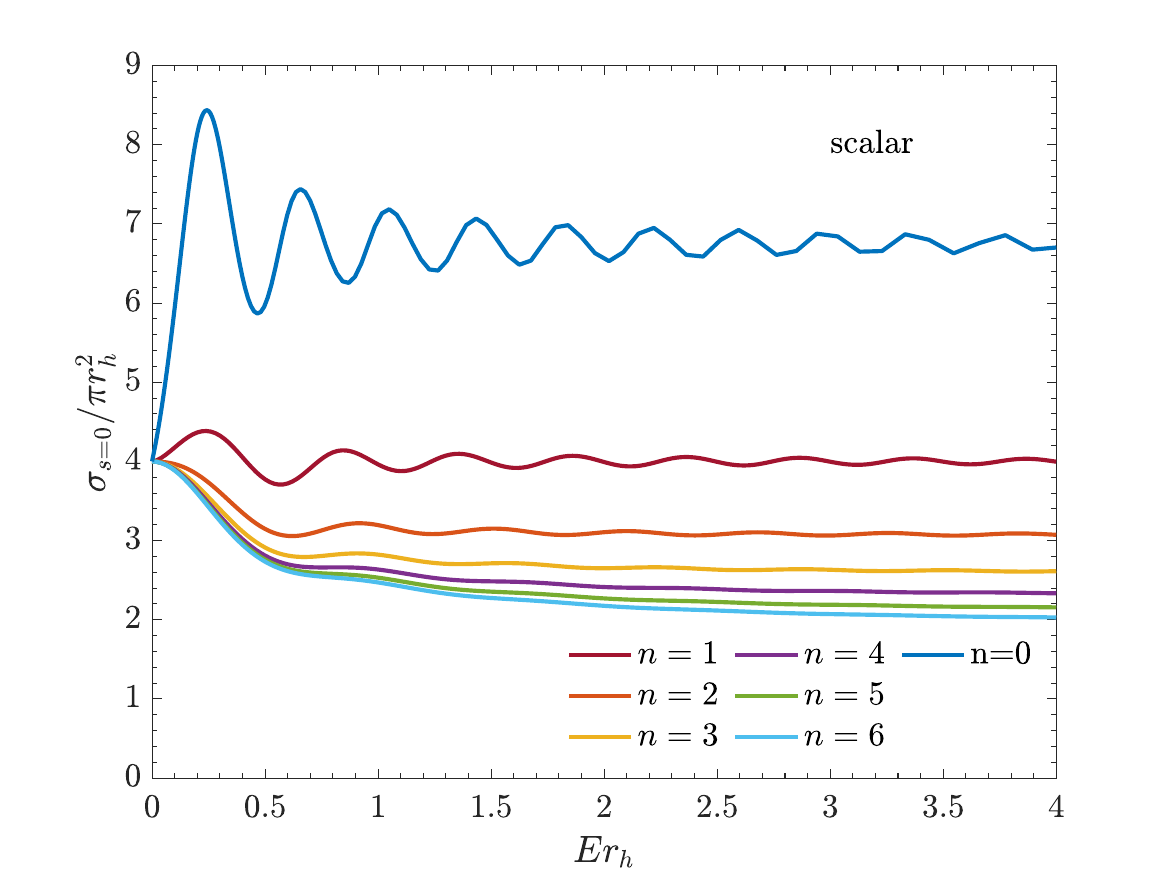}
    \includegraphics[width=0.48\textwidth]{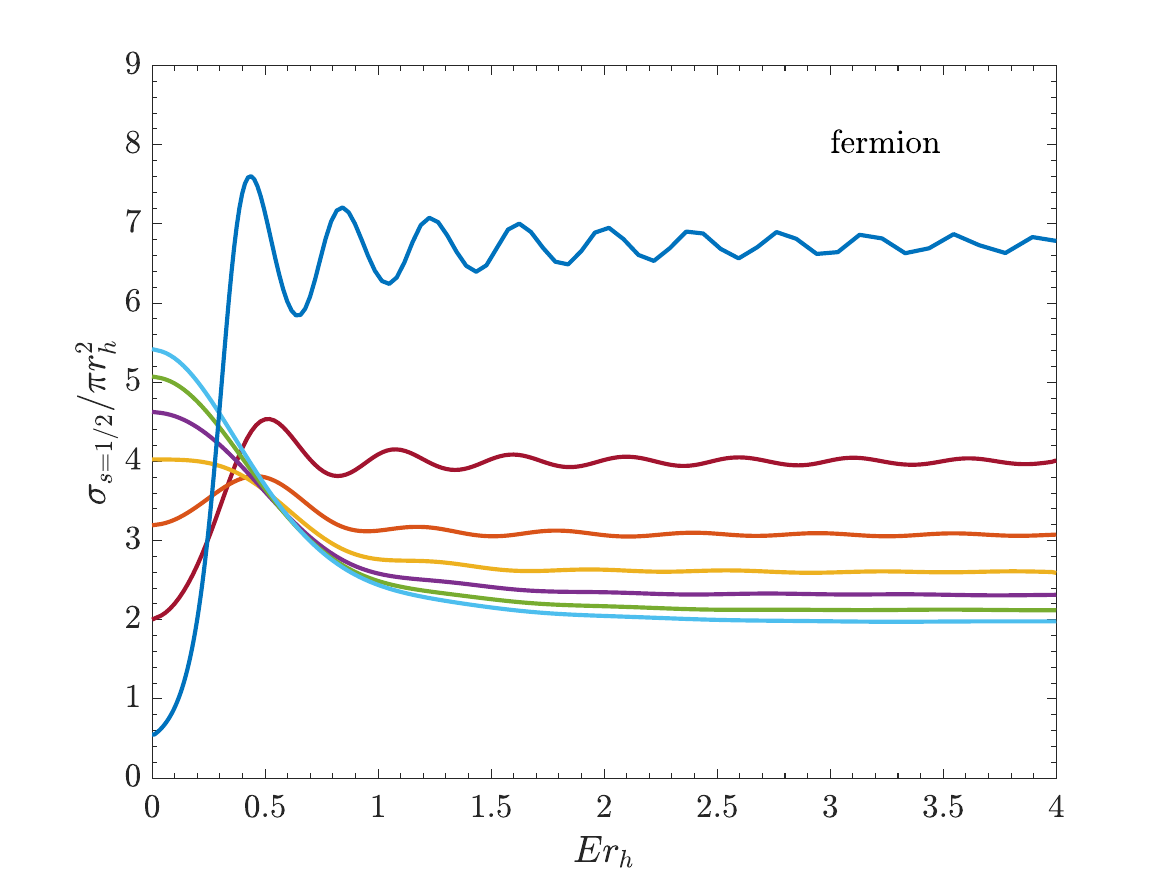}
    \includegraphics[width=0.48\textwidth]{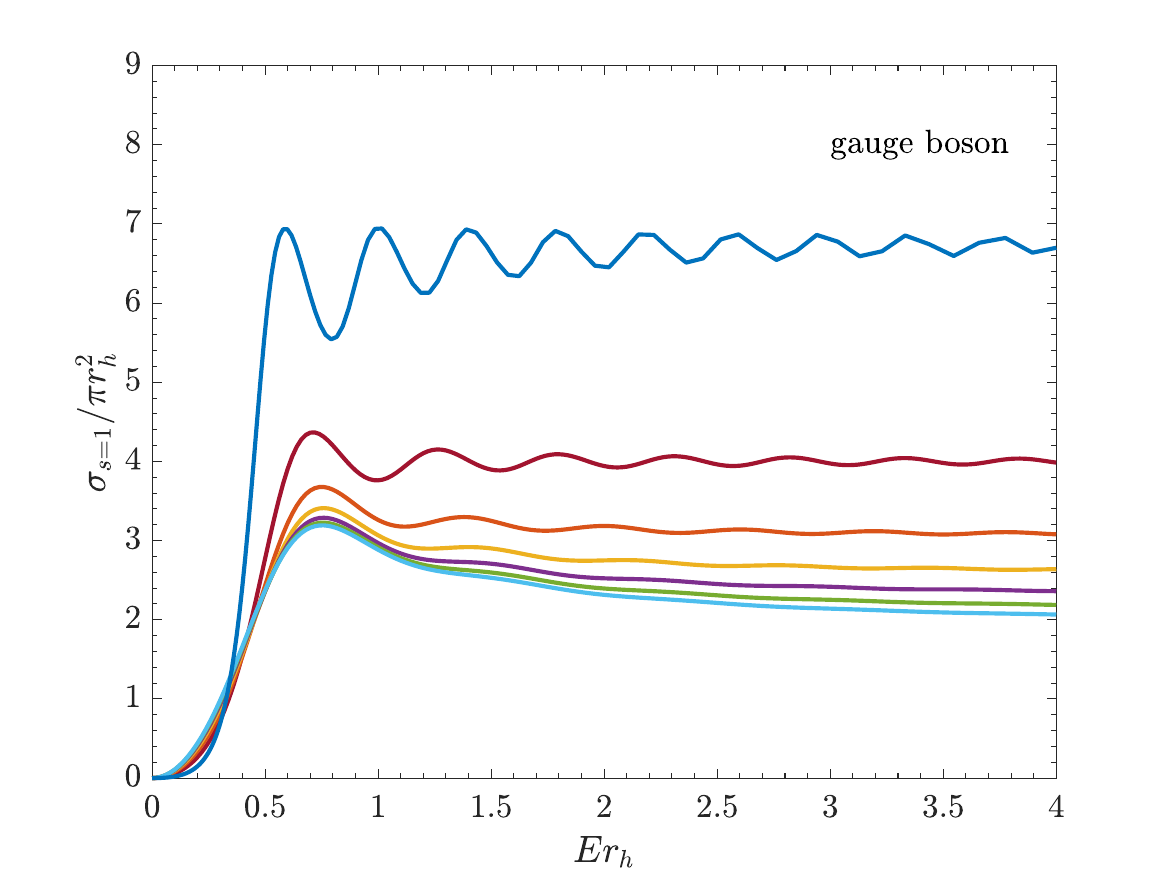}
    \includegraphics[width=0.48\textwidth]{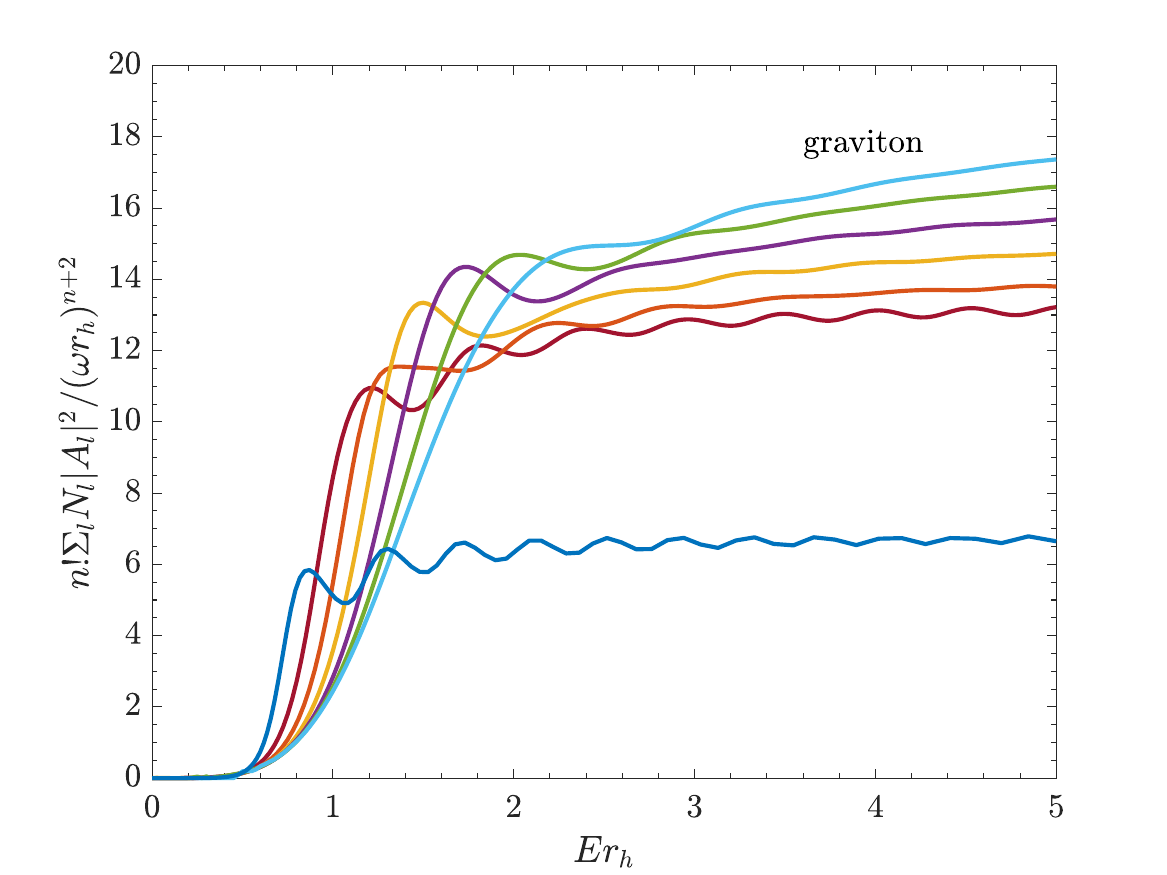}      
    \caption{Greybody spectra for the emission of scalars, fermions and gauge bosons in the brane, and the emission of gravitons in the bulk from the evaporation of higher dimensional black holes. Different colours correspond to $n=1$ to $n=6$ extra dimensions. Scaled absorption cross sections are depicted for scalar, fermions and gauge bosons, and the absorption probabilities are shown for gravitons where the contributions from scalar, vector and tensor perturbations are aggregated. The $n=0$ greybody spectra for all particle types are obtained from \texttt{BlackHawk}~\cite{Arbey:2019mbc,Arbey:2021mbl}.}
    \label{fig:greybody}
\end{figure}

BHs may evaporate into every degree of freedom that couples to gravity so long as it is not too thermally suppressed, \textit{i.e.}, the Hawking temperature is not too far below the mass of the particle. Since SM particles are confined to the brane, the emission of SM particles is limited to our three dimensional space. In contrast, gravitons are free to propagate in the bulk with significantly larger emission phase space.  The distribution of particles from BH evaporation resembles a black body spectrum, up to a correction due to the gravitational potential of the BH. The emission of an SM particle degree of freedom $j$ is given by
\begin{equation}
    -\dfrac{dM_{\bullet\rightarrow j}}{dt}=\sigma_j(E) \dfrac{E}{\exp(E/T_H)\mp 1}\dfrac{d^3p}{(2\pi)^3}\,,
    \label{eq:emissionSM}
\end{equation}
where $\sigma_j(E)$ is the absorption cross section, or \textit{greybody factor}, which quantifies the correction. Here, the energy of a single particle is $E=\sqrt{p^2+m_j^2}$. The greybody factor can be computed via partial wave scattering theory. It is obtained by solving the wave equation of a particle near the horizon and at infinity, and by summing up the contribution from all emission modes. Because the black hole horizon behaves as a black body, the ratio of ingoing radiation at the horizon to the ingoing radiation at infinity  yields the absorption coefficient $A_l$, which is related to the absorption cross section through
\begin{equation}
    \sigma_j=\sum\limits_l\dfrac{\pi}{E^2}(2l+1)|A_l^j(E)|^2\,,
\end{equation}
for brane-localized SM particles, where the sum runs over all angular momentum modes. We follow the numerical framework outlined in~\cite{Harris:2003eg,Harris:2004mf} and solve for the greybody spectrum for scalars~\cite{Kanti:2002nr}, fermions and gauge bosons~\cite{Kanti:2002ge} in the massless limit for non-rotating higher dimensional black holes. The effect of particle mass is mainly to introduce a lower limit for the emission spectrum~\cite{Page:1977um}. The greybody factors $\sigma_s$ for spin $s=0$, 1/2 and 1 are shown in Figure~\ref{fig:greybody}. We note that $M_\star$ does not appear in the wave equations explicitly, and the results remain valid for an arbitrary bulk Planck scale. At $E \rightarrow 0$, the scalar greybody factor $\sigma_0=4\pi r^2$ regardless of the number of extra dimensions. In contrast to scalars and fermions, the emission of gauge bosons is suppressed at low energies. In the high energy limit $E r_h\gg 1$, the greybody factors for all three particle types have the asymptotic value of $\sigma/(\pi r_h^2)\rightarrow4^{-1/(n+1)}(n+3)^{(n+3)/(n+1)}/(n+1)$. 

Unlike SM particles, gravitons may propagate in the bulk and thus have access to larger phase space. The emission spectrum of gravitons is more conveniently expressed by the absorption probability $|A_l|^2$ after integrating the angular distribution over the 3+n dimensional sphere
\begin{equation}
  - \dfrac{dM_{\bullet\rightarrow G}}{dt}=\sum\limits_l N_l |A_l|^2 \dfrac{E}{\exp(E/T_H)- 1}\dfrac{dp}{2\pi}\,,
    \label{eq:emissiongraviton}
\end{equation}
where the multiplicities of states $N_l$ are given in Ref.~\cite{Creek:2006ia}. Graviton emission in the bulk can be decomposed into a traceless symmetric tensor, a vector and a scalar mode. We solve for the radial parts of these three components separately and sum up their absorption probabilities. The total graviton absorption probability is displayed in the last panel of Figure~\ref{fig:greybody}. Our numerical results agree with the exact solutions in Refs.~\cite{Creek:2006ia,Cardoso:2005mh}, but differ from Ref.~\cite{Johnson:2020tiw} by a constant factor. Similarly to gauge bosons, the absorption probability is suppressed in the low energy region $E r_h\ll 1$. At high energies,  it scales asymptotically as $(E r_h)^{n+2}$.

The full BH mass loss rate is obtained by integrating the particle emission spectra in Eq.~\eqref{eq:emissionSM} and~\eqref{eq:emissiongraviton} while accounting for the particle degree of freedom $g_{\rm dof}$. For convenience, we define $\xi$ and $\alpha$, which are related to the BH evaporation rate by
\begin{equation}
    -\dfrac{d\Mbh^{\rm evap}}{dt}\equiv\sum\limits_j \dfrac{1}{2\pi}\dfrac{\xi_j}{r_h^2}\equiv\alpha(n,T_H) T_H^2\,,
    \label{eq:dMdt}
\end{equation}
where for SM particles
\begin{equation}
    \xi_j=g_{\rm dof,j}\int \dfrac{\sigma_j}{\pi r_h^2}\dfrac{E r_h^4}{\exp(E/T_H)\mp 1}p^2dp\,,
\end{equation}
and for gravitons
\begin{equation}
    \xi_G=\int \sum\limits_l N_l|A_l|^2\dfrac{E r_h^2}{\exp(E/T_H)-1}dp\,.
\end{equation}
It is evident that the emission probability of a particle depends on the ratio between particle mass and the Hawking temperature. When $m_j>T_H$, the emission will be exponentially suppressed. This is accounted for approximately by fitting $\xi_j$ with the functional shape 
\begin{equation}
    \xi_j=\xi_{j,0}\exp\left(-b_j(m_j /T_H)^{c_j}\right)\,,
\end{equation}
where $\xi_{j,0}$ is evaluated at $m_j=0$. Numerically, we obtain $b_j\simeq 0.3$ and $c_j\simeq 1.3$ for SM scalars, fermions and gauge bosons. The relevant parameters for different number of extra dimensions are given in Table~\ref{tab:emissionfit}. At high temperatures $T_H \gg m_j$, we may sum over all SM particles, gravitons and their helicity states to obtain an approximately constant value for  $\alpha(n,T_H) \simeq \alpha_0$, which is also listed in Table~\ref{tab:emissionfit}. In this limit, the contribution from the total emission power of gravitons in BH mass loss ranges from $0.1\%$ to $14.4\%$ for four dimensional ($n=0$) black holes to $n=6$ dimensional black holes, as also obtained in Ref.~\cite{Cardoso:2005mh}. 

\begin{table}[!htb]
\centering
\setlength\extrarowheight{3pt}
\begin{tabular}{ c  c  c  c  c  c  c c c c c c}
\hline\hline
$n$&\multicolumn{3}{c}{scalar}&\multicolumn{3}{c}{fermion}&\multicolumn{3}{c}{gauge boson}&graviton&total\\
&$\xi_{j,0}$&$b_j$&$c_j$&$\xi_{j,0}$&$b_j$&$c_j$&$\xi_{j,0}$&$b_j$&$c_j$&$\xi_j$&$\alpha_0$\\
0&0.00187&0.395&1.186&0.00103&0.337&1.221&0.000423&0.276&1.264&0.0000966&2.77\\
1&0.0167&0.333&1.236&0.0146&0.276&1.297&0.0115&0.220&1.361&0.00972&10.45\\
2&0.0675&0.283&1.291&0.0612&0.293&1.279&0.0611&0.264&1.311&0.0995&20.50\\
3&0.187&0.281&1.296&0.167&0.288&1.286&0.186&0.274&1.303&0.493&32.53\\
4&0.416&0.285&1.292&0.362&0.290&1.284&0.432&0.258&1.329&1.904&46.74\\
5&0.802&0.293&1.282&0.684&0.296&1.276&0.847&0.298&1.279&6.886&64.19\\
6&1.401&0.304&1.270&1.174&0.274&1.303&1.488&0.311&1.265&24.684&88.03\\
\hline\hline
\end{tabular}
\caption{$\xi_j$'s are given per dof for scalars, fermions and gauge bosons, while the emission accounting for all dof is included for gravitons. If all particles are massless, the BH evaporate rate is proportional to $\alpha=\alpha_0$ defined in Eq.~\eqref{eq:dMdt}.}
\label{tab:emissionfit}
\end{table}

From Eqs.~\eqref{eq:rh} and~\eqref{eq:TH} we find the BH mass loss rate $dM/dt\propto M^{-2/(n+1)}$. As the number of extra dimensions increases, BHs tend to evaporate faster. However, they remain substantially longer lived than 4D BHs, owing to $M_\star \ll M_{pl}$. The right panel of Fig.~\ref{fig:TBH} shows the lifetimes of BHs and Table~\ref{tab:Mmax} lists the lightest BHs that do not entirely evaporate before today. While an $n=6$ BH which does not saturate the bulk does not survive until today, $n=2$ BHs as light as $10^7$~g may still exist now. This has striking implications which will change the BH landscape we expect: BHs in the Universe might be lighter with a larger number density, and may thus escape gravitational lensing searches but still affect astrophysical and cosmological observations through evaporation or coalescence.

\subsection{Black hole formation in the early Universe}
\label{sec:BHformation}

The Hoop Conjecture~\cite{Thorne1995black,Banks:1999gd} posits that a black hole will be formed if the impact parameter $b$ of two colliding particles is smaller than twice the horizon radius $r_h$. Equivalently, a microscopic black hole of mass $M=E_{CM}$ can be created if the center of mass energy $E_{CM}$ is larger than $M_\star$\footnote{We neglect the mass loss in the formation stage and assume the minimum black hole mass $M_{\min}=M_\star$. For discussions see Ref.~\cite{Mack:2019bps} and the reference therein.}. The BH production cross section can thus be approximated by the geometric size of the scattering
\begin{equation}
    \sigma(\Mbh)\sim \pi b_{\max}^2=4\pi r_h^2\,.
\end{equation}
The high temperature primordial plasma consisted of quarks, leptons, higgs and gauge bosons. The kinetic energy of plasma particles is characterized by the reheating temperature $T_{\rm RH}$. The plasma temperature then drops due to expansion, and could also be affected by plasma loss from accretion. Given the thermal distribution of particles, $T_{\rm RH}$ need not exceed $M_\star$ in order for BH production to take place. During radiation domination, the BH formation rate per unit volume per unit mass is given by~\cite{Conley:2006jg}
\begin{equation}
    \dfrac{d\Gamma}{d\Mbh}=g_{\star}(T)^2\int \dfrac{d^3k_1}{(2\pi)^3}\dfrac{d^3k_2}{(2\pi)^3}e^{-k_1/T}e^{-k_2/T}\sigma(\Mbh)v_{rel}\delta\left(\sqrt{(k^\mu_1+k^\mu_2)^2}-\Mbh\right)\Theta(\Mbh-M_\star)\,,
    \label{eq:dGammadM}
\end{equation}
where $g_{\star}(T)$ is the effective number of relativistic particle species and we have approximated the phase space distribution as a Maxwell-Boltzmann distribution. The step function $\Theta$ is added to ensure $E_{CM}\geq M_\star$. If the plasma temperature $T\gtrsim 200$~GeV, then $g_{\star}=106.75$. To see the asymptotic behavior, we approximate the relative velocity $v_{rel}=|\vec{v}_1-\vec{v}_2|\simeq 1$ in radiation domination and carry out the integral explicitly. This yields
\begin{equation}
    \dfrac{d\Gamma}{d\Mbh}=\dfrac{g_{\star}(T)^2a_n^2}{8\pi^3}\Mbh T^2\left(\dfrac{\Mbh}{M_\star}\right)^{\frac{2n+4}{n+1}}\left[\dfrac{\Mbh}{T}K_1(\frac{\Mbh}{T})+2K_2(\frac{\Mbh}{T})\right]\Theta(\Mbh-M_\star)\,,
    \label{eq:BHprodrate}
\end{equation}
where $K_\nu(x) $ is the modified Bessel function of the second kind. In the low temperature limit $T\ll M_\star\lesssim M$, the Bessel function $K_\nu(\Mbh/T)\sim \sqrt{T/\Mbh}\exp(-\Mbh/T)$. This implies that there is a limited temperature window when BHs could be copiously produced. As the plasma temperature drops below $M_\star$, BH formation becomes  exponentially suppressed. Without the approximation $v_{rel}\simeq 1$ Eq.~\eqref{eq:dGammadM} is evaluated to be
\begin{equation}
    \dfrac{d\Gamma}{d\Mbh}=\dfrac{g_{\star}(T)^2a_n^2}{4\pi^3} T\left(\dfrac{\Mbh}{M_\star}\right)^{\frac{2n+4}{n+1}}\int dk e^{-k/T}\left[\Mbh e^{-\frac{\Mbh^2}{4kT}}+\sqrt{\pi kT}\ \mathrm{Erfc}\left(\frac{M}{\sqrt{2kT}}\right)\right]\Theta(\Mbh-M_\star)\,.
    \label{eq:BHprodrateexact}    
\end{equation}
The difference between Eq.~\eqref{eq:BHprodrate} and~\eqref{eq:BHprodrateexact} when integrating over $M$ is only fractional. Considering the BH production rate is extremely susceptible to the reheating temperature, the results are rather insensitive to the choice of the production formalism. To reduce computation cost, we therefore use Eq.~\eqref{eq:BHprodrate} in the numerical analysis.

\subsection{Black hole accretion and decay in an expanding universe}
\label{sec:BHevo}

If BHs are produced at a plasma temperature $T\lesssim M_\star$, most of them acquire a mass just above the Planck scale since more massive BH production is severely limited by kinematics. However, being immersed in the radiation bath of the primordial plasma, BHs are capable of trapping any particle that crosses the horizon and become progressively more massive. The accretion rate is proportional to the horizon area and the energy density of the plasma, with an $\mathcal{O}(1)$ accretion efficiency $f_{\rm acc}$ depending on the mean free path of the plasma particles and the peculiar velocity of the black holes~\cite{Bondi:1952ni,Nayak:2009wk,Masina:2020xhk}:
\begin{equation}
    \dfrac{d\Mbh^{\rm acc}}{dt}=f_{\rm acc} 4\pi r_h^2 \rho_r\,,
    \label{eq:dMdtacc}
\end{equation}
with the plasma radiation density
\begin{equation}
    \rho_r=\frac{\pi^2}{30}g_\star (T) T^4\,.
    \label{eq:rhor}
\end{equation}
 Combining the evaporation in Eq.~\eqref{eq:dMdt} and accretion, BH mass evolves as
\begin{equation}
    \dfrac{d\Mbh}{dt}=\left(-\alpha(n,T_H)+\beta\dfrac{T^4}{T_H^4}\right)T_H^2\,,
    \label{eq:dMdtcombine}
\end{equation}
where $\beta=\frac{\pi}{120}(n+1)^2f_{\rm acc}g_\star$ and $\alpha(n,T_H)$ is defined implicitly in Eq. \eqref{eq:dMdt}. Depending on the sign of the bracket on the right hand side of Eq.~\eqref{eq:dMdtcombine}, newly born BHs may either decay away or accrete and grow. Since $\alpha$ varies only mildly with $T_H$, $d\Mbh/dt$ is susceptible to the ratio $T/T_H$. If initially $d\Mbh/dt>0$, the Hawking temperature will decrease as accretion persists, further escalating the accretion rate. The accretion halts when the Universe becomes sufficiently cold to match the Hawking temperature again, at which time BHs may have accreted enough energy to appear macroscopic. For a BH created at the mass $\Mbh= M_\star$, the watershed plasma temperature between decay and accretion reads
\begin{equation}
    T_{\rm th}=\left(\dfrac{15\alpha_0}{32\pi^5f_{\rm acc}g_\star}\right)^{1/4}a_n^{-1}(n+1)^{1/2}M_\star\,,
    \label{eq:Tth}
\end{equation}
which ranges from $0.17M_\star$ to $0.62M_\star$ for $n=1$ to 6 extra dimensions assuming $g_\star=106.75$. For concreteness we have set $f_{\rm acc}=1$. A different accretion efficiency will slightly modify $T_{\rm th}$ as $T_{\rm th}\propto f_{\rm acc}^{-1/4}$. However, the formation of massive BHs is not shut down entirely at a temperature $T<T_{\rm th}$, as BHs that are born with a mass sufficiently higher than $M_\star$ may still have low enough Hawking temperature to ensure $d\Mbh/dt>0$. This amounts to the production of a BH with initial mass $M_i$ where
\begin{equation}
    M_i>M_{i,\min}=\max\left\{M_\star,\left[\dfrac{n+1}{4\pi a_n}\left(\dfrac{\alpha_0}{\beta}\right)^{1/4}\dfrac{M_\star}{T}\right]^{n+1}M_\star\right\}\,,
    \label{eq:Mimin}
\end{equation}
\textit{i.e.}, BHs that are created at a mass above $M_{i,\min}$ may accrete rather than decay immediately after formation. On the other hand, the production rate of $M_{i,\min}>M_\star$ BHs is exponentially suppressed by $\Mbh/T$ as seen in Eq.~\eqref{eq:BHprodrate}.

The mass evolution can be solved in a straightforward way assuming radiation dominates throughout. Relating the plasma temperature to time using the Friedmann equations in a radiation-dominated universe,
\begin{equation}
    -\dfrac{dT}{dt}=\sqrt{\dfrac{4\pi^3}{45}g_\star}\dfrac{T^3}{M_{pl}}\,,
    \label{eq:dTdt}
\end{equation}
Eq. \eqref{eq:dMdtcombine} becomes
\begin{equation}
    \dfrac{dM}{dT}=\sqrt{\dfrac{45}{4\pi^3g_\star}}M_{pl}\dfrac{T_H^2}{T^3}\left(\alpha-\beta\dfrac{T^4}{T_H^4}\right)\,.
    \label{eq:dMdTsimp}
\end{equation}

\begin{figure}
    \centering
    \includegraphics[width=0.48\textwidth]{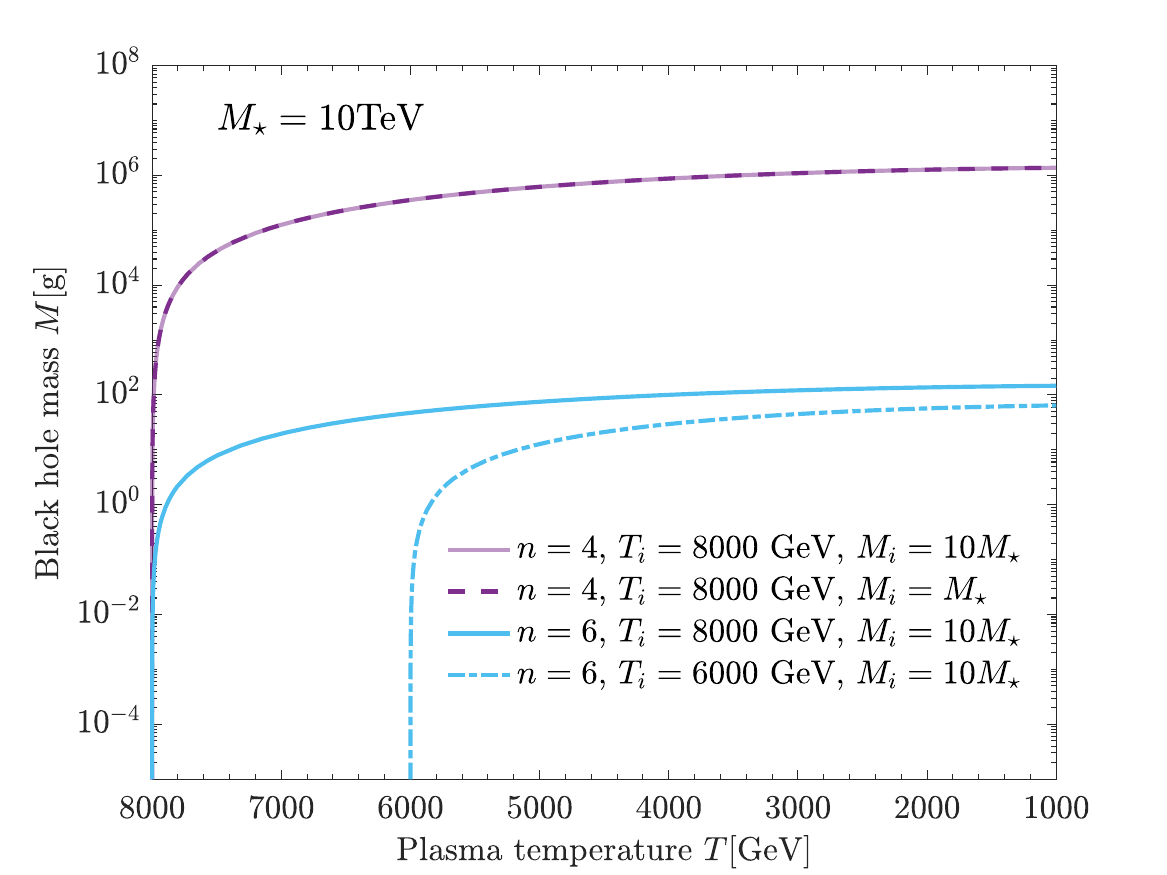}    
    \includegraphics[width=0.48\textwidth]{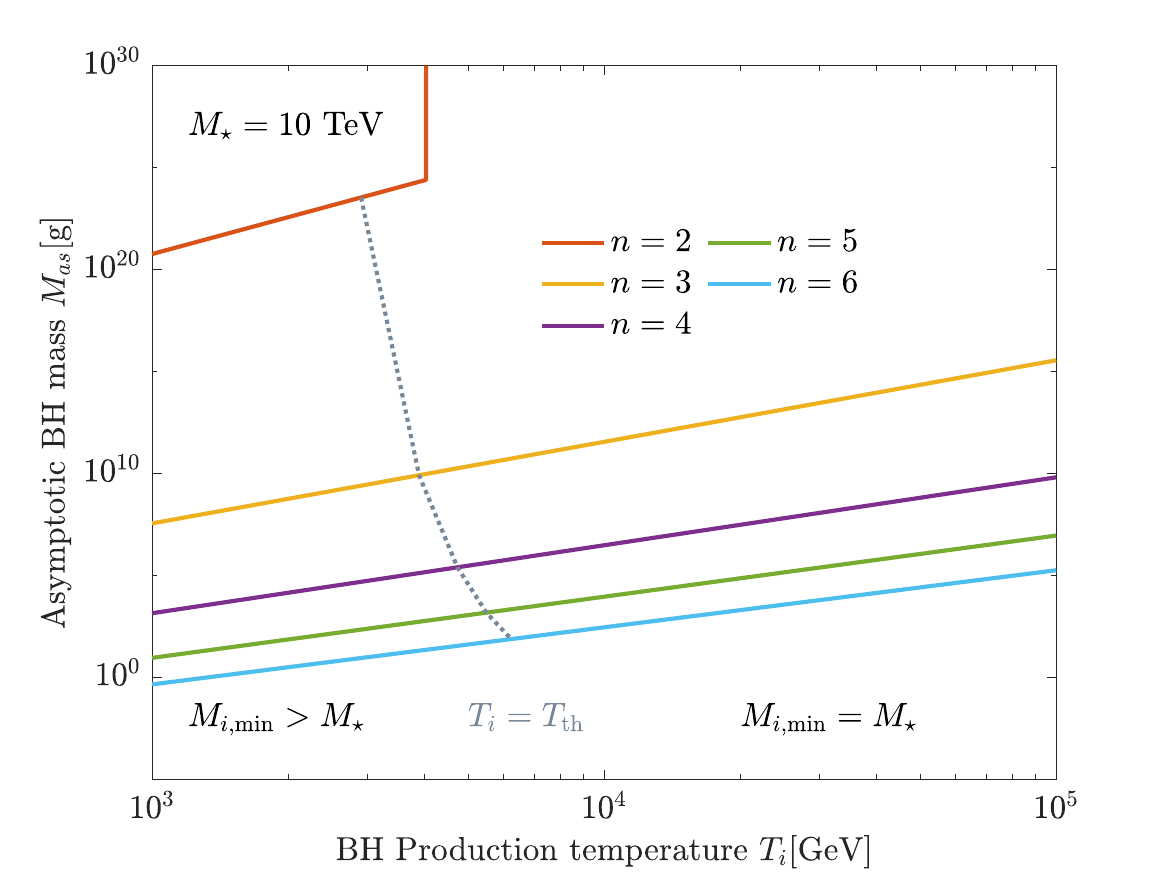}
    \caption{\textit{Left}: BH mass evolution as a function of the primordial plasma temperature, where time evolution proceeds from left to right as the Universe cools. Examples are given for $n=4$ and $n=6$ BHs created at different initial masses and initial temperatures. \textit{Right}: Asymptotic BH masses after accretion in the early Universe as a function of $T_i$, the BH production temperature, for $n = 2,3,4,5$ and $6$ extra dimensions. Left of the dotted line, the minimum initial BH mass has to be greater than $M_\star$, otherwise evaporation will be faster than accretion, and BHs will rapidly decay away. Right of the dotted black line, accretion is efficient and BHs grow to $\Mbh \simeq M_{as}$, regardless of their initial mass. For $n=2$ BHs may accrete to become 4D BHs and then keep growing. See text for details. In both panels the Planck scale $M_\star = 10$ TeV is assumed.}
    \label{fig:Mas}
\end{figure}

Some examples of the BH mass evolution are given in the left panel of Fig.~\ref{fig:Mas}, obtained by numerically solving  Eq.~\eqref{eq:dMdTsimp}. In cases where accretion wins out, the BH mass shoots up by many orders of magnitude at the initial stage of accretion. Because of this, the accreted matter contributes nearly all of the mass, and the final BH mass is independent of the initial BH mass $M_i$. However, the temperature dependence of the process means that the process is very sensitive to the temperature of the plasma at production, $T_i$. As the temperature falls $T(t) \ll T_i$, the black hole mass grows to its asymptotic value
\begin{equation}
    M_{as}= \left(\gamma_n\dfrac{M_{pl}T_i^2}{M_\star^3}\right)^\frac{n+1}{n-1}M_\star\,,
    \label{eq:asympmass}
\end{equation}
where $\gamma_n=f_{\rm acc}\sqrt{\frac{\pi^3}{20}g_\star}a_n^2\frac{n-1}{n+1}$. It is derived when the evaporation is neglected and $g_\star$ is assumed to be constant. The asymptotic BH masses are shown as a function of $T_i$ in the right panel of Fig.~\ref{fig:Mas}. A dotted grey line displaying $T_{\rm th}$ defined in Eq.~\eqref{eq:Tth} is also drawn in the middle of the panel. Right of the line, BHs of any mass above $M_\star$ will accrete and grow. To the left, $M_i$ has to exceed $M_{i,\min}$ to avoid immediate decay. The production of BHs at such temperatures is more kinematically suppressed. Special attention should be paid to $n=2$ BHs. For $M_\star=10$~TeV, if the production temperature is above 6.7~TeV, BH accretion will saturate the extra dimensions at some point. After that, they behave as four dimensional BHs and continue accreting material. As the 4D Hawking temperature drops more swiftly than that of LED BHs, Eq.~\eqref{eq:dMdTsimp} indicates that the accretion will become much more efficient, and an asymptotic mass is missing in this scenario. These BHs may keep accreting until the plasma density is almost exhausted, indicated by the vertical line in Fig.~\ref{fig:Mas}. 

Next, we proceed to solve for the mass and number density of BHs produced in the primordial plasma. For more precise solutions to BH evolution that do not assume radiation domination (\textit{e.g.}, the BH density could be large enough to affect the expansion rate of the Universe $H(t)$), we must solve a set of coupled integro-differential equations detailed in Appendix~\ref{sec:massspecapp}. Numerical study of these equations shows that, if the BHs are able to accrete, their mass distribution function will always be very close to a monochromatic spectrum. This can be understood qualitatively, as the evolution follows two broad scenarios. 

For high reheating temperatures ($T_{\rm RH}\gtrsim T_{\rm th}$), collisional production of BHs is efficient, and the high plasma density ensures rapid accretion. As seen in Fig.~\ref{fig:Mas}, BH masses quickly approach $M_{as}$ in a radiation dominated universe until they drain a significant fraction of energy density from the radiation bath, and the rapid cooling of the plasma suppresses the subsequent production of BHs. Here, the first BHs are created approximately with an initial number density $n_i\simeq \rho_{r}(T_{\rm RH})/M_{as}(T_{\rm RH})$. As they grow and dominate the energy budget of the Universe, the collisional production of lighter BHs is severely limited. The accreted BHs eventually decay and dump energy into the plasma. However, they must not imprint on any cosmological observations as the dominant component of the Universe. It follows that these BHs will decay before BBN and lead to an early matter domination era.

In the second scenario, the BHs initially produced at a mass $M_i>M_{i,\min}$ accrete but their energy density remains inferior to radiation density until $T\lesssim$~eV. 
In a radiation dominated universe all BHs are able to accrete to a mass close to $M_{as}$.  The second scenario usually happens at $T_{\rm RH}<T_{\rm th}$, otherwise BHs will be overpopulated. Similarly to the first scenario, as the expansion of the Universe cools the plasma, BH production will also cease quickly since it is kinematically suppressed  by $M_{i,\min}/T$. The initial BH number density is therefore given by $n_i=\int_{t(T_f)}^{t(T_{\rm RH})}dt\int d\Mbh d\Gamma/d\Mbh$. Because of the suppression, the choice of final production temperature $T_f$ does not change $n_i$ so long as $d\Gamma/d\Mbh(T_f)~\ll~d\Gamma/d\Mbh(T_{\rm RH})$. The transition between these two scenarios happens at a reheating temperature $T_{\rm RH}^c$ which satisfies
\begin{equation}
    \int_{T_f}^{T_{\rm RH}^c} dT \int_{M_{i,\min}}^\infty d\Mbh  \dfrac{d\Gamma}{d\Mbh}\left(\dfrac{dT}{dt}\right)^{-1}\simeq \dfrac{\rho_{r,\rm RH}}{M_{as}(T_{\rm RH}^c)}\,.
    \label{eq:nicond}
\end{equation}
Below $T_{\rm RH}^c$, $n_i$ is given by the left hand side of the equation, and above that $n_i$ is determined by the right. In both scenarios, the time or temperature window for BH production is extremely limited, and BHs created during that time always accrete to similar masses, leading to a distribution that is very near to a delta function. Consequently, the integro-differential equations in Appendix~\ref{sec:massspecapp} can be greatly simplified to
\begin{align}
     \dfrac{dn_\bullet}{dt}&=-3Hn_\bullet\,,\label{eq:dndtsimp}\\
     \dfrac{dT}{dt}&=-T\left(H+\dfrac{n_\bullet}{4\rho_r}\dfrac{d\Mbh}{dt}\right)\,,\label{eq:dTdt_delta}\\
      H^2&=\dfrac{8\pi}{3M_{pl}^2}(\rho_r+\rhobh)\,,\label{eq:Friedmann_delta}
\end{align}
with $d\Mbh/dt$ given by Eq.~\eqref{eq:dMdtcombine} and $\rhobh=\Mbh n_\bullet$. To solve the equations, we assume the instant production of BHs with number density $n_i$ determined by the left and right of Eq.~\eqref{eq:nicond}, contingent on the reheating temperature. We assume the all BHs are born with the minimum mass $M_i=M_{i,\min}(T_{\rm RH})$. We then evolve the BH mass and number density as a function of time, including both accretion and evaporation. \cref{eq:dndtsimp,eq:dTdt_delta,eq:Friedmann_delta} reproduce the BH mass and energy density quite precisely for low and intermediate reheating temperatures, as can be seen from Figs.~\ref{fig:rhooft_exact} and~\ref{fig:hM_exact} in the appendix. At very high reheating temperature, the production of microscopic BHs becomes more efficient than BH accretion, and BHs may not reach the asymptotic mass. The precise solution of BH spectrum and mass evolution in this scenario is quite involved, which we leave for future work.  Two caveats remain for this approach. First, the connection between the first and second scenarios may not be entirely smooth as we have assumed an abrupt transition. Second, Eq.~\eqref{eq:dTdt_delta} assumes the entropy from BH evaporation is all dumped to the radiation plasma and thermalizes instantaneously. A dedicated study, including the effects of particle decoupling and non-thermal injection, is left for future work.

%\cref{eq:dndtsimp,eq:dTdt_delta,eq:Friedmann_delta} reproduce the BH mass and energy density quite precisely, as can be seen from Figs.~\ref{fig:rhooft_exact} and~\ref{fig:hM_exact} in the appendix. Two caveats remain for this approach. First, the connection between the first and second scenarios may not be entirely smooth as we have assumed an abrupt transition. Second, Eq.~\eqref{eq:dTdt_delta} assumes the entropy from BH evaporation is all dumped to the radiation plasma and thermalizes instantaneously. A dedicated study, including the effects of particle decoupling and non-thermal injection, is left for future work.

\begin{figure}
    \centering
    \includegraphics[width=0.48\textwidth]{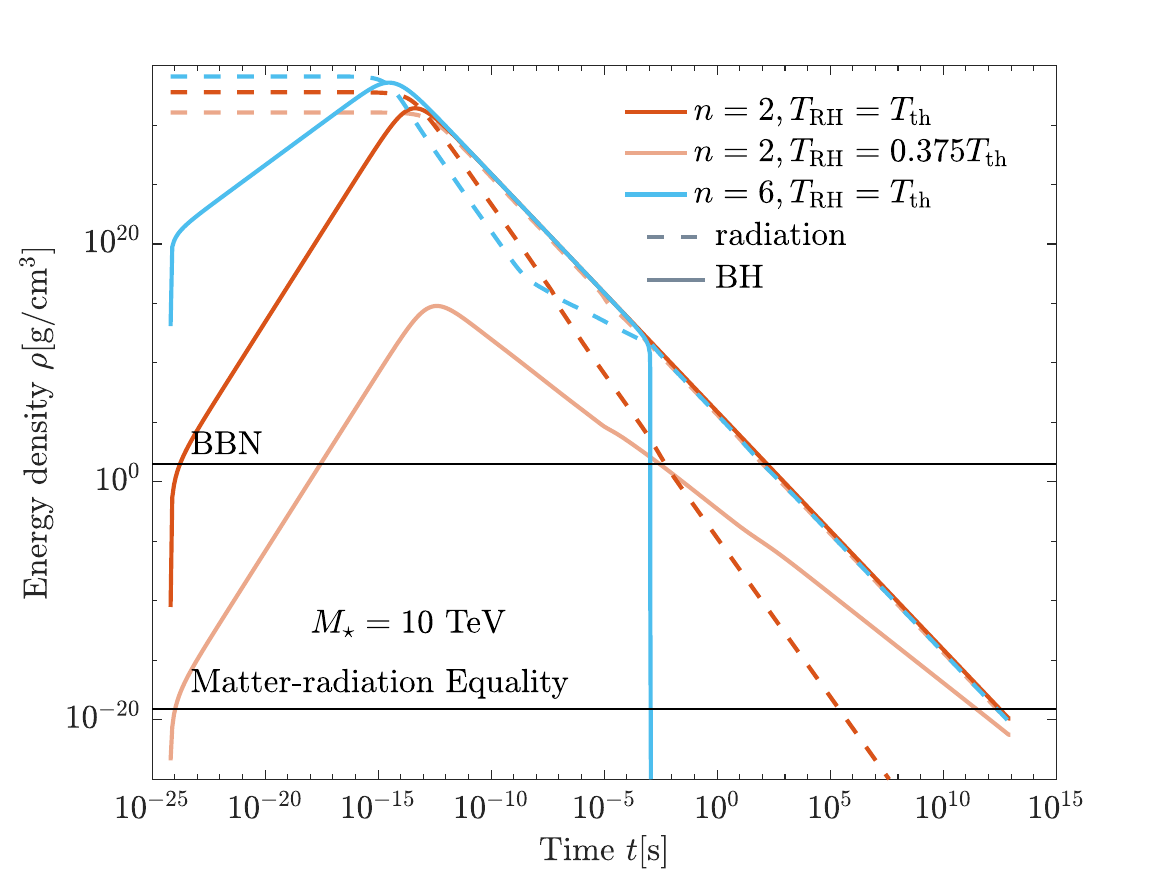}
    \includegraphics[width=0.48\textwidth]{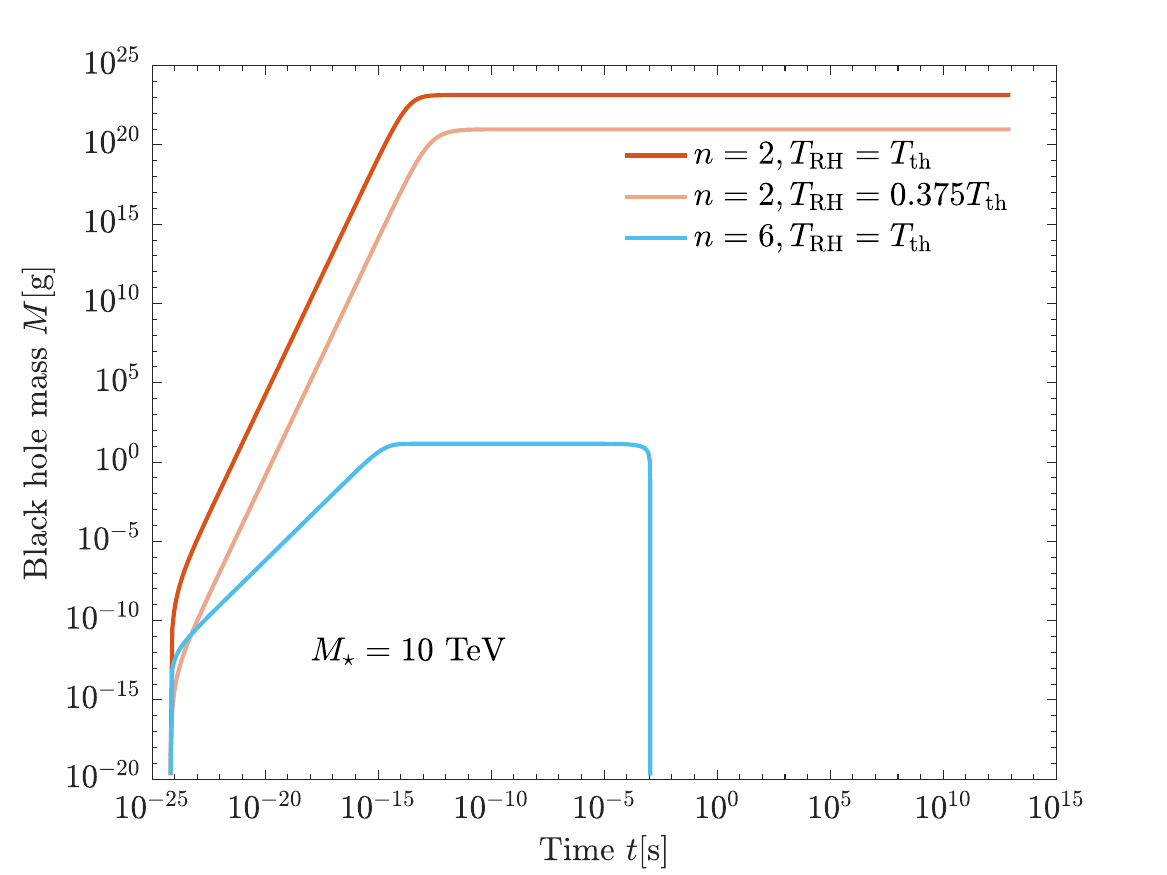}    
    \caption{\textit{Left}: Examples of BH and radiation energy density evolution over cosmological time. The orange and light orange lines correspond to $n=2$ extra dimensions with the reheating temperature $T_{\rm RH}=T_{\rm th}=2.9$~TeV and $T_{\rm RH}=0.375T_{\rm th}=1.09$~TeV respectively. The blue lines represent 6 extra dimensions and $T_{\rm RH}=T_{\rm th}=6.2$~TeV. Solid and dashed lines depict the energy density of BHs and radiation, respectively. Horizontal lines show the expected radiation density of the plasma when BBN and matter-radiation equality take place in the standard $\Lambda$CDM scenario. After matter-radiation equality, dark matter should dominate the expansion of the Universe; this is not included in this plot. We assume the fundamental scale $M_\star=10$~TeV. \textit{Right}: Evolution of BH mass as a function of time for the same scenarios as in the left panel.}
    \label{fig:rhooft}
\end{figure}

Examples of the solutions are shown in Figure~\ref{fig:rhooft}, in the presence of radiation and black holes only. For reference, we include horizontal lines that indicate the density at which BBN and matter-radiation equality occur in the standard $\Lambda$CDM scenario. For $n=2$ extra dimensions, if the reheating temperature $T_{\rm RH}=T_{\rm th}$, BHs dominate the Universe after a mere $10^{-15}$s, then their number density drops as the scale factor $a^{-3}$ while radiation is washed away, preventing standard Big Bang cosmology from unfolding. However, if $T_{\rm RH}=0.375T_{\rm th}$, the BH energy density remains subdominant until $10^{12}$s, when it becomes close to the radiation density near matter-radiation equality, behaving as cold dark matter should. We have not shown evolution past this time, since these illustrative models do not include a realistic treatment of baryons, dark energy, or an additional CDM component. 

For $n=6$ extra dimensions and $T_{\rm RH}=T_{\rm th}$, we still expect BHs to exhaust the radiation density promptly. However, these BHs do not survive until matter-radiation equality, their decay at about $10^{-3}$s replenishes the thermal bath, causing the temperature of the plasma to decrease less efficiently. BH production and decay lead to an era of {\it early matter domination}. 

Early matter domination before BBN typically does not leave any detectable features. However, the decay may produce gravitational waves which do not thermalize but still contribute to $N_{eff}$, or to the stochastic gravitation wave background to be discovered at more sensitive gravitational wave observatories. Only gravitons that are localized to the brane instead of propagating in the bulk will contribute to the stochastic gravitational waves. The greybody factor of these gravitons can be obtained by solving the wave equations of gravitons on the brane, which we leave for future work.

\begin{figure}
    \centering
    \includegraphics[width=0.7\textwidth]{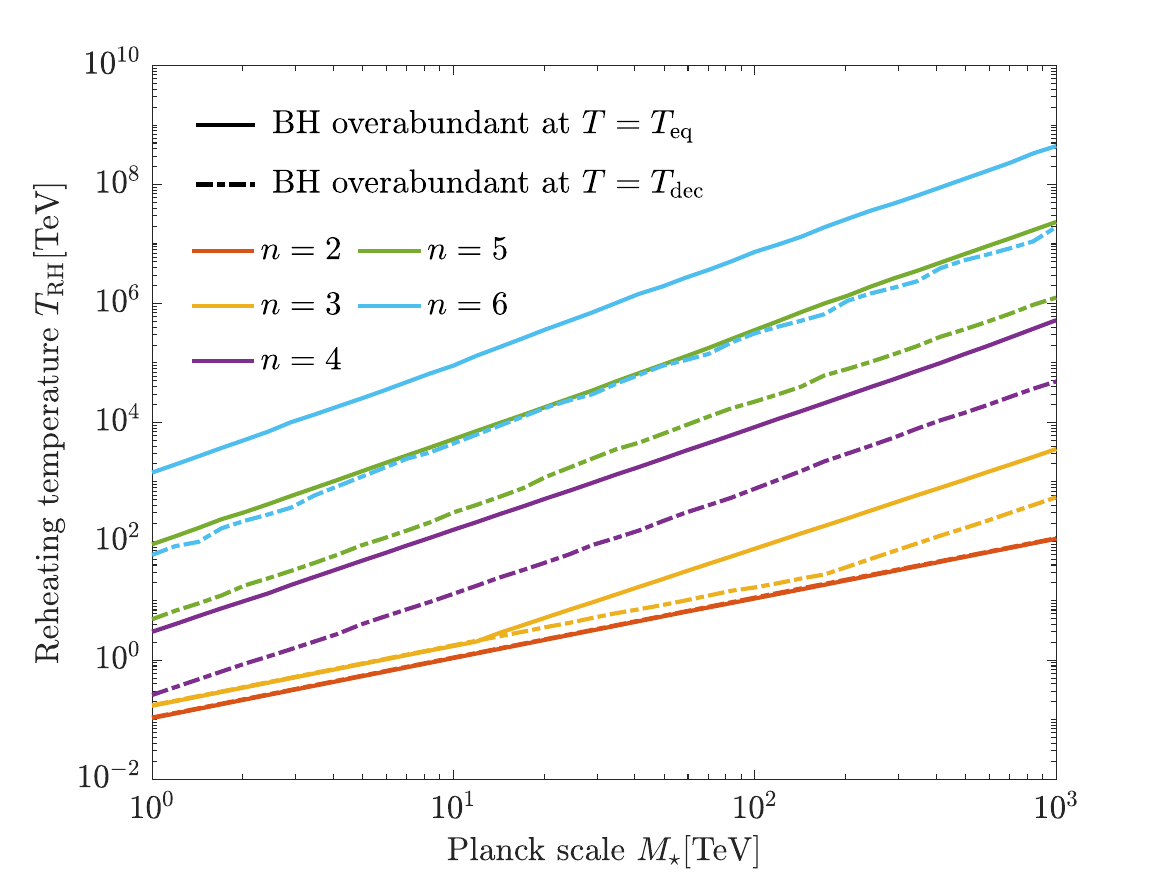}
    \caption{Constraints on the reheating temperature as a function of the fundamental Planck scale $M_\star$. Above the solid lines, the energy density of BHs $\rho_\bullet$ exceeds that of radiation at a photon temperature $T_{\rm eq}=0.75$~eV, when matter-radiation equality is expected. Above the dash-dotted line, $\rho_\bullet>10^{-3}\rho_r$ at the neutrino decoupling temperature $T_{\nu,\rm de}\simeq 2.33$~MeV. The regions above the lines are excluded due to BH distortions to BBN or CMB. For the Planck scales considered, $n=2$ BHs along the solid line will always survive until today and make the correct relic abundance, while $n=3$ BHs with $M_\star<1.4$~TeV still exist. The latter $n = 3$ range has been excluded by collider searches.}
    \label{fig:TRHvsMstar}
\end{figure}

Next, we vary the bulk Planck scale $M_\star$ and solve the evolution of BHs under different reheating temperatures $T_{\rm RH}$. We derive the constraints on $T_{\rm RH}$ based on two conditions: 1) As we will find more precisely in Sec.~\ref{sec:alterbbn}, if BHs survive until BBN terminates, the fraction of BH energy density must be less than $10^{-3}$ in the Universe at the neutrino decoupling temperature $T_{\rm dec}\simeq 2.33$~MeV. We conservatively require these BHs not to have evaporated significantly until 1~keV, far below the temperature when all nucleosynthesis processes freeze out. In other words, if BHs do not live long enough, they are not subject to this BBN constraint. 2) If BHs survive until the plasma temperature drops to about 0.75~eV, when matter radiation equality is expected in standard cosmology, BHs must remain subdominant in order not to change the sound horizon in a significant way, which would contradict CMB observations. The results are shown in Fig.~\ref{fig:TRHvsMstar}. The dash-dotted line corresponds to condition 1) and the solid line stems from condition 2). The regions above the lines are excluded. For $n=2$ these two conditions yield very similar constraints, while the BBN constraint tends to be stronger starting from $n=3$ when $T_{\rm RH}\gtrsim 30$~TeV. This behaviour can be understood intuitively from Fig.~\ref{fig:TBH}. As $n$ rises, BH lifetime decreases sharply. The reheating temperature has to be high enough to produce massive BHs that live until matter-radiation equality, rendering weaker constraints. The same applies to the BBN condition where the constraints on  $T_{\rm RH}$ are weaker for larger number of extra dimensions. For $n=2$, the solid line produces the right amount of BHs as $100\%$ dark matter, which remain until today. For $n\geq 3$, no reheating temperature is found such that BHs can dominate the dark matter density today for $M_\star\lesssim 10$~TeV. 

\begin{figure}
    \centering
    \includegraphics[width=0.48\textwidth]{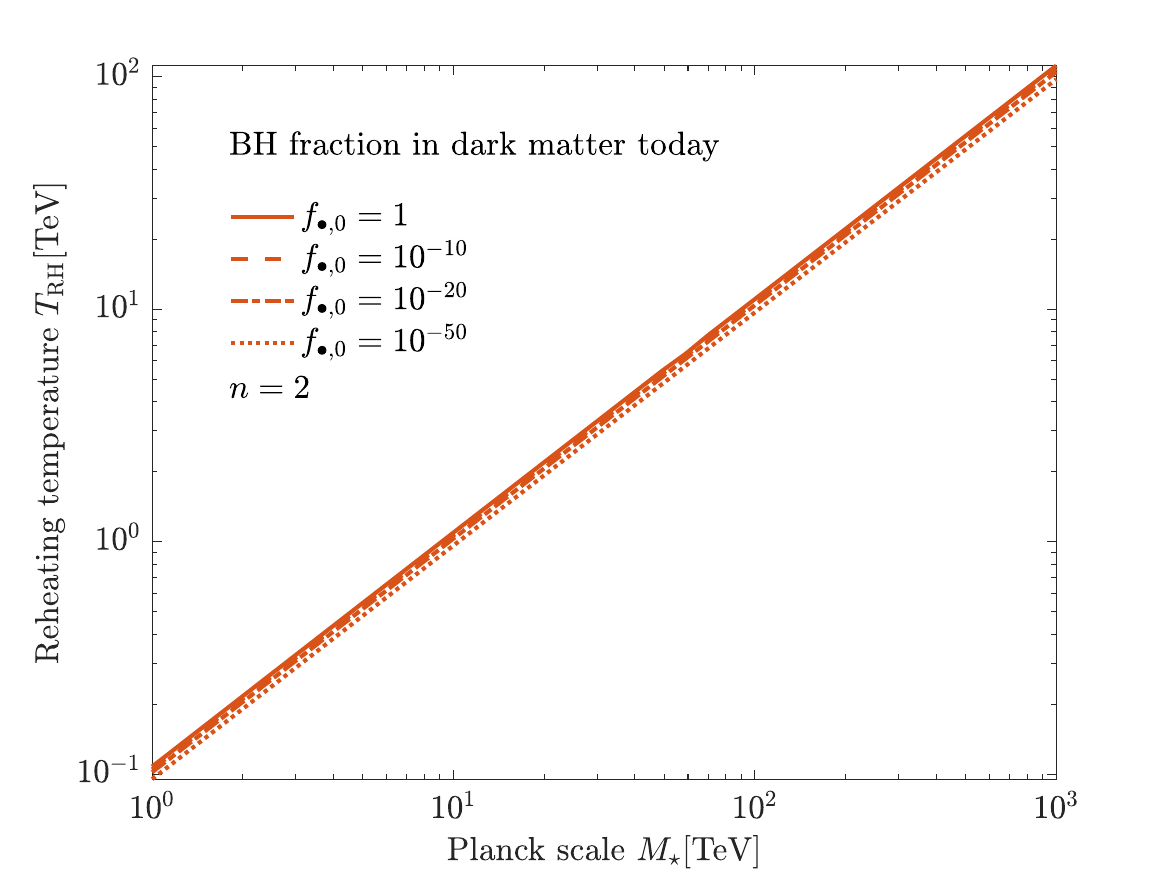}
    \includegraphics[width=0.48\textwidth]{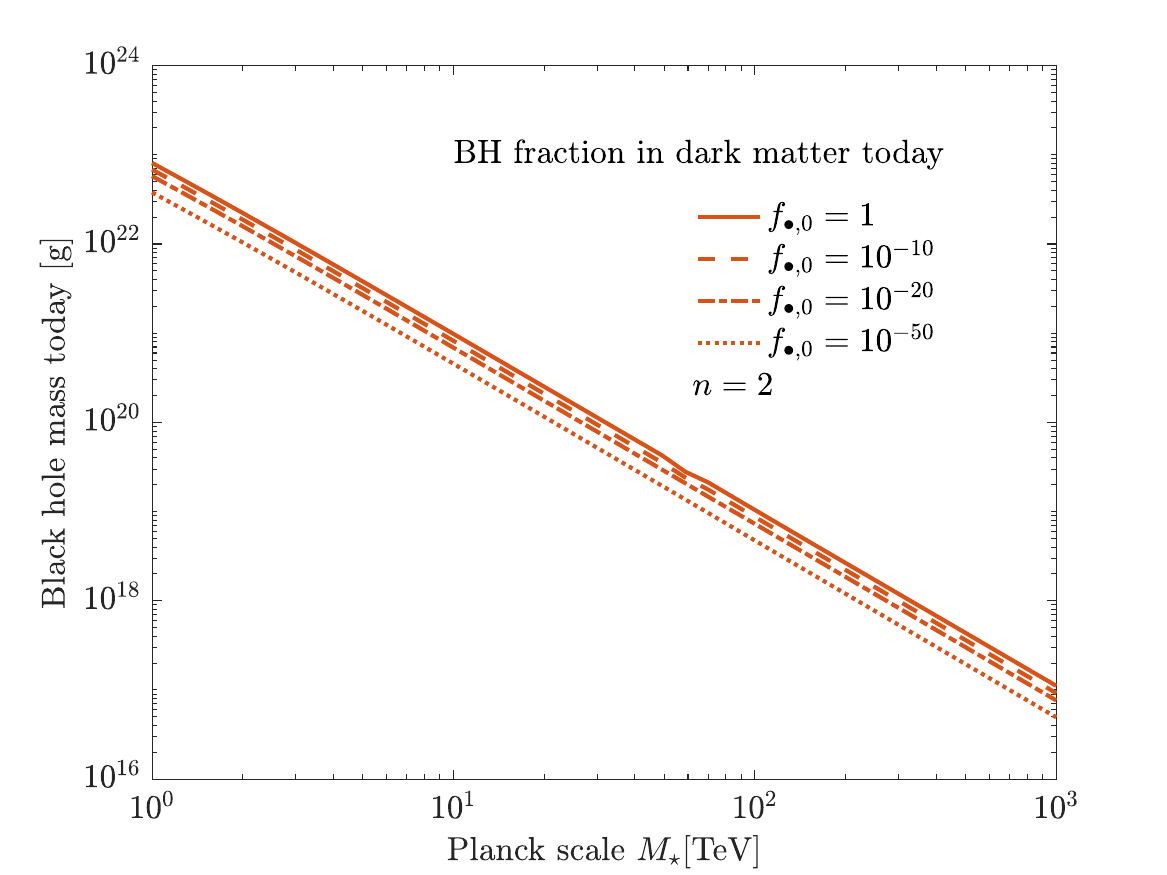}    
    \caption{{\it Left:} The reheating temperature $T_{\rm RH}$ as a function of the fundamental Planck scale $M_\star$, that produces primordial BHs as a fraction $\fbhToday$ of dark matter today, for $n = 2$ large extra dimensions. $\fbhToday$ corresponds to the observed relic density inferred by \textit{Planck}, $\Omega_{c}h^2 = 0.120$.   {\it Right:} Same as left, but showing the BH mass today instead. There is a one-to-one correspondence between $T_{\rm RH}$ in the left panel and BH mass today in the right panel.}
    \label{fig:BHtoday}
\end{figure}

To investigate LED BHs as part of the dark matter today, we therefore focus on $n=2$. We also add a flexible non-BH dark matter component to~\cref{eq:dndtsimp,eq:dTdt_delta,eq:Friedmann_delta} and evolve the energy density of dark matter over time. The non-BH dark matter energy density is adjusted such that the total cold dark matter density matches observations, $\Omega_c h^2=0.120$ while fixing the dark energy and baryon density to the Planck 2018 best fit~\cite{Planck:2018vyg}. We then solve for the fraction of dark matter today that is comprised of BHs, $\fbhToday\equiv \Omega_\bullet/\Omega_c$. We show the reheating temperature and BH mass today in Fig.~\ref{fig:BHtoday} that corresponds to a specific $\fbhToday$ by varying $M_\star$. Since the BH production rate is exponentially suppressed when $T_{\rm RH}\ll M_\star$, a minuscule change in the reheating temperature will alter $\fbhToday$ remarkably. The required reheating temperature to produce BH dark matter is roughly proportional to $M_\star$. As indicated in Eq.~\eqref{eq:asympmass}, the asymptotic BH mass, and hence the BH mass today $M\propto M_\star^{-2}$ with $n=2$. Indeed, the fit to Fig.~\ref{fig:BHtoday} reveals

\begin{equation}
    T_{\rm RH}=0.11 M_\star,\ \ M= 10^{23}{\rm g}\left(\dfrac{M_\star}{\rm TeV}\right)^{-2}\,.
\end{equation}
For $M_\star\gtrsim 10$~TeV, which evades collider constraints, the primordial BH mass today ranges from  $10^{17}$g to $10^{21}$g for Planck scales below a PeV.

\subsection{Observable evaporation products}
\label{sec:evapproducts}

In Sec.~\ref{sec:BHbasics} we have described the primary particles from BH evaporation. If the only important observable effects of BH evaporation are the change in BH mass and injection of energy into the plasma of the early Universe, then Eqs. \eqref{eq:emissionSM} and \eqref{eq:emissiongraviton} are sufficient. However, observable stable particles (here, photons, electrons and positrons) are also produced as decay or hadronization products from heavy and coloured primary states. To correctly account for production of these \textit{secondary} observable particles, we consider the contribution from several sources. As a first step, we use tabulated spectra from PPPC4DMID \cite{Cirelli:2010xx} to compute the secondary particle spectra generated from primary particles above $E_P=5$~GeV, which arises from the limitation of particle energy in \texttt{PYTHIA}~\cite{Sjostrand:2014zea}, used in PPPC4DMID for the production of tabulated values. Below this energy, the unstable states that we include are the $\tau^\pm$ leptons, muons and pions. We extrapolate the $\tau$ decay spectra from PPPC4DMID down to $E = m_\tau$. The $e^\pm$ and $\gamma$ spectra from $\pi$ and $\mu$ decay are computed and boosted to the lab frame in a similar way to Ref. \cite{Coogan:2019qpu}, taking care to include the electron mass where appropriate---see Appendix~\ref{sec:pimudecay} for details.
These are added to the primary electrons and photons below $E_P$ produced by the evaporating BH. 
Overall, the secondary spectra are computed as
\begin{equation}
    \dfrac{d^2N}{dE_idt}=\left.\dfrac{d^2N}{dE_idt}\right\vert_{E_i<E_P}+\sum\limits_j\int_{E_P}^{\infty}\dfrac{df_j}{dE_i}\dfrac{d^2N}{dE_jdt}dE_j+\int_{m_\tau}^{\infty}\dfrac{df_\tau}{dE_i}\dfrac{d^2N}{dE_\tau dt}dE_\tau+\sum\limits_k\int_{m_k}^{E_P}\dfrac{df_k}{dE_i}\dfrac{d^2N}{dE_kdt}dE_k\,,
    \label{eq:flux_sec}
\end{equation}
where $i=\{e^\pm,\gamma\}$, $j=\{e^\pm,\mu^{\pm},q\bar{q},W^\pm,Z,g,\gamma,h,\nu_e,\nu_\mu,\nu_\tau\}$, and $k=\{\mu^\pm,\pi^\pm,\pi^0\}$. The BH primary emission spectrum is
\begin{equation}
    \dfrac{d^2N}{dE_jdt}=-\dfrac{1}{E_j}\dfrac{dM^2_{\bullet\rightarrow j}}{dE_jdt}Q_j\,.
    \label{eq:flux_prim}
\end{equation}
To account for QCD confinement transition, we adopt the same prescription as in Ref.~\cite{Stocker:2018avm} and include a factor
\begin{equation}
    Q_j=\left[1+\exp\left(\pm\dfrac{1}{\sigma} \log_{10}\dfrac{T_H}{\Lambda_{\rm QCD}}\right)\right]^{-1}\,,
\end{equation}
where the plus sign applies for $\pi^\pm$ and $\pi^0$, and the minus sign for quarks and gluons. For all other species, $Q_j=1$. We take the confinement scale $\Lambda_{\rm QCD}\simeq 300$~MeV and $\sigma=0.1$. Below $\Lambda_{\rm QCD}$, the emission of quarks and gluons from BH will be exponentially suppressed and the emission of hadrons is preferred.

\begin{figure}
    \centering
    \includegraphics[width=0.48\textwidth]{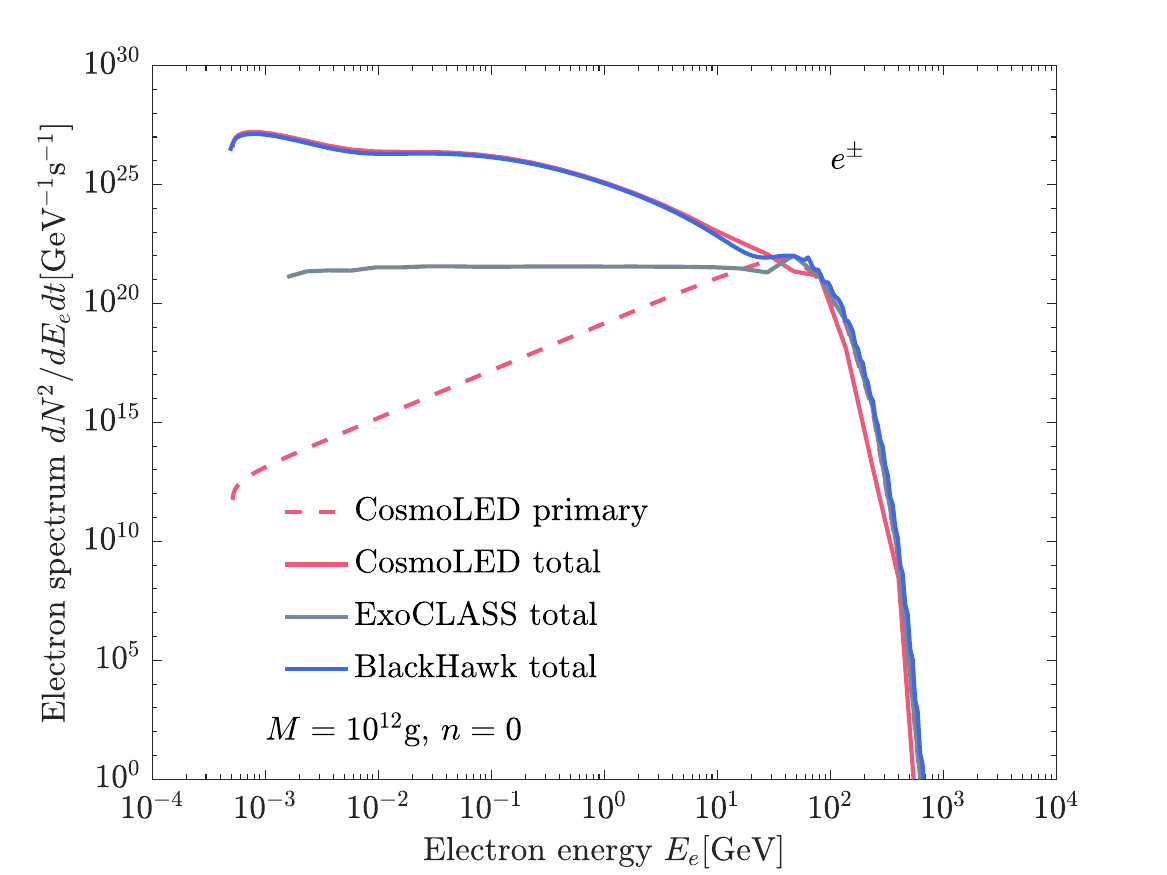}
    \includegraphics[width=0.48\textwidth]{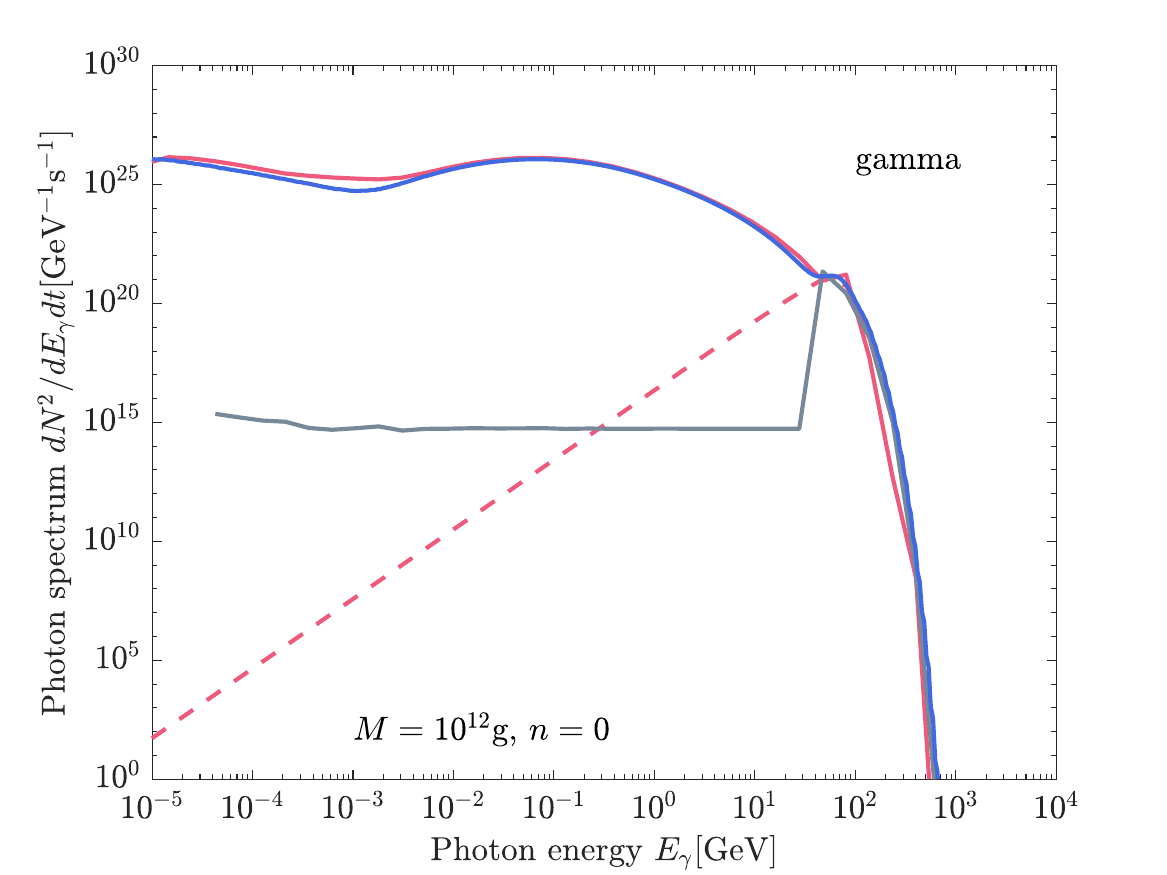}    
    \caption{The emission spectra of electron/positrons (left) and gamma (right) from the evaporation of a 4D BH with mass $\Mbh=10^{12}$~g. The dashed lines show the primary particles obtained directly from the greybody spectrum. The solid red, grey and blue lines show the total emission of $e^\pm$ and $\gamma$ including hadronization and decay, computed using \texttt{CosmoLED} (this work), \texttt{ExoCLASS}~\cite{Stocker:2018avm} and  \texttt{BlackHawk v2.1}~\cite{Arbey:2019mbc,Arbey:2021mbl}.}
    \label{fig:gammaande}
\end{figure}

For comparison with prior work, we show the emission spectra of $e^{\pm}$ and $\gamma$ in Fig.~\ref{fig:gammaande}, for 4D ($n = 0$, $M_\star = M_{pl}$) black holes. Our code, \texttt{CosmoLED}, computes the spectra of observable products from BHs, and the cosmological constraints. The dashed lines are the primary spectra obtained from Eqs.~\eqref{eq:flux_prim} and~\eqref{eq:emissionSM}. The solid lines depict the total spectra of $e^{\pm}$ and $\gamma$ by considering the decay and hadronization of more energetic particles. The \texttt{CosmoLED} total spectra are computed using Eq.~\eqref{eq:flux_sec}. For comparison, we also show the spectra obtained directly from the \texttt{ExoCLASS} package~\cite{Stocker:2018avm}, and \texttt{BlackHawk v2.1}~\cite{Arbey:2019mbc,Arbey:2021mbl}. Note that the \texttt{ExoCLASS} BH module computes the secondaries from muon and pion decay only, and \texttt{BlackHawk} cascades down from 5 to $10^5$~GeV primary particles with the ``\texttt{PYTHIA}'' hadronization option at the present epoch. Our results agree well with \texttt{BlackHawk} at almost all energies, while \texttt{ExoCLASS} tends to underestimate the secondary spectra. As BH mass increases, the difference between \texttt{CosmoLED} and \texttt{ExoCLASS} spectra becomes less dramatic as fewer primary particles are produced above 5~GeV. However, the \texttt{CosmoLED} spectra remain to be larger in most of the energy range. Throughout, we assume that BHs evaporate only to standard model particles and gravitons.

\section{Observational constraints on LED black holes}
\label{sec:constraints}
Once produced, primordial black holes born of microscopic collisions in the early Universe will exhibit similar phenomenology to their four-dimensional cousins. In addition to affecting the energy budget of the Universe, their evaporation products will affect cosmological evolution and can interfere with Big Bang Nucleosynthesis (BBN) and the CMB, as well as produce a detectable flux of galactic and extragalactic X-rays. These constraints will not probe 4D BH masses larger than $\sim 10^{18}$ g, and thus do not overlap with constraints from lensing and dynamical disruption of gravitational systems. In this section, we compute the dominant constraints from X-rays (Secs. \ref{sec:galactic} and \ref{sec:extragalactic}), the CMB (Sec. \ref{sec:CMB}) and BBN (Sec. \ref{sec:BBN}), first describing the physics, and then producing constraints from observational data. We then discuss the combined constraints (Sec. \ref{sec:fullconstraints}) as well as previous PBH constraints not studied in this work (Sec. \ref{sec:otherConstraints}).

In order to consistently compare constraints on PBHs with differing lifetimes, we define the parameter $\fbh$ as
\begin{equation}
    \fbh \equiv \frac{\rho_\bullet(z_i)}{\rho_{DM,0}(1+z_i)^3}
    \label{eq:fbh}
\end{equation}
where $\rho_\bullet$ is the density of PBHs at an initial redshift, $z_i$, before the PBHs evaporate any significant fraction of their mass and $\rho_{DM,0}$ is the observed dark matter density today. With this definition, $\fbh$ describes the fraction of dark matter in the early Universe comprised of PBHs. For certain observable constraints, other parameters are used to describe the abundance of PBHs. When studying galactic centre constraints we use $\fbhToday$, the fraction of dark matter comprised of PBHs today and when studying the impact PBHs have on the expansion history near BBN we use $\beta_{\rm dec}$, the fraction of the total energy density comprised of PBHs at the time of neutrino decoupling.

%----------------------------------------------------------
%----------------------------------------------------------
%----------------------------------------------------------

\subsection{Galactic constraints}
\label{sec:galactic}
For PBHs that survive until the present, the Milky Way halo is a promising source of evaporation products. Detectable sub-GeV evaporation products can consist of gamma rays, and cosmic ray electrons, positrons, protons and antiprotons. The ``prompt'' gamma ray flux is given by:
\begin{equation}
    \frac{d \Phi_\gamma}{dE d\Omega} = \frac{1}{4\pi} \frac{dN}{dEdt} \frac{\fbhToday}{\Mbh} \frac{1}{\Delta \Omega} \mathcal{D}(\Omega),
    \label{eq:GCgammas}
\end{equation}
where the $\mathcal{D}$-factor is defined as an integral over the dark matter density $\rho_{DM}(\vec x)$:
\begin{equation} \label{eq:Ddefinition}
    \mathcal{D}(\Omega) \equiv \int_{\mathrm{l.o.s.} \Delta \Omega} \rho_{DM}(\vec{x}) d\Omega dx,
\end{equation}
where the integral in $x$ is over the line of sight (l.o.s.) and $\Delta \Omega$ is the solid angle of interest. $\rho_{DM}$ is the DM density in the Milky Way. We take it to follow an NFW profile
\begin{equation}
    \rho_{DM}(r) = \rho_s \frac{2^{3-\gamma}}{\left(\frac{r}{r_s}\right)^\gamma \left(1+ \frac{r}{r_s}\right)^{3-\gamma}}.
    \label{eq:NFW}
\end{equation}
where $r$ is the galactocentric distance, $r_s$ is the DM halo scale radius, and the conventional the factor of $2^{3-\gamma}$ ensures that $\rho_s \equiv \rho(r_s)$. We employ parameters consistent with kinematic data \cite{deSalas:2019}\footnote{In Ref.~\cite{deSalas:2019}, best fit values for the Milky Way halo profile for two separate models of the baryonic component of the galaxy. For this work we adopt the best fit values that correspond to modelling the stellar disk, dust, and gas components as a double exponential. It should also be noted that there is a large uncertainty on the dark matter halo parameters, especially $\gamma$ and $r_s$. A complete analysis should marginalize over the posterior likelihood of the halo density distribution however for the purpose of setting constraints we have held all halo parameters fixed at their best fit values. For an overview of the various determinations of $\rho_0$, see the review in Ref.~\cite{deSalas:2020hbh}.} , $r_s = 9$ kpc and $\gamma = 1.2$. The DM density at the Sun's position is $\rho_0 = \rho(R_\odot) = 0.3$ GeV cm$^{-3}$, where we use recent measurements from GRAVITY \cite{Abuter:2018drb} for the distance to the galactic centre $R_0 = 8.127$ kpc.

In addition to the gamma ray flux from Eq. \eqref{eq:GCgammas}, low-energy positrons produced by BH evaporation will lead to a gamma ray line signal at $E_\gamma = 511$ keV from $e^+e^-$ annihilation in the interstellar medium. The flux of photons from in-situ $e^+e^-$ annihilation is:
\begin{equation}
     \frac{d \Phi_{511}}{ d\Omega}  = 2(1-0.75 f_P) \frac{dN_{e^+}}{dt}  \frac{1}{4\pi} \frac{1}{\Mbh} \frac{1}{\Delta \Omega} \mathcal{D}(\Omega)
\end{equation}
where $f_P$ is the positronium formation fraction and ${dN_{e^+}}/{dt}$ is the total positron production rate per BH integrated over energy.

We employ data from  INTEGRAL/SPI, the X/gamma-ray spectrometer onboard the ESA INTEGRAL satellite, launched in 2003. A full analysis of SPI data requires a template-based likelihood analysis, as there is no way to reconstruct the direction of a single photon event. Rather, SPI uses a coded mask, for which each individual photon recorded on the detector corresponds to a number of possible trajectories. This means that an image cannot be reconstructed, and one must instead compare templates using a maximum likelihood method. To sidestep this cumbersome process, we use previously-processed data reported in Ref. \cite{Bouchet2011}. Although this is based on only 6 years ($\sim10^8$ s) of data, it is the only published reference to include a binned reconstruction of the diffuse flux as a function of energy and galactic latitude and longitude. We follow a similar method to Ref. \cite{Laha:2020ivk}, who used this data to constrain 4D primordial black holes in the Milky Way. We employ the 5 energy bins in Figure 5 of Ref. \cite{Bouchet2011} (digitized from \cite{Cirelli:2020bpc}),  corresponding to 27-49 keV, 49-90 keV, 100-200 keV, 200-600 keV and 600-1800 keV. These each consist of 21 latitude bins within $-90^\circ < b < 90^\circ$, integrated over longitudes $-23.1 < \ell < 23.1^\circ$, with the exception of the 800-1800 keV range, which is presented in 15 bins, within $-60^\circ < \ell < 60^\circ$. We do not employ the results from Figure 4, as they are drawn from the same data, but binned over latitude instead.  
We construct a one-sided chi-squared statistic, and obtain 95\% confidence limits assuming one degree of freedom. Our limits agree with those presented by Laha \textit{et al.} \cite{Laha:2020ivk} in the $n = 0$ case, who instead ask that the predicted flux in every bin does not exceed the measurement by more than 2 times the reported error in that bin; using both methods, we have checked that our chi-squared approach yields identical results to the Laha \textit{et al.} method, except above $M \sim 1.2 \times 10^{17}$,  where our constraints are stronger by a factor of a few. At lower masses, small differences with respect to the Laha \textit{et al.} results can be attributed to a different choice of dark matter halo parameters. Our results are also similar to the very recent \cite{Auffinger:2022dic}. While their addition of Fermi and EGRET data may strengthen bounds at lower masses, they may still be superseded by the 511 keV bounds that we discuss next.

For the 511 keV signal, we may use more recent data. We have taken the binned 511 keV flux shown in Fig. 5a of Ref.~\cite{Siegert:2019tus} (black crosses). These correspond to the total 511 keV flux within galactic latitudes $-10.5^\circ < b < 10.5^\circ$, in 5 equally-spaced longitude bins within $-30^\circ < \ell < 30^\circ$. We again produce a one-sided chi-squared, in order to establish $2\sigma$ limits on the BH fraction via Wilks' theorem. As in Ref.~\cite{DeRocco:2019fjq}, we conservatively only consider positrons with energies less than 1 MeV, as high-energy particles may not annihilate in-situ.

\begin{figure}
    \centering
    \includegraphics[width= \textwidth]{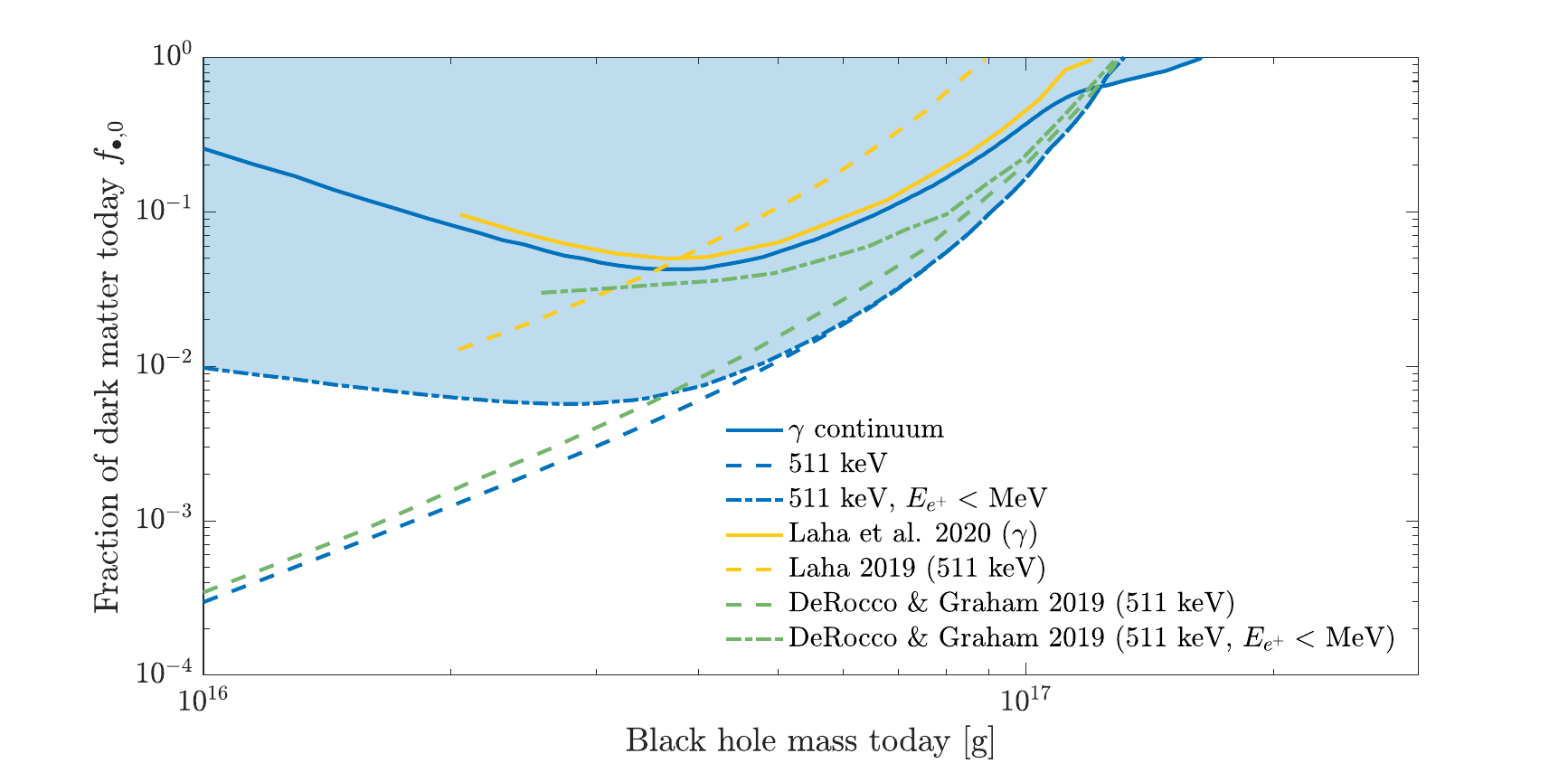}
    \caption{Updated constraints on ``ordinary'' four-dimensional primordial black holes from INTEGRAL/SPI gamma-ray data. Solid blue: using gamma-ray continuum data from \cite{Bouchet2011}; dashed blue: 511 keV line from $e^+e^-$ annihilation, using data reported in \cite{Siegert:2019tus}; dot-dashed blue: 511 keV constraints, but omitting the flux from positrons with energies higher than $1$ MeV which may not annihilate in-situ.  Prior results are shown from Laha \textit{et al.} 2020 \cite{Laha:2020ivk}, Laha 2019 \cite{Laha:2019ssq} and DeRocco \& Graham 2019 \cite{DeRocco:2019fjq}. }
    \label{fig:4Dgalactic}
\end{figure}

Since our method slightly improves on previous results, we first show the resulting limits for the $n = 0$, ordinary 4D PBH case in blue, in Fig. \ref{fig:4Dgalactic}. Continuum gamma-ray constraints are presented as solid lines, dash-dotted lines show the 511 keV limits from positron annihilation, and the dashed lines present the same limits, but without the $E_{e^+} < 1$ MeV requirement. We also show the aforementioned gamma-ray limits of Laha et al. \cite{Laha:2020ivk} (solid yellow), as well limits based on evaporation to positrons obtained by Laha \cite{Laha:2019ssq} and DeRocco \& Graham \cite{DeRocco:2019fjq}. When using the full range of positron energies, we attribute the slight improvement over  DeRocco \& Graham to the use of more recent data and angular information from \cite{Siegert:2019tus}. The stronger improvement comes when comparing the $E_{e^+} < $MeV cases: here, our inclusion of secondary particles leads to a sizeable flux of low-energy positrons not present when only primary thermal particles are accounted for---as can be read e.g. from the left-hand panel of Fig.~\ref{fig:gammaande}.

\begin{figure}[ht]
    \centering
    \includegraphics[width=0.5\textwidth]{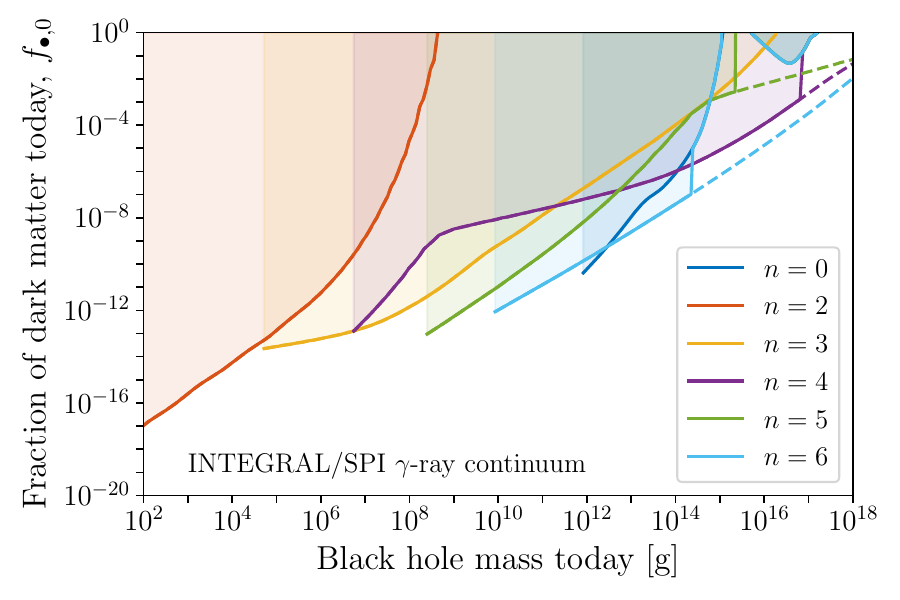}\includegraphics[width=0.5\textwidth]{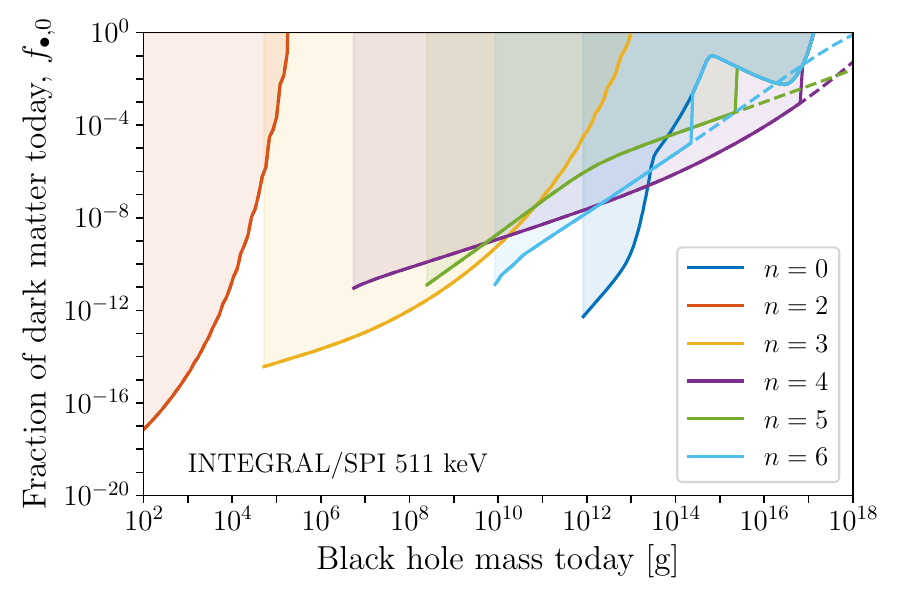}
    \caption{Constraints on the fraction of LED primordial black holes \textit{today} in LED scenarios as a function of their mass, based on data from INTEGRAL/SPI as discussed in Sec. \ref{sec:galactic}, where $n$ is the number of extra dimensions, and $n = 0$ corresponds to classical 4D black holes. The dashed lines indicate what the constraints would be if large black holes did not act like 4D black holes. \textit{Left}: constraints from continuum emission of primary and secondary photons. \textit{Right}: constraints from the 511 keV line flux produced from annihilation of positrons produced by decaying BHs.}
    \label{fig:GCconstraints}
\end{figure}

Constraints for $n \ge 0$ are shown in Fig. \ref{fig:GCconstraints}. We arbitrarily cut the mass range to include BHs that would survive for at least 10 years, the approximate duration of the INTEGRAL mission (hence the large difference Fig.~\ref{fig:4Dgalactic}, which corresponds to BHs that would live for $\sim$ the age of the Universe or longer). Masses in the lower range are obviously ``tuned'' to end their lifetimes around the present day and correspond to a small sliver of initial BH masses. We will translate these constraints into cosmologically-consistent bounds in Sec. \ref{sec:fullconstraints}. 

Depending on whether the Hawking temperature is high enough to produce positrons, and where the gamma-ray spectrum peaks, gamma-ray (left panel) and positron (right panel) constraints dominate for different values of $\Mbh$ for different $n$.  The sharp vertical jump at the right-hand side of some constraints corresponds to the transition from $4+n$-dimensional to 4-dimensional behaviour of the PBHs as they saturate the extra dimensions---\textit{i.e.} masses above $\Mbh = M_{\rm 4D}$ \eqref{eq:m4d}. We indicate with dashed lines the constraints that would be attainable in the absence of such a transition. 

%----------------------------------------------------------
%----------------------------------------------------------
%----------------------------------------------------------

\subsection{Isotropic background light}
\label{sec:extragalactic}
The isotropic photon spectrum can be split into two observationally indistinguishable components. One component is the extragalactic background light (EBL) produced by extragalactic PBHs homogeneously distributed throughout the Universe. The EBL component  has previously been used to constrain the abundance of extra-dimensional PBHs \cite{Johnson:2020tiw}. The other component is the isotropic part of the galactic signal, produced by PBHs within the galactic halo. Despite, the galactic halo being anisotropic (as discussed in the previous section), there is a non-zero flux in all directions. Therefore, there appears to be an isotropic component equivalent to the flux in the direction with the smallest contribution from galactic PBHs. This isotropic galactic signal has recently been used to constraint the abundance of long-lived four-dimensional PBHs \cite{Iguaz:2021irx,Chen:2021ngo}.

\subsubsection{Extragalactic photon flux} \label{sec:EBL}
The sum of the evaporation products from all extragalactic PBHs  could produce a significant isotropic flux of X-rays or gamma rays. This signal depends on the primary spectrum of photons, electrons and positrons described in Eq.~\eqref{eq:emissionSM} as well as the secondary spectrum described in Sec.~\ref{sec:evapproducts}. As the evaporation products travel from the point of evaporation to Earth, the flux changes due to the photons redshifting, being absorbed, and scattering with IGM material. By taking into account all of these processes, whose relative importance is a function of energy and redshift, we will obtain a predicted EBL flux that may be constrained by observations. 

The EBL contribution to the isotropic photon flux can be found by evolving the photon spectrum over time starting at recombination. At any given redshift, $z$, the change in the flux of photons of energy $E$ can be parameterized by
\begin{equation} \label{eq:photonFluxTotalChange}
\frac{d\Phi_{\gamma,\textrm{EBL}}}{dEdz}(E,z) =\sum_i \frac{d\Phi_{\gamma,{i}}}{dEdz}(E,z),
\end{equation}
where $\Phi_{\gamma,\textrm{EBL}}$ is the extragalactic isotropic photon flux and $i$ denotes the four different channels for energy injection and loss: Universe expansion, photon absorption, Compton scattering, and photon injection.

The {\it expansion of the Universe}  redshifts photon energy and dilutes their number density. As shown in Appendix \ref{sec:RSFluxChangeDerivation}, these effects may be combined into:
\begin{equation} \label{eq:photFluxChangeRS}
\frac{d\Phi_{\gamma,\textrm{exp}}}{dEdz}(E,z) = \frac{2}{1+z}\frac{d\Phi_{\gamma,\textrm{EBL}}}{dE}(E,z) - \frac{E}{1+z}\frac{d^2\Phi_{\gamma,\textrm{EBL}}}{dE^2}(E,z).
\end{equation}
This results in the flux per unit energy being diluted as $(1+z)^2$, as the photon number density is diluted as $(1+z)^3$ while the spectral density removes a factor of $(1+z)$. Although Eq.~\eqref{eq:photFluxChangeRS} depends on the derivative of $d\Phi/dE$, the discretized method that we use (Appendix \ref{sec:numericEBL}) does not actually require numerical differentiation.

The processes that cause the {\it absorption of photons} are: photoionization of neutral gas, pair production from atoms and ions, photon-photon scattering, and pair production off the CMB. All of these processes either absorb a photon or remove almost all of a photon's energy. The change in photon flux due to these absorption processes is
\begin{equation} \label{eq:photFluxChangeABS}
\frac{d\Phi_{\gamma,\textrm{abs}}}{dEdz}(E,z) = \frac{d\tau}{dz}(E,z)\frac{d\Phi_{\gamma,\textrm{EBL}}}{dE}(E,z).
\end{equation}
where $\frac{d\tau}{dz}(E,z)$, as determined in \cite{zdziarski1989absorption}, is the optical depth of a photon of energy $E$ over a differential redshift step at redshift $1+z$.

Absorption of photons causes an initial flux of photons starting at redshift $z_i$ and travelling to a final redshift $z_f$ with final energy $E$  to be suppressed by an exponential factor of $e^{-\tau(E,z_i,z_f)}$ where
\begin{equation}
    \tau(E,z_i,z_f) = \int_{z_i}^{z_f} dz \frac{d\tau}{dz}(E \frac{1+z}{1+z_f}, z).
\end{equation}

For photons with energies between $\sim$ 1 keV and $\sim$ 10 GeV the Universe is transparent ($\tau < 1$) up to redshifts of order $z\sim100$. However, for photon fluxes that originate at higher redshifts, a large fraction of the photons may be absorbed.

High-energy photons can also {\it Compton scatter} with electrons, losing some amount of energy, without being entirely absorbed. The instantaneous change in photon flux due to Compton scattering is calculated as the sum of a negative loss term that accounts for the attenuation of photons of a given energy and a positive source term that accounts for all the higher-energy photons downscattered to that energy. This is given as
\begin{equation} \label{eq:ComptonNetChange}
\frac{d\Phi_{\gamma,\textrm{comp}}}{dEdz}(E,z) = \frac{1}{(1+z)H(z)}\bigg( n_e \sigma_c(E) \frac{d\Phi_{\gamma,\textrm{EBL}}}{dE}(E,z) - n_e \int d\tilde{E} \frac{d\sigma_c(\tilde{E})}{dE} \frac{d\Phi_{\gamma,\textrm{EBL}}}{d\tilde{E}}(\tilde{E},z)\bigg),
\end{equation}
where $H(z)$ is the Hubble parameter, $n_e$ is the total electron density, which includes electrons bound in hydrogen and helium as the small ionization potentials do not distinguish those from free electrons (see e.g. \cite{Chen:2003gz,Sunyaev:1996} for more discussion), $\sigma_c$ is the total Compton cross section, and $\frac{d\sigma_c(\tilde{E})}{dE}$ is the differential cross section of an incoming photon with energy $\tilde{E}$ scattering and losing energy so that it ends up with an outgoing energy $E$. 

Solving this integro-differential equation is computationally slow, and the effect of Compton scattering is often approximated either as an absorption process which contributes to Eq.~\eqref{eq:photFluxChangeABS} or as a process that causes all photons to continuously lose some fraction of their energy in a similar way to the expansion of the Universe. For scenarios where Compton scattering is important we utilize the full integro-differential equation. A discussion of the different computation schemes and more details on how Compton scattering was numerically calculated in this work can be found in Appendix \ref{sec:numericEBL}.

The differential Compton cross section is typically given in the rest frame of the electron in terms of the scattering angle $\theta$ by the Klein-Nishina equation 
\begin{equation} \label{eq:KleinNishina}
    \frac{d\sigma_c (\tilde{E})}{d\cos\theta} = \frac{\pi \alpha^2}{m_e^2} \frac{E}{\tilde{E}} \bigg(\frac{E}{\tilde{E}} + \frac{\tilde{E}}{E} - \sin^2\theta    \bigg),
\end{equation}
whereas $d \sigma_c/dE$ is required to solve Eq.~\eqref{eq:ComptonNetChange}. Here, $\alpha$ is the fine-structure constant, $m_e$ is the electron mass, and the outgoing photon energy $E$ is related to the incoming energy $\tilde{E}$ and  $\theta$ via
\begin{equation} \label{eq:ComptonEtheta}
    E = \frac{\tilde{E} }{ 1 + \frac{\tilde{E}}{m_e}(1 - \cos\theta)  }.
\end{equation}
The differential Compton cross section with respect to outgoing photon energy is thus
\begin{equation}
    \frac{d\sigma_c (\tilde{E})}{dE} =      \frac{m_e}{E^2}    \frac{d\sigma_c (\tilde{E})}{d\cos\theta}.
\end{equation}
The integration bounds in Eq. \eqref{eq:ComptonNetChange} are found by noting $-1 \le \cos\theta \le 1$ and translating that to a range of $\tilde{E}$ using Eq. \eqref{eq:ComptonEtheta}.

The total Compton cross section at a given energy, $E$, is \cite{rybicki2008radiative} 
\begin{equation}
    \sigma_c(E) = \sigma_T \frac{3}{4}\bigg[ \frac{1+x}{x^3}\bigg(\frac{2x(1+x)}{1+2x} - \ln(1+2x) \bigg) +\frac{\ln(1+2x)}{2x} - \frac{1+3x}{(1+2x)^2}   \bigg]~,
\end{equation}
where $\sigma_T$ is the Thomson cross section and $x = E/m_e$.

Finally, {\it photon injection} from BH decay yields
\begin{equation} \label{eq:eblInjSpecChange}
\frac{d\Phi_{\gamma,\textrm{inj}}}{dEdz}(E,z) = -\frac{d^2N_{\gamma}}{dEdt}(E,\Mbh(z), z) \frac{n_{\bullet}(z)}{H(z)(1+z)},
\end{equation}
where $n_{\bullet}$ is the black hole number density and $\frac{d^2N_\gamma}{dEdt}$ is the spectrum of produced photons from a single black hole of mass $\Mbh$.

The photons are produced as primaries and secondaries directly from black hole evaporation, annihilation of positrons, and inverse Compton scattering (ICS) of high-energy electrons and positrons. Therefore, the rate of photon production per black hole can be split into:
\begin{equation}\label{eq:eblInjRate}
    \frac{d^2N_\gamma}{dEdt}(E,\Mbh) = \frac{d^2N_{\gamma,\textrm{evap}}}{dEdt}(E,\Mbh) + \frac{d^2N_{\gamma,\textrm{pos}}}{dEdt}(E,\Mbh) + 
    \frac{d^2N_{\gamma,\textrm{ics}}}{dEdt}(E,\Mbh).
\end{equation}
The photon production rate due to evaporation, ${d^2N_{\gamma,\textrm{evap}}}/{dEdt}$, is calculated as the sum of the photon greybody spectrum as expressed in Eq.~\eqref{eq:emissionSM} and the secondary photons produced by the annihilation of unstable massive particles as discussed in Sec.~\ref{sec:evapproducts}.

Sufficiently hot black holes also produce high-energy electrons and positrons. As these cool down, they yield additional X-rays by upscattering CMB photons via ICS. The production rate of photons due to ICS is given by the convolution of the electron and positron evaporation spectrum with the secondary photon spectrum produced by the cooling of a single electron or positron with a given energy. This can be expressed as
\begin{equation} 
    \frac{d^2N_{\gamma,\textrm{ics}}}{dEdt}(E,\Mbh) = 2 \int_0^\infty dE_e \frac{d^2N_{e^-}}{dEdt}(E_e,\Mbh) \frac{d\tilde{N}_{\gamma,\textrm{ics}}}{dE}(E, E_e, T_{\rm CMB}),
\end{equation}
where $E$ is the photon energy, $E_e$ is the electron energy, $T_{\rm CMB}$ is the CMB temperature, $\frac{d^2N_{e^-}}{dEdt}$ is the production rate of electrons from black hole evaporation, and $\frac{d\tilde{N}_{\gamma,\textrm{ics}}}{dE}$ is the secondary photon spectrum from a single electron or positron cooling down. The factor of 2 accounts for the fact that both electrons and positrons contribute to the ICS signal. The secondary photon spectrum from electron cooling was determined by interpolating a table calculated using \texttt{DarkHistory} \cite{Liu:2019bbm}.

After an energetic positron quickly loses most of its energy via ICS and other cooling processes, it will find a partner and annihilate to photons. First,  positronium is formed in either the singlet or triplet state. One quarter of the positrons form positronium in the singlet (parapositronium, $j = 0$) state, which annihilates to two photons with $E_\gamma = m_e$. The remaining three quarters of the positrons form the triplet (orthopositronium, $j = 1$) state, which produces three photons with a spectrum first calculated in \cite{Ore1949} and expressed in \cite{Liu:2019bbm} as
\begin{equation}
    \frac{d\tilde{N}_\gamma^\textrm{ann}}{dE}\bigg|_{\textrm{triplet}} = \frac{6}{(\pi^2 - 9)m_e}\bigg(\frac{2-x}{x} + \frac{x(1-x)}{(2-x)^2} + 2\log(1-x)\bigg[ \frac{1-x}{x^2} - \frac{(1-x)^2}{(2-x)^3} \bigg]\bigg),
\end{equation}
where $x=E/m_e$ and $0\le x\le 1$.

Assuming 100\% positronium formation, the photon yield per positron is thus
\begin{equation} \label{eq:positroniumSpec}
    \frac{d\tilde{N}_\gamma^\textrm{ann}}{dE}(E) = \frac{1}{2}\delta(E-m_e) + \frac{3}{4}\frac{dN_\gamma^\textrm{ann}}{dE}\bigg|_{\textrm{triplet}}.
\end{equation}
Numerically, the Dirac delta function is modelled as a Gaussian with a width of 1 keV, which is a realistic approximation for the peak shape from galactic positronium annihilations \cite{Guessoum:2005cb}. Although Ref.~\cite{Guessoum:2005cb} does not address extragalactic positron annihilation, cosmic expansion causes the integrated signal from all extragalactic annihilations to form a continuum below 511 keV.  The resulting observed EBL flux is therefore insensitive to how the initial annihilation peak is parameterized.

The production rate of photons due to positron annihilation can be found by multiplying Eq.~\eqref{eq:positroniumSpec} by the positron production rate ${dN_{e^+}}/{dt}$, including primaries and secondaries:
\begin{equation} \label{eq:positroniumPhotonFlux}
    \frac{d^2N_{\gamma,\textrm{pos}}}{dEdt}(E,\Mbh) = \frac{dN_{e^+}}{dt}(\Mbh) \frac{d\tilde{N}_\gamma^\textrm{ann}}{dE}(E).
\end{equation}

Starting from recombination, the photon flux can be evolved forward in time using Eq.~\eqref{eq:photonFluxTotalChange} to calculate  the extragalactic contribution to the isotropic X-ray and gamma-ray spectrum today. Further details on how this equation was solved numerically are in Appendix~\ref{sec:numericEBL}.

\subsubsection{Galactic contribution} \label{sec:galIsotropic}
While the flux of evaporating black holes within the Milky Way halo would be highly anisotropic, because there is a non-zero flux in all directions, the flux in the direction that produces the smallest flux contributes an irreducible isotropic component on top of the extragalactic flux \cite{Iguaz:2021irx}. This flux can be calculated by evaluating Eq.~\eqref{eq:GCgammas} in the direction with the minimum flux, directly away from the galactic centre. Then, Eq.~\eqref{eq:GCgammas} simplifies to
\begin{equation} \label{eq:galIsoGammas}
    \frac{d\Phi_{\gamma,\textrm{gal}}}{dE} = \frac{\fbhToday}{4\pi \Mbh}  \frac{d^2N_\gamma}{dEdt}(E,\Mbh) \mathcal{D}_\textrm{min},
\end{equation}
where $\fbhToday$ is the fraction of dark matter comprised of PBHs today, ${d^2N_\gamma}/{dEdt}$ is calculated in the same way as in the EBL case except only accounting for evaporation to photons and positronium annihilation (the flux from ICS was not included in the galactic calculation), and $\mathcal{D}_\textrm{min}$ is the integral of the Dark Matter density along the line of sight opposite to the galactic centre
\begin{equation}
    \mathcal{D}_\textrm{min} = \int_{R_0}^\infty dr \rho_{DM}(r).
\end{equation}

\subsubsection{Observational constraints}
\begin{figure*}[hbt]
	\centering	\includegraphics[width=0.7\textwidth]{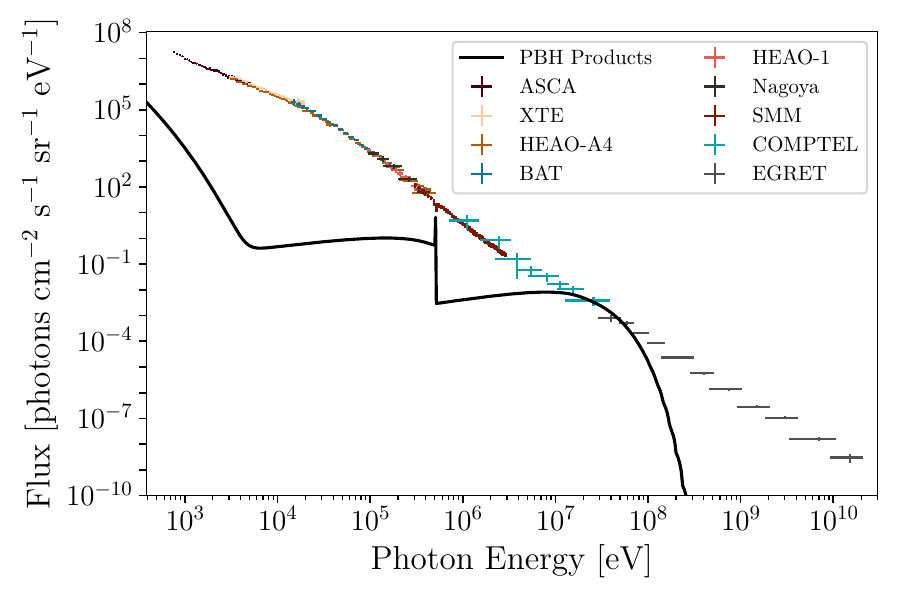}
	\caption{The observed fluxes from various X-ray and gamma-ray telescopes compared to the maximum allowed calculated flux from black holes with a mass of $2\times10^{15}$ g in $n = 5$ extra dimensions and a scale of quantum gravity  $M_\star = 10$~TeV. The experiments in the legend are ordered from lowest energy to highest energy. See main text for references.}
	\label{fig:IBLFlux}
\end{figure*}

The total expected isotropic photon flux can be calculated by adding together the extragalactic contribution found by solving Eq.~\eqref{eq:photonFluxTotalChange} and the galactic contribution from Eq.~\eqref{eq:galIsoGammas}. That calculated photon flux was compared to measurements of the isotropic X-ray and gamma-ray signal compiled in \cite{Ajello:2008xb}. The experiments included are, from lowest to highest energy: ASCA \cite{Tanaka:1994mt}, RXTE \cite{Revnivtsev:2003wm}, HEAO-1\cite{Kinzer_1997} , HEAO-A4 \cite{Gruber:1999yr}, Swift/BAT \cite{Ajello:2008xb}, Nagoya \cite{Fukada_1975}, SMM \cite{Watanabe1997}, CGRO/COMPTEL \cite{weidenspointner2000cosmic}, and CGRO/EGRET \cite{Strong:2004ry}. When the widths of the energy bins was not provided, it was assumed that bin widths extended to the midpoint with neighbouring bins. Although a measurement from the instruments on INTEGRAL (JEM-X, IBIS, SPI) are available \cite{Churazov:2006bk} we do not include them, as they are less precise, and overlap with other data used here.  The observed fluxes as well as a sample calculated spectrum are shown in Fig. \ref{fig:IBLFlux}.

To account for sharp features such as the 511 keV peak from Milky Way positronium annihilations, the calculated flux was averaged over each bin width to determine  the expected flux for each experiment. Constraints were then set by ensuring that the expected flux does not exceed the observed flux by more than $2\sigma$ in any energy bin. This approach leads to conservative constraints on the PBH abundance because no assumptions are made about other astrophysical sources of X-rays and gamma-rays. Including models of astrophysical X-ray and gamma-ray sources can currently strengthen PBH constraints by more than an order of magnitude \cite{Iguaz:2021irx,Chen:2021ngo} and have an even larger effect when projecting the discovery potential of future X-ray telescopes \cite{Ghosh:2021gfa}.

\begin{figure*}[hbt]
	\centering	\includegraphics[width=0.75\textwidth]{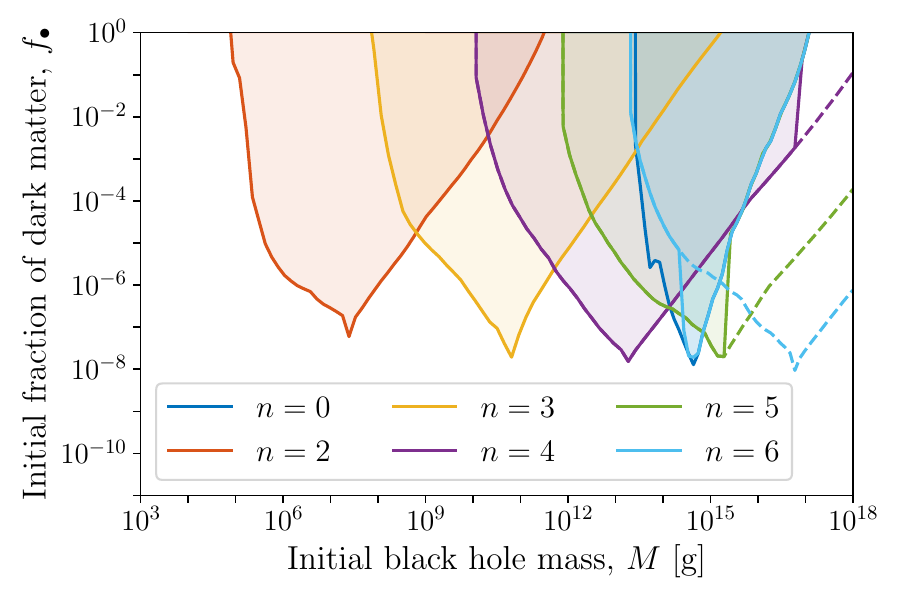}
	\caption{Constraints on the fraction of dark matter comprised of LED black holes in the early Universe, $\fbh$, from ensuring that the isotropic X-ray and gamma ray flux produced by evaporation does not exceed the observed flux in any energy bin by two standard deviations. The constraints keep the scale of quantum gravity, $M_\star$, fixed at 10 TeV (except for $n=0$ where $M_\star=M_{pl}$) while varying the initial black hole masses, $\Mbh$ and numbers of extra dimensions, $n$. The dashed lines indicate what the constraints would be if large black holes did not act like 4D black holes.}
	\label{fig:iblConstraint}
\end{figure*}

Constraints from isotropic background light are shown in Fig. \ref{fig:iblConstraint}. The shapes of the $n=0$ and $n=2\textrm{-}6$ constraints are generally similar. The low mass cutoff of the constraints is given by the black hole mass that leads to evaporation before the time of recombination (taken to be $z=1100$), as photons from these BHs cannot propagate freely until today. At slightly higher masses, BHs evaporate completely between recombination and today. The largest signal comes from the high-temperature emission at the end of their lives; more massive black holes evaporate closer to today such that their emitted photon spectrum has redshifted less, and the observed spectrum has a higher energy, where observed fluxes are lower. This leads to constraints strengthening with increasing initial BH mass. This trend continues until the black holes are massive enough to survive until today. Beyond this point, more massive black holes have lower temperatures and there are fewer black holes for a given energy density, causing the trend to reverse.

For $n=4,5\textrm{, and } 6$, as the mass increases, limits weaken sharply as the BHs Schwartzschild radius exceeds the size of the extra dimensions (as in Eq. \ref{eq:m4d}), leading them to mimic the $n = 0$ limits. Since this transition depends on the details of the compactification, the true behaviour would not be as sharp. 

\begin{figure*}[hbt]
	\centering	
	\includegraphics[width=0.47\textwidth]{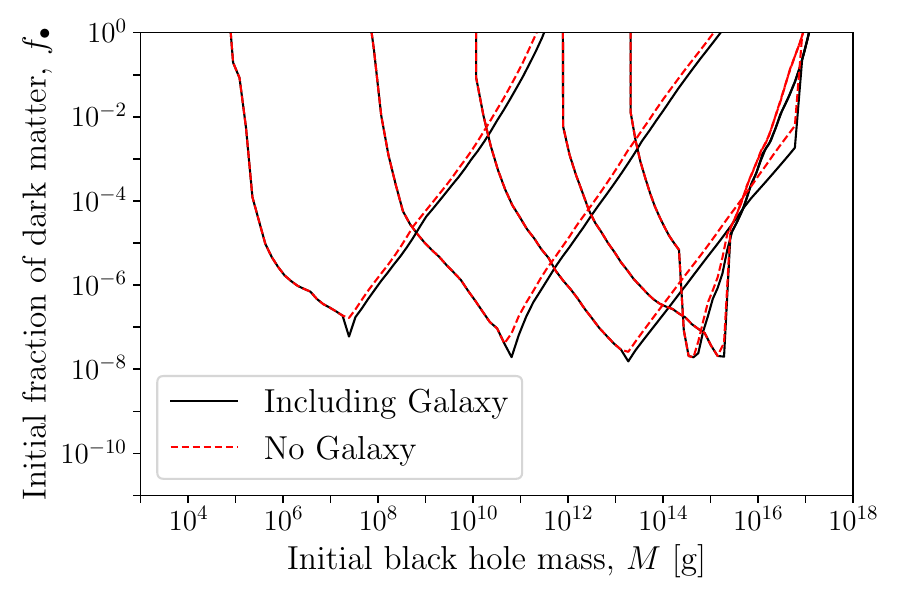}
	\includegraphics[width=0.47\textwidth]{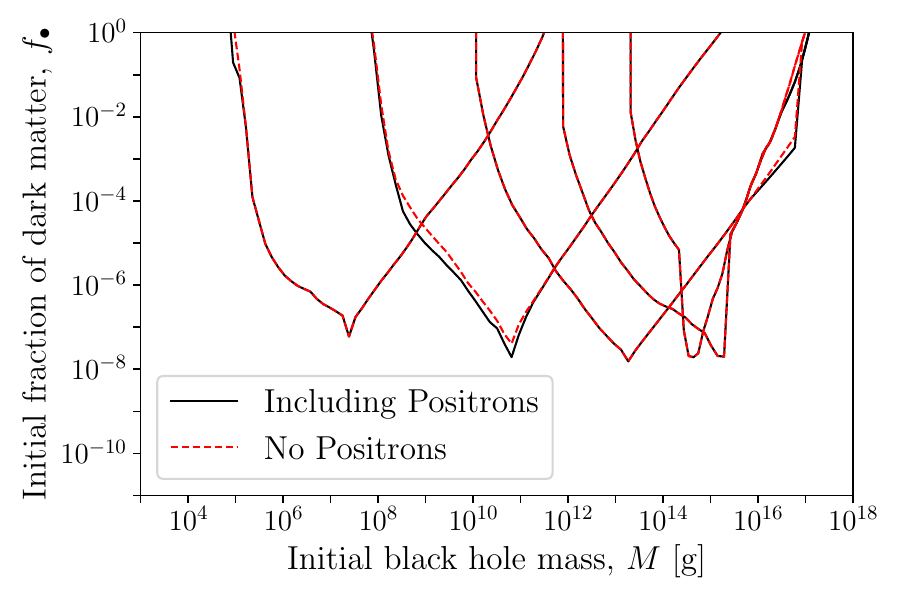}
	\caption{A comparison of the isotropic background light constraints on the initial fraction of dark matter comprised of LED black holes with or without different components. In both plots the black line is the full constraint as seen in Fig. \ref{fig:iblConstraint}. The various lines in each plot represent different numbers of extra dimensions (2-6 going left to right) and the scale of quantum gravity, $M_\star$, is fixed at 10 TeV. \textit{Left}: The dashed red lines show the strength of the constraints without the isotropic component coming from the milky way halo so that only the extragalactic background light is included. \textit{Right}: The red line shows the strength of constraints ignoring the effect of positron annihilation described in Eq.~\eqref{eq:positroniumPhotonFlux}.}
	\label{fig:posAndGalWOWY}
\end{figure*}

We found that there are no BH masses where the inclusion of photons produced by inverse Compton scattering improves the constraints. As shown in Fig. \ref{fig:posAndGalWOWY}, the galactic isotropic flux strengthens the constraints set on black holes that survive until today and including the flux from positron annihilations increases the strength of the constraints for black holes with temperatures close to $511$ keV.

\subsection{Cosmic microwave background}
\label{sec:CMB} 
Evaporation of primordial black holes during and after recombination can lead to high-energy electrons and photons producing heating and ionization---an effect first discussed in the context of decaying heavy neutrinos \cite{Adams:1998nr} and later adapted to annihilating dark matter \cite{Chen:2003gz}. A higher ionization floor will rescatter CMB photons. During the dark ages, this has the effect of ``blurring'' the last scattering surface (LSS), suppressing the angular power spectrum on small scales (large $\ell$). For ionization at lower redshifts, this rescattering additionally enhances power at lower multipoles in the EE polarization power spectrum, because Thomson scattering is polarized \cite{Slatyer:2009yq}. 

As part of the \texttt{CosmoLED} package, we modify the public \texttt{ExoCLASS} code \cite{Stocker:2018avm}, a branch of the \texttt{CLASS} linear anisotropy solver \cite{Blas:2011rf} which deals with the energy injection from WIMPs or primordial black holes. To be specific, we change the \texttt{DarkAges} module to incorporate LED BHs with $n=$1---6 and a flexible Planck scale $M_\star$. 4D BH remains a choice when $n$ is set to 0. The electron and gamma spectrum required for the module is now computed as described in Sec. \ref{sec:evapproducts}. We improve \texttt{ExoCLASS} in the following aspects: 1) We implement the complete greybody spectrum for all particles, instead of cutting the spectrum at $E=3T_H$ and approximate the absorption cross section as $\sigma_s=27\pi G^2M^2_\bullet$. 2) We include the secondary particles from primary particles at energies above 5~GeV using the PPPC4DMID tables. 3) At low energies, we use \texttt{Hazma} and our own code to calculate the decay of pions and muons as a function of particle energy, instead of using the fixed decay table in \texttt{ExoCLASS}. A comparison of secondary particle spectra from \texttt{CosmoLED} and \texttt{ExoCLASS} can be found in Fig.~\ref{fig:gammaande}.  We have also altered the black hole mass evolution of a function of time in \texttt{DarkAges} module and \texttt{CLASS} main code. Apart from these changes, we follow the approaches in \texttt{ExoCLASS} to compute the energy deposition from LED BHs, which we briefly summarize below.

The injection of energy from decaying black holes with initial mass $M_i$ and initial fraction $f_\bullet$, relevant for CMB observation is given by
\begin{equation}
    \left.\dfrac{d^2E}{dVdt}\right\rvert_{\rm inj}=\dfrac{f_\bullet f_{\rm e.m.}\rho_c \Omega_{\rm CDM}(1+z)^3}{M_{i}}\dfrac{d\Mbh}{dt}\,,
\end{equation}
where $f_{\rm e.m.}$ is the fraction of BH evaporation that ends up with $e^{\pm}$ and $\gamma$, and $\rho_c\Omega_{\rm CDM}$ is the cold dark matter energy density today. In \texttt{CosmoLED}, this is computed from
\begin{equation}
    f_{\rm e.m.}\dfrac{d\Mbh}{dt}=\int dE_{\rm e.m.}\dfrac{d^2N}{dE_{\rm e.m.}dt}\,,
\end{equation}
with the right hand side given by Eq.~\eqref{eq:flux_prim}. The injected energy is then deposited at different redshift $z$, in the form of ionization, excitation of the Lyman-$\alpha$ transition and heating of the intergalactic medium. The energy deposition is therefore connected to the energy injection by
\begin{equation}
    \left.\dfrac{d^2E}{dVdt}\right\rvert_{{\rm dep},c}(z)=h_c(z)\left.\dfrac{d^2E}{dVdt}\right\rvert_{\rm inj}(z)\,,
\end{equation}
and the energy deposition functions in the three channels denoted by $h_c$ can be obtained by convolving the injected electromagetic particle spectra with a transfer function that models streaming and absorption of electromagnetic products in the high-redshift IGM. We follow the treatment in \texttt{ExoCLASS} and employ the transfer functions precomputed in Refs.~\cite{Slatyer:2015jla,Slatyer:2015kla}.

\begin{figure}
    \centering
    \includegraphics[width=0.75\textwidth]{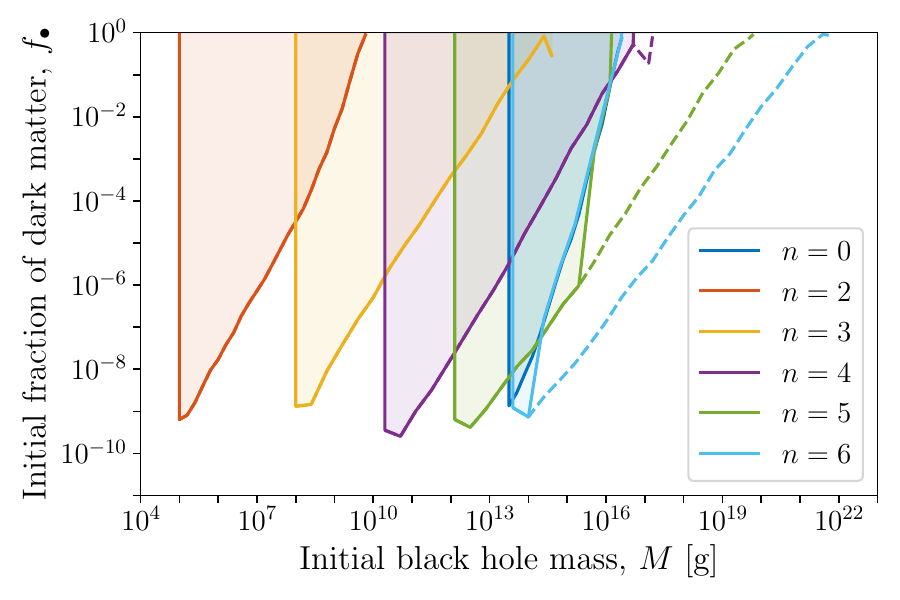}
    \caption{Constraints on the fraction of dark matter comprised of LED black holes in the early Universe, $\fbh$, based on their impact on the CMB angular power spectrum, computed using \texttt{ExoCLASS}, modified to include LED BHs, with updated greybody factors, secondary spectra, and precise evaporation rates. Each colour corresponds to a different number of extra dimensions, $n$. The constraints for PBHs with $n\ge2$ keeps the scale of quantum gravity, $M_\star$, fixed at 10 TeV while the case of 4D ($n=0$) PBHs, maintains $M_\star=M_{pl}$. The dashed lines indicate what the constraints would be if large black holes did not act like 4D black holes.}
    \label{fig:cmb}
\end{figure}

To constrain the initial fraction of BHs in dark matter, we use \texttt{MontePython} \cite{Audren:2012wb,Brinckmann:2018cvx} to run a Markov Chain Monte Carlo (MCMC), which interfaces with the modified version of \texttt{ExoCLASS} in \texttt{CosmoLED}. For each PBH initial mass and $n$, we impose flat prior on the initial fraction of BHs, and six $\Lambda$CDM parameters $\{\omega_b,\ \omega_{\rm cdm},\ \theta_s,\ {\rm ln}(10^{10}A_s),\ n_s,\ \tau_{\rm reio}\}$. We adopt the Planck high-$l$ TT,TE,EE+low $l$ TT, EE+Planck lensing 2018 \cite{Planck:2018nkj} likelihoods, with standard Planck nuisance parameters marginalized over. Fig. \ref{fig:cmb} shows the boundary of each 95\% one-dimensional credible interval on the initial fraction of PBHs as DM, $f_\bullet$, as a function of the initial black hole mass $\Mbh$. For each $n$, the excluded region cuts off abruptly at low mass, where BH evaporation occurs before recombination and thus does not affect the ionization floor. The cutoff of $n=6$ BHs coincides with that of 4D BHs, as from Fig.~\ref{fig:TBH} $\Mbh \simeq 10^{14}$~g BHs disappear at CMB in both cases. As the mass increases, sensitivity is gradually reduced as the Hawking temperature of the PBH population falls with mass. 

Even though our inclusion of secondary particles leads to a larger $\gamma$ and $e^\pm$ flux than in the default \texttt{ExoCLASS} scenario, our constraints for $n = 0$ are slightly weaker than those presented in Ref. \cite{Stocker:2018avm}. These differences may be due to their implementation of a prior on $\tau_{\rm reio}$, with which $f_\bullet$ is degenerate, or the use of different Planck data sets.

\subsection{Big Bang nucleosynthesis}
\label{sec:BBN}
Big Bang nucleosynthesis (BBN) presents a critical evolutionary epoch of the early Universe. The expansion-driven cooling of the Universe leads to the formation of the first light elements as the temperature of the background photons drops below the binding energy of said light nuclei. The final abundances of light elements synthesised during this era, (in conjunction with the relevant nuclear cross sections and cosmological framework) therefore also provide a fruitful testing ground for physics beyond the Standard Model. Constraints on mechanisms that modify either the expansion rate or balance of the synthesis processes during this era have been explored previously in, for example Refs.~\cite{Carr2010, Sarkar1996, Jedamzik:2009,Pospelov:2010,Hufnagel:2018,Huang:2018,Forestell:2019,Depta:2019,Kawasaki:2018}. 

In a similar spirit, the presence and evaporation of black holes leading up to, during, and beyond BBN, can impact the resulting relic abundances in a number of ways. Weak interactions freeze out around temperatures of $\sim 1$~MeV, just before the onset of BBN, setting the neutron-to-proton ratio which is critical to the eventual formation of helium. An additional black hole density component may alter the expansion history of the Universe and the subsequent freeze-out of this ratio. More specifically, an increase in the expansion rate will lead to an earlier weak interaction freeze-out, an enhanced neutron-proton ratio and eventually, a greater helium-4 abundance (see Sec.~\ref{sec:alterbbn} for further discussion.) Black hole evaporation products, namely pions, may also alter the neutron-proton fraction after freeze-out via direct conversion. In addition, if the temperature of the black holes is sufficiently high, the resulting evaporation products will be able to directly contribute to the dissociation of the forming nuclei. 

In order to incorporate black holes and their evaporation products correctly into the relic calculation, a complex system of reactions needs to be solved self-consistently. As most public codes do not allow for non-thermal energy injection, we will deal with these two effects separately.\footnote{Recently, photodisintegration of light elements due to distorted photon phase space distribution from exotic entropy injection has been implemented in the \texttt{ACROPOLIS} code~\cite{Depta:2020mhj,Depta:2020zbh,Hufnagel:2018bjp}, which is yet to be employed to study LED BHs. A dedicated analysis with \texttt{ACROPOLIS} is left for future work.}  In Sec.\,\ref{sec:photohadro} we recast prior results following the method of Ref. \cite{Keith:2020jww}; in Sec.\ref{sec:alterbbn} we adapt the \texttt{AlterBBN} code to produce the light abundances from the appropriately modified expansion histories.

\subsubsection{Photo- and hadrodissociation}
\label{sec:photohadro}
If the bulk of BH evaporation occurs during or shortly after BBN, the production of high-energy particles can lead to dissociation of nuclei, affecting the relic abundance of D, He and Li. The addition of a non-thermal component to existing BBN codes is non-trivial. Kawasaki \textit{et al.}~\cite{Kawasaki:2018} performed a detailed numerical analysis, deriving constraints on the lifetime of decaying dark matter during the BBN epoch as a function of its mass and density. They utilized updated reaction rates, newly implemented interconversion of energetic protons and neutrons by inelastic scattering off background nuclei, as well as the incorporation of energetic antiprotons and antineutrons. Their results use the observed relic abundance of light elements,  including the primordial mass fraction of $^4$He, $\mathrm{Y}_{p} \equiv \rho(^{4}\mathrm{He})/\rho_{b} = 0.2449 \pm 0.0040$~\cite{Aver:2015iza}, the primordial deuterium to hydrogen ratio $\left( \mathrm{D/H} \right)_{p} = (2.53 \pm 0.04) \times 10^{-5}$~\cite{Cooke:2013cba} and the upper limit on the primordial $^3$He to deuterium ratio $(^3{\rm He}/{\rm D})_p<0.83+0.27$~\cite{geiss2003isotopic}. Keith \textit{et al. }\cite{Keith:2020jww} pointed out that evaporating black holes modify BBN abundances in a similar manner to decaying massive particles and recast the results of Kawasaki \textit{et al.} to derive equivalent constraints for black holes. We will mostly follow the procedure outlined in Ref.~\cite{Keith:2020jww} to recast the results in Ref.~\cite{Kawasaki:2018} for the LED BHs described in this article. The method, assumptions, limitations and results are presented below.

Ref.~\cite{Keith:2020jww} broadly distinguishes between two  phases of nuclear dissociation due to BH evaporation products: the \textit{hadrodissociation} era at high plasma temperatures, and the \textit{photodissociation} era at later times. Both of them lead to the dissociation of $^4$He and the production of D and $^3$He. We follow the same approach as Ref.~\cite{Keith:2020jww} to account for the photodissociation of $^4$He caused by BH evaporation, while for hadrodissociation, we adopt a different procedure which better captures the total number of hadrons injected by BHs. In both cases, we use the precise greybody spectrum to compute the average quark energy, instead of assuming a thermal Fermi-Dirac distribution.

If decays happen at late enough times, when the plasma temperature is lower than $T \lesssim 0.4$ keV, all electromagnetic final states contribute to dissociation. Because a majority of SM degrees of freedom---and thus evaporation products---are in the hadronic sector, this can be mapped to previous bounds on dark matter decay to quark-antiquark pairs. Neglecting the quark masses and averaging over the quark greybody spectrum, the mean energy $\langle E_q\rangle_M$ for a given BH mass is
\begin{equation}
    \langle E_q\rangle_M=\dfrac{\bigintsss  E\dfrac{d^2N}{dEdt} dE}{\bigintsss \dfrac{d^2N}{dEdt}  dE}\,,
    \label{eq:EqM}
\end{equation}
where ${d^2N}/{dEdt}$ is the radiated quark energy distribution, given by Eq.~\eqref{eq:flux_prim}. Since quarks are typically produced above the QCD transition scale, the mean quark energy is obtained by averaging the emission over the lifetime of a BH when the Hawking temperature is high enough, \textit{i.e.}
\begin{equation}
    \langle E_q\rangle =\dfrac{ \bigintsss_{M_i}^0 \langle E_q\rangle_M \dfrac{dN}{dM}\Theta(T_H-\Lambda_{\rm QCD})dM}{\bigintsss_{M_i}^0 \dfrac{dN}{dM}\Theta(T_H-\Lambda_{\rm QCD})dM}=k_q \mathrm{max}( T_{H,i},\Lambda_{\rm QCD})\,,
    \label{eq:Eq}
\end{equation}
 where $dN/dM$ is the number of quarks produced per change in BH mass, and can be inferred from Eq.~\eqref{eq:flux_prim} (after integrating over $E$) and Eq.~\eqref{eq:dMdt} considering quarks and gluons. $T_{H,i}$ is the initial BH Hawking temperature.

The total energy injection, which is relevant for photodissociation of $^4$He, of a BH with initial mass $M_i$, should thus yield a similar effect to the decay of DM particles with mass $M_X\simeq 2\langle E_q\rangle$ into quark pairs. The step function in Eq. \eqref{eq:Eq} ensures that quarks are not produced below the QCD scale. This approach is conservative, in that it ignores evaporation for Hawking temperatures below $\Lambda_{\rm QCD}$ to other states.

At higher temperatures ($T\gtrsim 0.4$~keV), $e^+e^-$ pair production from photons is efficient, and the dissociation of $^4$He primarily expected to be from hadrons produced from quark and gluon jets, which builds up with the injection of more hadrons. The number of hadrons in a quark jet scales as $E_q^{0.3}$, and therefore on average, the number of quarks produced from the greybody spectrum is approximated by quarks with a single energy $\langle E_h\rangle_M$ which satisfies
\begin{equation}
    \langle E_h\rangle_M^{0.3}=\dfrac{\bigintsss  E^{0.3}\dfrac{d^2N}{dEdt} dE}{\bigintsss \dfrac{d^2N}{dEdt}  dE}\,.
    \label{eq:EhM}
\end{equation}
Again averaging over the evaporation lifetime of a BH, the number of hadrons per unit energy, proportional to $E_q^{0.3}/E_q$, is computed as
\begin{equation}
    \langle E_h\rangle^{-0.7} =\dfrac{ \bigintsss_{M_i}^0 \langle E_h\rangle_M^{0.3} \dfrac{dN}{dM}\Theta(T_H-\Lambda_{\rm QCD})dM}{\bigintsss_{M_i}^0 \langle E_q\rangle_M \dfrac{dN}{dM}\Theta(T_H-\Lambda_{\rm QCD})dM}\,.
    \label{eq:Eh}
\end{equation}
The numerator gives the total number of hadrons emitted during the lifetime of a BH, and the denominator shows the total hadronic energy. This can again be mapped to dark matter which decays to quark-antiquark pairs, with the number of hadrons per unit energy given by $(M_X/2)^{-0.7}$. Therefore, we have the relation 
\begin{equation}
    \langle E_h\rangle=k_h \mathrm{max}( T_{H,i},\Lambda_{\rm QCD})=\dfrac{M_X}{2}\,.
    \label{eq:Ehkh}
\end{equation}

The values of $\langle E_q\rangle_M$ and $\langle E_h\rangle_M$, as well as the $k_q$ and $k_h$ coefficients are computed and listed in Table~\ref{tab:BBNfactors}. Note that these differ from values presented in Ref.~\cite{Keith:2020jww} as we use the full greybody spectra to model the quark phase space distributions, and a different method for hadrodissociation.

To obtain the constraints on the enregy density of BHs, we find the correspondence between BHs and decaying dark matter that causes the same amount of dissociation to light elements. Conservatively we only consider the hadrons and photons produced from quarks and gluons, not other particles. If BHs initially  have a Hawking temperature above the QCD transition scale, i.e. $M_i<M_{\rm QCD} (T_H=\Lambda_{\rm QCD})$, the entire mass of BHs is injected to the plasma in the form of quarks (and gluons), up to an order 1 number $f_q$ which quantifies the fraction of hadronic injection. Therefore, roughly the same amount of quarks are produced in BH evaporation and dark matter decay, provided that they start from the same energy density. However, if $M_i>M_{\rm QCD}$, quarks are only emitted when BH mass reduces to $M_{\rm QCD}$, and the early stage of the BH mass dump does not dissociate any nuclei. To match the number of quarks injected, the initial fraction of BHs $\fbh$ that we constrain is related to the fraction of dark matter made of decaying particles, $f_X$ constrained by Kawasaki by
\begin{equation}
    f_\bullet=\begin{cases}
  f_X/f_q\,,&M_i<M_{\rm QCD}\,,\\      
  f_XM_i/(f_qM_{\rm QCD})\,,&M_i\geq M_{\rm QCD}\,.
\end{cases}
\label{eq:fscaling}
\end{equation}
4D BHs always have $M_i<M_{\rm QCD}$ in the relevant mapping mass range. However, LED BHs can have longer lifetimes and lower Hawking temperatures, rendering the $M_i/M_{\rm QCD}$ factor important. The fraction of hadronic energy injection $f_q$ mildly depends on BH mass, running from 76\% for 4D BHs, to 65\% for $n=6$ BHs, due to differences in the greybody spectra, as well as a growing fraction of graviton emission.

To complete the translation of constraints from decaying dark matter to BH evaporation, we must determine the appropriate correspondence between the lifetime of BHs $\tau_\bullet$ and dark matter decay time $\tau_X$ . While we expect that $\tau_X\simeq \tau_\bullet$, these processes are fundamentally different in that DM decay represents a steady injection of energetic particles, while BH evaporation products increase in energy until a dramatic spike at $\tau_\bullet$, after which no BHs remain. As done by Keith \textit{et al.} in Ref.~\cite{Keith:2020jww}, we match BHs and decaying dark matter at a time when half of the energy is injected. For decaying dark matter, this happens at a time $t=\tau_X \ln 2$. For BHs with initial mass $M_i<M_{\rm QCD}$, injecting half of the total energy takes the time $t=f_t \tau_\bullet$, and $f_t\simeq 0.5$. This yields the relation $\tau_X=f_t\tau_\bullet/\ln 2$. If however $M_i\gg M_{\rm QCD}$, the lifetime of $M_{\rm QCD}$ BH is negligibly small, and we have $\tau_X=\tau_\bullet/\ln 2$ instead.

\begin{table}[!htb]
\centering
\setlength\extrarowheight{3pt}
\begin{tabularx}{0.7\textwidth}{X | X | X | X | X | X | l}
\hline\hline
	$n$ & 0 & 2 & 3 & 4 & 5 & 6\\ \hline
    $\langle E_q\rangle_M/T_H$ & 4.23 & 3.05 & 2.95 & 2.90 & 2.89 & 2.88\\
    $\langle E_h\rangle_M/T_H$ & 3.97 & 2.74 & 2.62 & 2.56 & 2.54 & 2.53 \\
    $k_q$ & 8.46 & 4.04 & 3.68 & 3.48 & 3.36 & 3.29 \\
    $k_h$ & 9.27 & 4.27 & 3.90 & 3.68 & 3.56 & 3.49\\
    $M_{\rm QCD}$[g] & $3.53\times 10^{13}$ & $1.88\times10^{-8}$ & $6.96\times 10^{-4}$ & $28.8$ & $1.31\times 10^6$ & $6.40\times 10^{10}$\\
\hline \hline
\end{tabularx}
\caption{The factors relevant for BBN constraints assuming $M_\star=10$~TeV for LED BHs, and $M_\star=M_{pl}$ for 4D BHs, as defined in \cref{eq:EqM,eq:Eq,eq:EhM,eq:Eh,eq:Ehkh}. The last row shows the BH mass at which the Hawking temperature matches the QCD transition scale $\Lambda_{\rm QCD} \simeq 300$ MeV.}
\label{tab:BBNfactors}
\end{table}

\begin{figure*}[hbt]
	\centering	\includegraphics[width=0.75\textwidth]{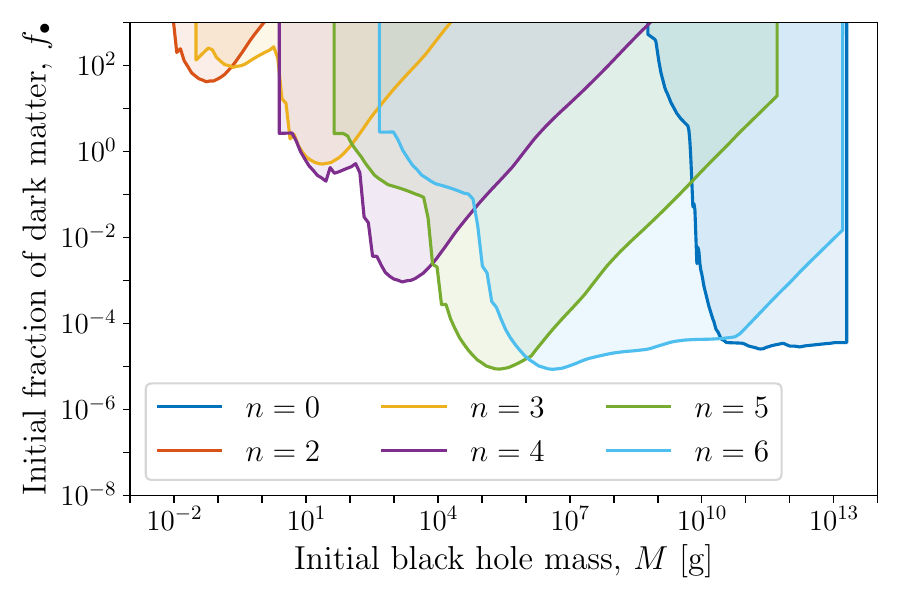}
	\caption{
		Constraints on the initial fraction of dark matter  $\fbh$ comprised of LED black holes in the early Universe, due to dissociation of primordial nuclei from the products of BH evaporation during BBN. These are recast from the decaying dark matter constraints of Ref. \cite{Kawasaki:2018}, using the method of \cite{Keith:2020jww}. The constraints keep the scale of quantum gravity, $M_\star$, fixed at 10~TeV (except for $n=0$ where $M_\star=M_{pl}$) while varying the initial black hole masses, $\Mbh$ and numbers of extra dimensions, $n$.}
	\label{fig:BBNConstraint}
\end{figure*}

Our constraints are presented in Fig. \ref{fig:BBNConstraint}. For BH with mass $M$, we find the dark matter lifetime that matches BH lifetime, and the dark matter mass $M_X$ that reproduces the dissociation effects of a BH, using the method outlined above. We then interpolate the constraint lines in Ref.~\cite{Kawasaki:2018} according to $M_X$, using $X\rightarrow u\bar{u}$ decay channel. The interpolation works well for 4D BHs. However, for LED BHs, the corresponding dark matter mass is below the smallest mass considered of 0.03~TeV in most of the parameter space due to the low Hawking temperature. Noting that the constraints on the energy density of dark matter get stronger for lighter dark matter mass, as hadrodissociation depends on the number of emitted hadrons proportional to $M_X^{0.3}$, and photodissociation is roughly determined by the total energy injection. We therefore use the $M_X=0.03$~TeV constraint line for any mapped dark matter mass below 0.03~TeV, to produce a conservative bound on the energy density of BHs. We present results in terms of the initial fraction of dark matter made up of black holes, $\fbh$. This can be equated to $\beta \equiv \rho_\bullet / \rho_{\rm tot}$ and $MY$, the decaying particle mass times their number density per unit entropy, used in Ref.~\cite{Keith:2020jww} and Ref.~\cite{Kawasaki:2018}  respectively, via
\begin{equation}
    \fbh = \beta \frac{\Omega_{r}}{\Omega_{DM}} \frac{T_\mathrm{form}}{T_0} = \frac{4}{3} \frac{\Omega_{r}}{\Omega_{DM}} \frac{MY}{T_0}
    \label{eq:fconversion}
\end{equation}
where $T_0$ and $T_\mathrm{form}$ are the CMB temperatures today and the plasma temperature at black hole formation respectively. As in previous figures, red, yellow, purple, green and light blue curves correspond to the $n=2-6$ extra dimensional cases respectively. The rightmost dark blue curve shows the 4D results, which are well-matched to those derived in Ref.~\cite{Keith:2020jww}, though the inclusion of the relevant greybody factors and the updated method leads to some small differences at lower masses. The different mass range covered by the LED BHs also leads to a number of qualitative modifications of the 4D results. As seen in Table~\ref{tab:BBNfactors}, the maximum 4D BH mass translated from decaying dark matter is below $M_{\rm QCD}$. However, for any $n\geq 2$ and $M_\star=10$~TeV, in some part of the mass ranges BHs have initial Hawking temperatures that fall below the QCD transition scale. The correction due to $M>M_{\rm QCD}$ is more pronounced for lower number of extra dimensions, and starts to severely restrict the parameter space that can be constrained above about $10^{11}$~g for $n=6$ BHs. For all $n\geq 2$, this accounts for the $\fbh\propto M$ loss of sensitivity at higher masses.

There are a number of assumptions underwriting the validity of this methodology. They mostly pertain to being able to match both the spectral and the temporal distribution of the injected energy from an evaporating BH to that of a decaying particle. 

Firstly, it is assumed that the spectral shape does not significantly vary the impact on BBN, provided the average energy of the injected particles is the same. Similarly, the temporal spread of injected energy from BHs can be treated as equivalent to that of a decaying particle, as long the averaged energy is injected at approximately the same time. Keith \textit{et al.} note that the spread of particle energy around the mean for the 4D case could lead to an error of around $30$\% for BHs evaporating after $\approx 10^7$s. The effect is larger for BHs with shorter lifetimes where errors of up to a factor of $2$ are possible.

\subsubsection{Altered expansion history}
\label{sec:alterbbn}

 The method described above accounts for the catastrophic injection of nonthermal energy during or after nucleosynthesis leading to nuclear dissociation. 
In addition to this effect, the presence of extra matter in the form of black holes during BBN, as well as the smooth injection of entropy leads to an altered expansion history, baryon-to-photon ratio, and ratio of neutrino-to-plasma temperatures, which all contribute to altering the freeze-out abundances of the primordial elements. It will turn out that only the former effect has an impact on nucleosynthesis. We treat these effects separately from the dissociation discussed above, as it pertains to a slightly earlier epoch---and publicly available software allows for a more exact treatment. We modify \texttt{AlterBBN}~\cite{Arbey:2011nf,Arbey:2018zfh} to include BHs as additional species. In the code, BHs alter BBN in two ways: 1) the energy density of BHs contributes to the expansion of the Universe, and 2) BHs dump entropy to the plasma, increasing the temperature of photons and neutrinos. As described above, only effect 1) will turn out to be constraining, though these constraints will be subdominant to those presented in Sec. \ref{sec:photohadro}. Details of the implementation in the code, and resulting constraints, are described below.

\begin{figure}
    \centering
    \includegraphics[width=0.75\textwidth]{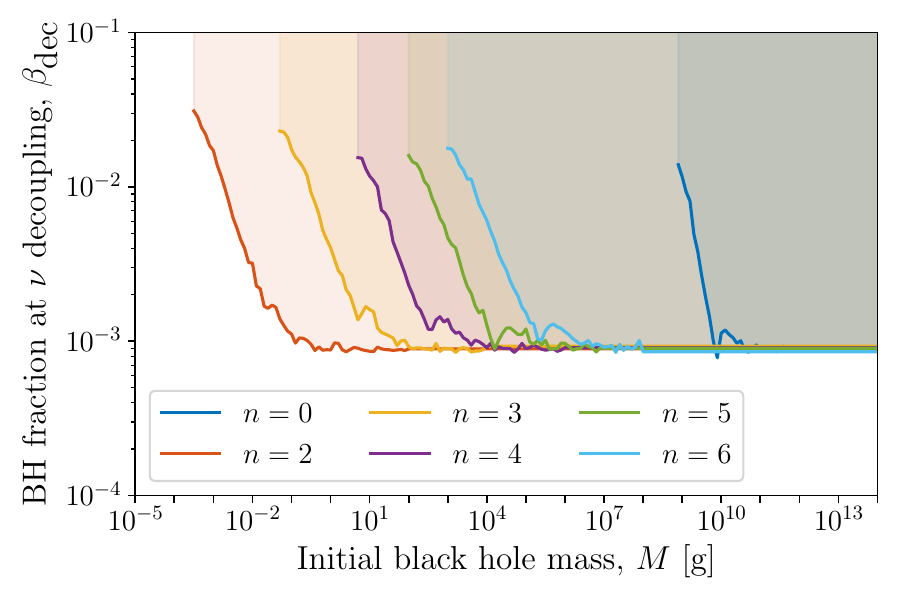}
    \caption{$2\sigma$ limits on the energy density fraction of BHs in the Universe $\beta_{dec} = {\rho_\bullet^{\rm dec}}/{\rho_{\rm tot}^{\rm dec}}$ at the temperature $T_{\rm dec}=2.33$~MeV where neutrino decoupling is expected in standard cosmology, as described in Sec.~\ref{sec:alterbbn}. The limits are obtained with modified \texttt{AlterBBN}~\cite{Arbey:2011nf,Arbey:2018zfh}, accounting for the BH contribution to the expansion of the Universe, and the BH entropy dump to the plasma. The Planck scale $M_\star=10$~TeV is assumed for LED BHs. Because BBN occurs deep in radiation domination, these constraints all correspond to a fraction of the matter content $f_\bullet \gg 1$, and are thus subleading.}
    \label{fig:limit_alterbbn}
\end{figure}

With BHs, the energy density of the Universe during BBN is given by
\begin{equation}
    \rho_{\rm tot}=\rho_\gamma+\rho_\nu+\rho_e+\rho_b+\rho_\bullet\,,
\end{equation}
where we include the energy density of photons, neutrinos, electron and positrons, baryons and BHs. The energy density of $e^\pm$ is connected to the photon temperature, parametrized with a series of Bessel functions~\cite{Arbey:2018zfh}. The baryon density is fixed by the baryon-to-photon today, and we assume $\eta_0=6.1\times 10^{-10}$. We start evolving the code from the neutrino decoupling temperature $T_{\rm dec} =2.33$~MeV. We assume neutrinos and photons are in thermal equilibrium separately with the temperature $T_\gamma$ and $T_\nu$ after neutrino decoupling. The neutrino energy density $\rho_\nu=\frac{7\pi^2}{120}N_\nu T_\nu^4$, where we fix $N_\nu=3.046$, and the photon energy density $\rho_\gamma=\frac{\pi^2}{15}T_\gamma^4$. For each decoupled species, the continuity equation implies
\begin{equation}
    \dfrac{d\ln a^3}{dT}=-\dfrac{d\rho/dT}{\rho+P-\frac{T}{3H}\left(\frac{\partial s}{\partial t}+3Hs\right)}\,,
    \label{eq:dlna3dT}
\end{equation}
where $\rho$, $P$ and $s$ are the energy density, pressure and entropy of the species. BH evaporation will dump entropy to the plasma containing photon, baryons and electrons, as well as the neutrino sector. We assume the two sectors keep thermal equilibrium efficiently and separately. The net effect of the entropy dump is to raise the temperature of the plasma and neutrinos, which in turn increases the energy density of photons and neutrinos through the expressions given above. Considering BH evaporation, the neutrino entropy follows
\begin{equation}
    \dfrac{\partial s_\nu}{\partial t}=-3Hs_\nu+\dfrac{n_\bullet}{T_\nu}\dfrac{dM_{\bullet\rightarrow\nu}}{dt}\,,
\end{equation}
and the plasma entropy
\begin{equation}
    \dfrac{\partial s_p}{\partial t}=-3Hs_p+\dfrac{n_\bullet}{T_\gamma}\dfrac{dM_{\bullet\rightarrow ({\rm SM}-\nu)}}{dt}\,,
\end{equation}
which are employed in Eq.~\eqref{eq:dlna3dT} to determine the evolution of the plasma and neutrino temperatures. With \texttt{AlterBBN}, we compute the abundances of He and deuterium and confront them with observations. We use the most updated primordial deuterium to hydrogen abundance ratio in PDG 2020~\cite{smoot2020review} where
$
     \left(\mathrm{D/H} \right)_{p} = (2.547 \pm 0.025) \times 10^{-5}\,,
$
 which reflects the weighted average of the most precise measurements. Similarly, the primordial $^4$He abundance is determined to be
$
     \mathrm{Y}_{p} \equiv \rho(^{4}\mathrm{He})/\rho_{b} = 0.245 \pm 0.003\,.
$
The numbers are slightly different from the abundances quoted in Ref.~\cite{Kawasaki:2018}, but the results remain robust regardless of the subtleties. To have a sizeable effect, the BH abundance must be $\sim 10^{-3}$ or larger at BBN, which is significantly larger than the matter density expected during BBN, i.e. $f_\bullet \gg 1$. We thus present the constraints on the fraction of BH energy density in the Universe at the neutrino decoupling temperature,
\begin{equation}
    \beta_{\rm dec}=\dfrac{\rho_\bullet^{\rm dec}}{\rho_{\rm tot}^{\rm dec}}\,,
\end{equation}
and show the $2\sigma$ limits on $\beta_{\rm dec}$ Fig.~\ref{fig:limit_alterbbn}. To the left of the lines BHs evaporate significantly before the plasma temperature drops below $T_{\rm de}$. If BHs survive through BBN, a BH fraction as low as $10^{-3}$ will modify the expansion of the Universe substantially, resulting in He and D abundances that are inconsistent with observation. The same bound holds for higher mass BHs which barely evaporate during BBN, but becomes weaker for lighter BHs that disappear before BBN ends. 

\subsection{Combined constraints}
\label{sec:fullconstraints}
\begin{figure}
    \centering
    \includegraphics[width=0.48\textwidth]{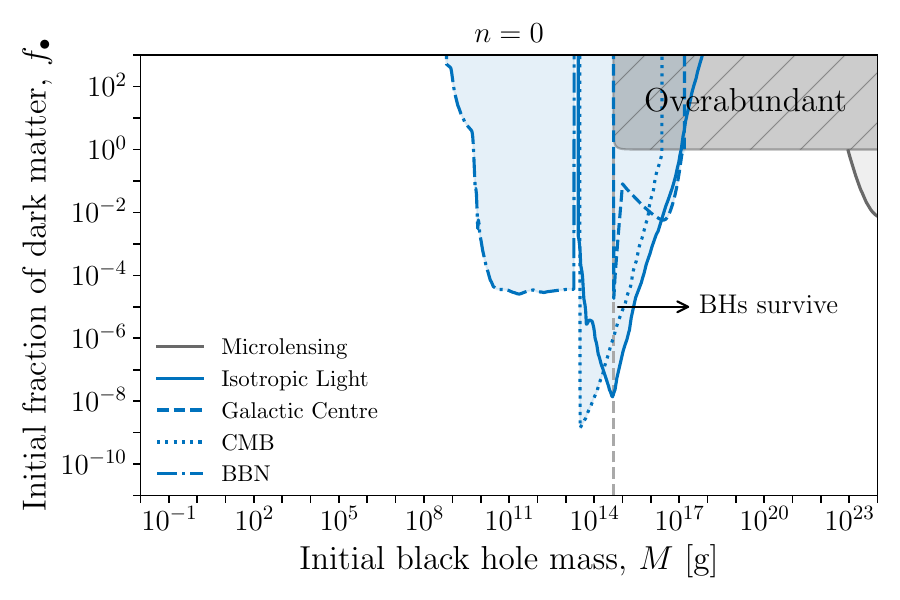}
    \includegraphics[width=0.48\textwidth]{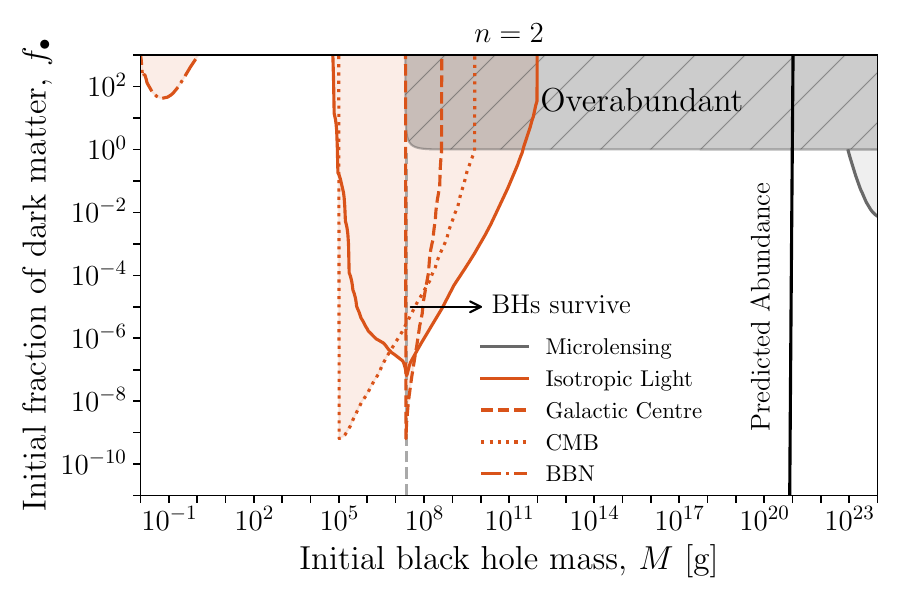}
    \includegraphics[width=0.48\textwidth]{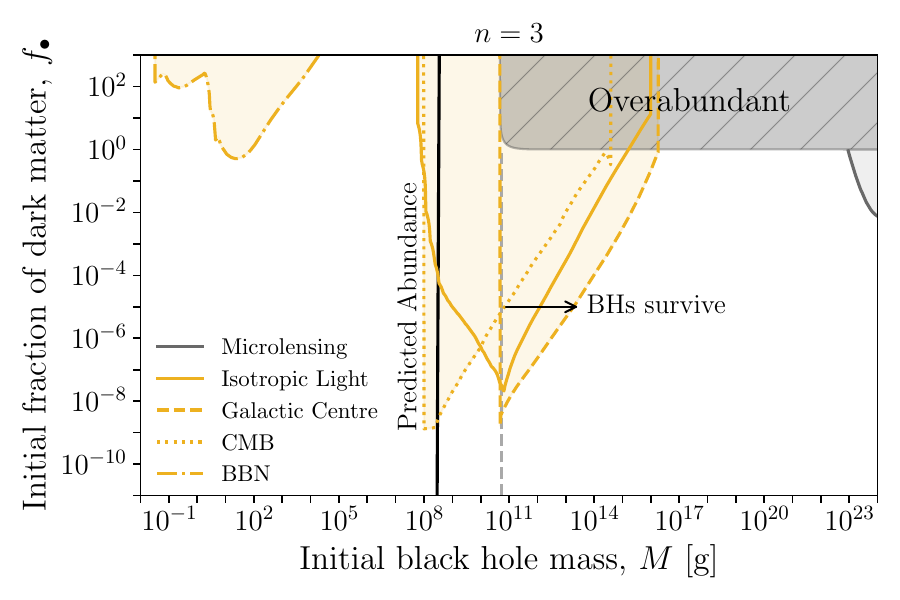}
    \includegraphics[width=0.48\textwidth]{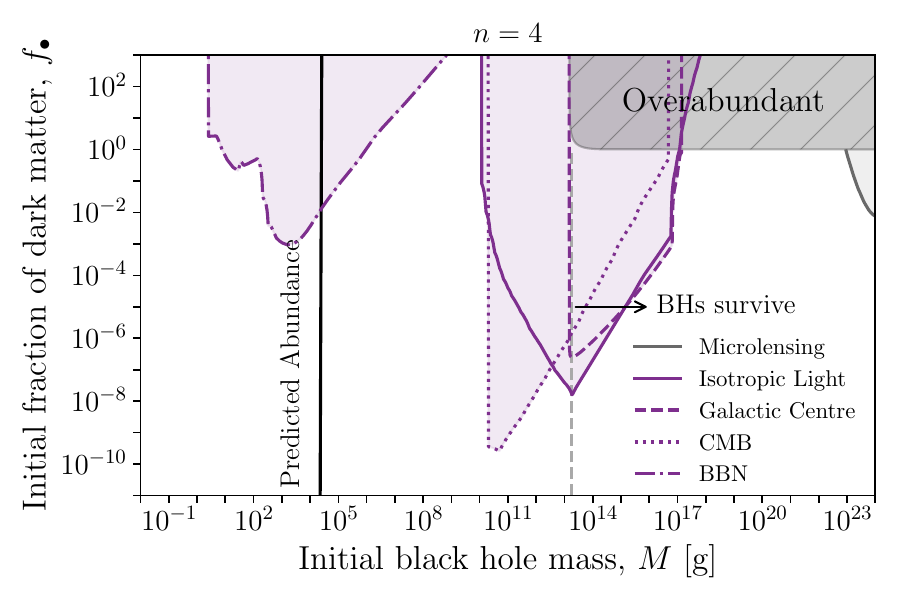}
    \includegraphics[width=0.48\textwidth]{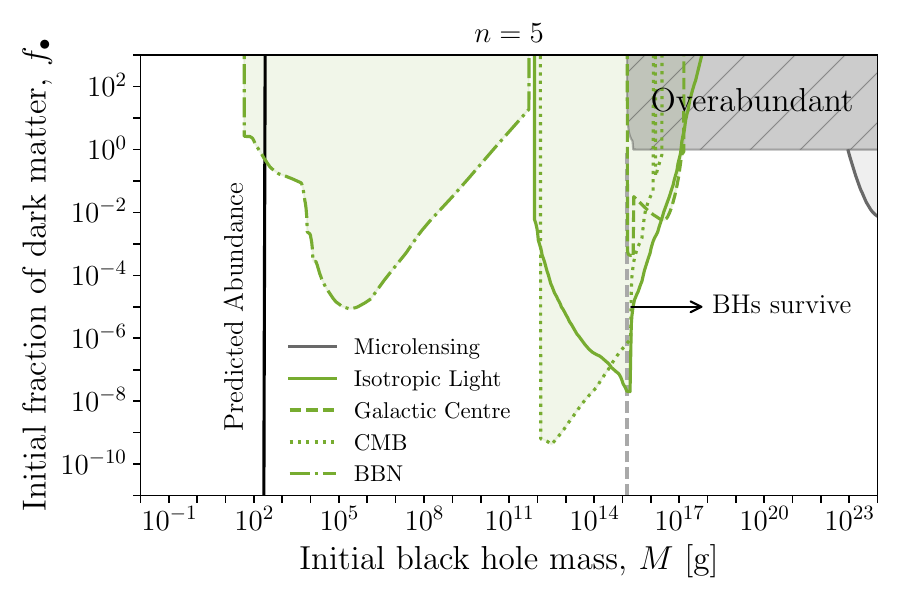}
    \includegraphics[width=0.48\textwidth]{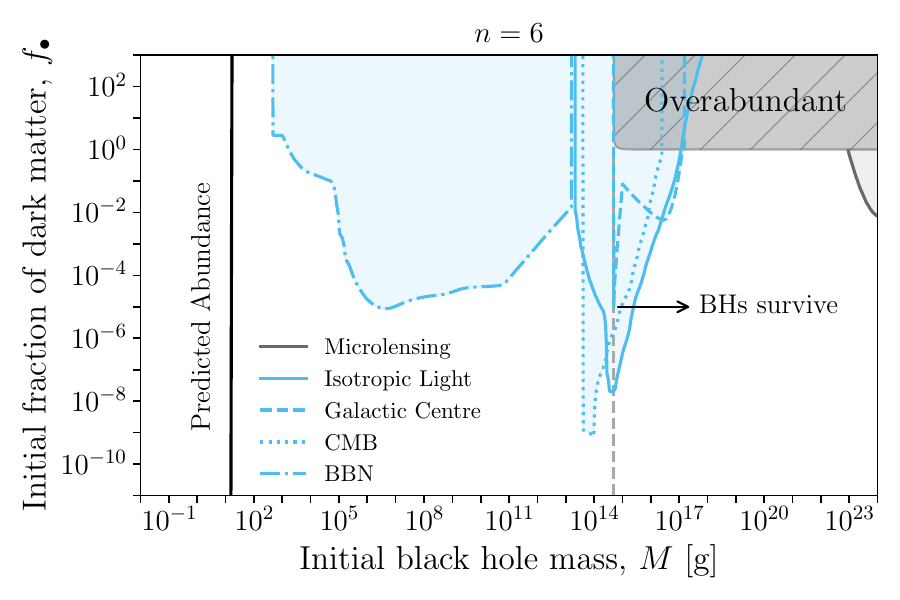}
    \caption{Combined constraints on the fraction of dark matter comprised of LED black holes in the early Universe, $\fbh$. The solid coloured lines show the isotropic background light constraints from Fig. \ref{fig:iblConstraint}, the dotted lines show the CMB constraints from Fig. \ref{fig:cmb}, the dashed coloured lines show the galactic-centre gamma-ray constraints from Fig. \ref{fig:GCconstraints}, and the dot-dashed lines show the BBN constraints from Fig. \ref{fig:BBNConstraint}. They grey area at the far right shows the constraints on macroscopic PBHs from microlensing of M31~\cite{Smyth:2019whb}. The grey hatched region in the top right of each plot corresponds to the parameter space for which PBHs would survive until today with an energy density greater than the observed  dark matter density. The shaded region covers the parameter space which is ruled out by at least one  of the constraints. The first plot shows our updated constraints for 4D PBHs while each of the other five plots show the constraints for a different number of extra dimensions, $n$, while keeping the scale of quantum gravity, $M_\star$, fixed at 10 TeV. The black solid lines in the plots with $n>0$ show the different masses and abundances of PBHs produced by high energy collisions in the early Universe assuming $M_\star=10$ TeV and allowing the reheating temperature to vary. The black lines are very steep because for a given value of $M_\star$ only a small range of $\Mbh$ can be produced. The vertical grey dashed line shows the cutoff point for a given number of extra dimensions where PBHs heavier than that line survive until today and PBHs lighter would have entirely evaporated at some point in the past. }
    \label{fig:combinedConstraints}
\end{figure}

The combined constraints on the initial fraction of dark matter in the form of PBHs are shown in Fig.~\ref{fig:combinedConstraints} for $M_\star = 10$ TeV, and $n = 2,3,4,5$ and $6$ extra dimensions as well as for the regular 4D scenario, where $M_\star \equiv M_{pl}$, denoted $n=0$. Features are qualitatively similar for different $n$. 
 
At low masses, rapid evaporation leads to excessive injection of high-energy hadrons and photons during and after BBN. at higher masses ($\gtrsim 10^8$--$10^{14}$ g), longer lifetimes allow BH decay to take place after recombination, leading to strong constraints from the rescattering of CMB photons on the higher ionization floor. BHs that survive longer still produce an isotropic extragalactic signal, as well as a flux of gamma rays from the Milky Way halo. As $n$ rises from 2 to 6, the Galactic flux of gamma rays moves into, and then out of, the INTEGRAL/SPI energy window, explaining how the isotropic background light and Galactic constraints trade places as the dominant limits with varying $n$. 

For $n \ge 5$ and the 4D case of $n=0$, there is a small gap in the combined constraints between the BBN constraints on low-mass PBHs and the constraints on PBHs that survive until after recombination. This gap is due to the limited range over which decaying DM BBN bounds can be recast as PBH constraints. It is expected that the BBN constraints could be extended to higher-masses, closing the gap, by directly calculating the primordial nuclei abundances in the presence of LED PBHs.

Similarly, for $2 \le n \le 4$, the BBN bounds weaken significantly for black holes which  evaporate completely between the time of BBN and recombination. This creates a window of PBH masses for which the constraints allow $\fbh \gg 1$ corresponding to a scenario where in the early universe dark matter is dominated by PBHs but those PBHs completely evaporate leaving the stable dark matter density observed today. This window of weak constraints is due in part to our conservative approach to estimating the photodissociation of light nuclei expected from PBHs with temperatures greater than the QCD scale as discussed in Sec.~\ref{sec:photohadro}. A dedicated calculation of the effect of PBHs on primordial abundances may be able to set stronger constraints in this mass range and to some extent close this weakly constrained window.

The galactic centre constraints in Fig. \ref{fig:combinedConstraints} have a somewhat different shape and cover a smaller mass range compared to those presented in Fig. \ref{fig:GCconstraints}. This is because Fig. \ref{fig:GCconstraints} shows the constraints in terms of the PBH mass and abundance today, whereas Fig. \ref{fig:combinedConstraints} shows the constraints in terms of the PBH properties before evaporation occurred. For massive PBHs that have only evaporated a negligible fraction of their mass since their formation, their relative abundance is unchanged since the early Universe so $\fbhToday \approx \fbh$. Therefore, for larger masses, the galactic centre constraints in Fig. \ref{fig:GCconstraints} and Fig. \ref{fig:combinedConstraints} look the same. However, the smaller masses presented in Fig. \ref{fig:GCconstraints} correspond to a very narrow range of initial PBH masses in Fig. \ref{fig:combinedConstraints} for which the PBHs just happen to be in the final stages of evaporation today. This leads to the leftmost end of the galactic constraints being ``compressed'' in Fig.~\ref{fig:combinedConstraints}.

The highest mass PBHs in the range presented in Fig. \ref{fig:combinedConstraints} are constrained by microlensing of stars in the M31 galaxy \cite{Smyth:2019whb}. These constraints, shown in grey, are the same for all values of $n$ and therefore were not recalculated in this work because the presence of LEDs would not affect microlensing. The mass range between the grey microlensing constraints and the coloured evaporation constraints is completely unconstrained so long as PBHs are not more abundant than dark matter, corresponding to $\fbh \le 1$. The region of parameter space where PBHs survive until today and would be more abundant than the observed dark matter abundance is shown in Fig.~\ref{fig:combinedConstraints} by the grey hatched region.

\subsubsection{Relic abundance of LED black holes as dark matter}

For $n\ge2$, the solid black lines in Fig.~\ref{fig:combinedConstraints} show the predicted abundance and mass of PBHs produced by energetic collisions in the early Universe assuming that $M_\star=10$ TeV. These lines appear vertical because they have an extremely steep slope, where each point along the line corresponds to the expected mass and fraction obtained by varying the reheating temperature, $T_{\rm RH}$. As shown in Fig. \ref{fig:BHtoday}, a small change in $T_{\rm RH}$ corresponds to a very large change in the abundance of PBHs.  We solve for the energy density of BHs as described in Sec.~\ref{sec:BHevo}, and define the initial fraction $\fbh$ in Eq.~\eqref{eq:fbh} at the time when BH mass is largest under accretion and evaporation.  Due to the strong dependence of abundance on $T_{\rm RH}$, and a comparatively weak dependence of PBH mass on $T_{\rm RH}$, the predicted abundance lines are very steep. Therefore, for a given number of LEDs, there is a narrow predicted mass range for PBHs that would be produced with a fixed $M_\star$.

In the case of $n=2$, the PBHs produced are sufficiently heavy that they would only have evaporated a negligible fraction of their mass. These surviving PBHs can comprise all of the dark matter. This scenario where $n=2$ and the fraction of dark matter made up of PBHs, $\fbh=1$, would correspond to BHs with a mass of $\sim 10^{21}$ g. These BHs are too heavy to be constrained by evaporation bounds and lighter than any lensing constraints, making them a viable unconstrained dark matter candidate.

PBHs produced in theories with $n>2$ and $M_\star=10$ TeV would have entirely evaporated before today and are therefore not dark matter candidates. However, the narrow mass window does make specific predictions about when they finish evaporating, pointing at their most promising paths to discovery. For $n=3$, the PBHs finish evaporating after recombination, so that the most likely cosmological method of discovering them is from their impact on CMB anisotropies or through the isotropic X-ray and gamma ray signal they produce.  PBHs with $n=4\textrm{ or } 5$ complete their evaporation earlier, before recombination such that their most apparent cosmological imprint would stem from the destruction of primordial nuclei formed during BBN. For scenarios with $n=6$, the PBHs complete evaporating so early that they would be entirely gone before BBN begins. This makes it very difficult to constrain the existence of $n=6$ PBHs with $M_\star = 10$ TeV. However, the possibility of a very large abundance of PBHs forming and evaporating to all particle types in the very early Universe raises the intriguing scenario that the PBHs may have evaporated to stable dark matter particles, yielding a non-thermal (in the cosmological sense) relic abundance production mechanism. Evaporation to dark matter particles can be incorporated into any of the scenarios with $n \ge 3$, but the short hot lifespan of $n=6$ makes them especially interesting scenarios to explore in future work. 

\subsubsection{Comparison with prior four-dimensional results}

\begin{figure}
    \centering
    \includegraphics[width=1\textwidth]{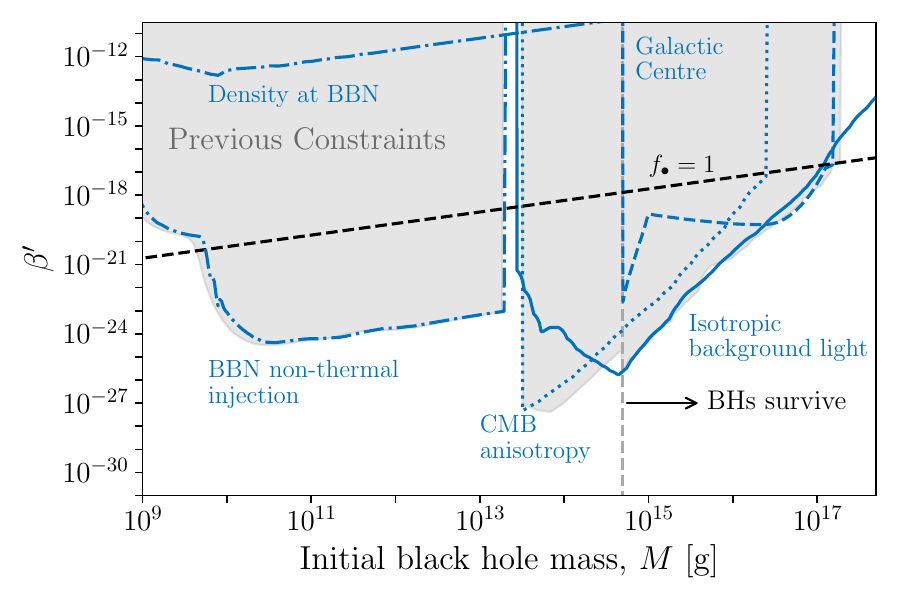}
   \caption{Updated constraints on 4D PBHs in terms of the BH abundance parameter $\beta'$ used in previous studies of 4D PBHs, defined in \eq~\eqref{eq:betaPrime}. The isotropic background light, galactic centre, CMB, and BBN non-thermal injection constraints are the same as those in Fig. \ref{fig:combinedConstraints} except converted from $\fbh$ to $\beta'$. The weaker ``Density at BBN'' constraint is from ensuring that PBHs are not overabundant at the time of neutrino decoupling as shown in Fig. \ref{fig:limit_alterbbn}. The dashed black line indicates the value of $\beta'$ that corresponds to the energy density of PBHs equalling that of dark matter. The vertical grey dashed line shows the cutoff point where PBHs heavier than that line survive until today. The grey shaded region shows the constraints on 4D PBHs set in previous work. These include BBN \cite{Keith:2020jww}, CMB anisotropies \cite{Stocker:2018avm}, isotropic background light \cite{Chen:2021ngo}, galactic centre photons \cite{Auffinger:2022dic}, and positron annihilations in the galactic centre \cite{DeRocco:2019fjq}.
   }
    \label{fig:4DBetaPrimecombinedConstraints}
\end{figure}

Many previous studies of 4D PBHs (see Ref.~\cite{Carr:2020gox} for a review) set constraints on the abundance of PBHs not in terms of $\fbh$, but instead in terms of $\beta'$ which is defined as
\begin{equation}\label{eq:betaPrime}
    \beta' \approx 7.06\times10^{-18}\Omega_{PBH} \bigg(\frac{M}{10^{15} \textrm{ g}}\bigg)^{1/2} = 7.06\times10^{-18}\fbh \Omega_{DM} \bigg(\frac{M}{10^{15} \textrm{ g}}\bigg)^{1/2} .
\end{equation}
To make comparisons between the PBH constraints computed in this and previous work simpler, the 4D PBH constraints are shown in terms of $\beta'$ in Fig. \ref{fig:4DBetaPrimecombinedConstraints}.  The combined constraints in Fig. \ref{fig:4DBetaPrimecombinedConstraints} also includes the BBN constraints due to changes in the expansion of the Universe as shown in  Fig. \ref{fig:limit_alterbbn}, converted from $\beta_{\rm dec}$ using Eq.~\eqref{eq:fconversion}.

The grey shaded region in Fig. \ref{fig:4DBetaPrimecombinedConstraints}, shows a selection of the strongest constraints on low-mass 4D PBHs from previous work. This region combines constraints set with BBN \cite{Keith:2020jww}, CMB anisotropies \cite{Stocker:2018avm}, isotropic photons \cite{Chen:2021ngo}, galactic centre photons \cite{Auffinger:2022dic}, and galactic centre positron annihilation \cite{DeRocco:2019fjq}. These are generally very similar to the strongest constraints set in this work, although there are a few differences worth noting. As discussed in Sec.\ref{sec:galactic}, the 4D constraints we have set using positron annihilation in the galactic centre are stronger than those previously set in Ref. \cite{DeRocco:2019fjq}. It should also be noted that Refs. \cite{Iguaz:2021irx,Chen:2021ngo} have set stronger constraints using the isotropic X-ray and gamma ray flux by modelling astrophysical sources. However, these  are dependent on the astrophysical source model used, though our results are stronger than the conservative background-agnostic constraints of Ref. \cite{Chen:2021ngo} The isotropic background light bounds set here  are stronger than those in Ref. \cite{Chen:2021ngo} for lower mass PBHs and weaker for higher mass PBHs. For lighter PBHs that would have completely evaporated this difference is driven by different approaches in calculating the secondary spectrum of photons from unstable evaporation products. For PBHs that survive until today, the difference is driven by differing assumptions for the parameterization of the Milky Way halo. Finally, the CMB constraints due to energy injection during the dark ages are a factor of $\sim 6$ weaker than those presented by the authors of \texttt{ExoCLASS} in Ref. \cite{Stocker:2018avm}. However, even with a fresh installation of \texttt{ExoCLASS} we were unable to exactly reproduce their results---our inclusion of more precise secondary spectra yields a factor of 2 improvement over constraints found using the public code as-is. This discrepancy is possibly attributable to the updates to \texttt{ExoCLASS} since Ref. \cite{Stocker:2018avm} was published, or a different choice of priors or nuisance parameters.  

\subsection{Other Constraints} \label{sec:otherConstraints}
Previous analyses have set constraints on the existence of light 4D PBHs using more methods than we have employed in this article. In this section we discuss some of those constraints and whether they are expected to be important for the study of LED black holes.

Positrons directly injected into the interstellar medium (ISM) from BH evaporation can contribute to the local cosmic ray flux. Since these are predominantly at sub-GeV energies, they are strongly affected by solar modulation and associated uncertainties. Ref. \cite{Boudaud2019}  placed constraints on 4D PBH evaporation for $\Mbh \sim 10^{14}$--$10^{17}$~g, using data from the Voyager I spacecraft, which has recently crossed the heliopause. These are subdominant to the more recent constraints from gamma ray emission using INTEGRAL data derived by Ref.~\cite{Laha:2020ivk}. Since our INTEGRAL/SPI galactic constraints use the same dataset as Ref.~\cite{Laha:2020ivk}, we anticipate that the Voyager I constraints would be similarly subdominant in the LED scenario.

Dwarf spheroidal galaxies are a prime target for gamma ray searches for dark matter decay or annihilation signatures thanks to their high mass-to-light ratio, which implies a low standard model background and a large prospective signal. Ref.~\cite{Siegert:2021upf} recently analyzed $\sim 1$ Ms of observations of the galaxy Reticulum II with INTEGRAL/SPI over energies 25-8000 keV. Though this leads to improved limits on DM annihilation, the resulting limits on PBH decay are weaker than galactic centre analyses. 

Radio data from the inner Galactic Centre have been used to constrain 4D PBHs \cite{Chan:2020zry}. Large magnetic fields cause ultrarelativistic electrons and positrons to cool via synchrotron radiation thus producing an observable radio signal. In the case of LED PBHs this is most likely to be a viable observational channel for $n=5$ where PBHs that survive until today can be hot enough to produce ultrarelativistic electrons. However, constraints on 4D PBHs from radio data are always weaker than constraints based on X-ray and gamma-ray observations and therefore including radio data in this analysis is unlikely to improve the constraints we have set on LED PBHs.
    
Ultra-light PBHs could dominate the very early Universe and entirely evaporate before BBN evading all bounds presented in this work. However, these PBHs and associated curvature perturbations could source a measurable stochastic gravitational wave background (SGWB) \cite{Papanikolaou:2020qtd}. Recently, that SGWB has been used to produce constraint forecasts for future space-based gravitational wave interferometers \cite{Kozaczuk:2021wcl}. Due to the different lifetime and production mechanism of LED PBHs, these forecasts must be recomputed for the case of LEDs. Some of these constraints would apply to PBHs with masses lower than the BBN constraints presented here.

The evaporation of PBHs during the epoch of star formation and reionization could leave imprints in the high-redshift 21cm signal by heating and ionizing intergalactic gas. Several studies have presented current or future limits, considering the evaporation of 4-dimensional PBHs, either motivated by the recent detection of a deep 21cm absorption trough by the EDGES \cite{Bowman:2018yin} experiment \cite{mittal2021constraining,Clark:2018ghm,Cang:2021owu,Halder:2021rbq,Tashiro:2021kcg,Tashiro:2021xnj} or looking ahead to large-scale experiments such as the Square Kilometer Array \cite{Mack:2008nv}. Many other studies have examined the impact of matter accretion onto macroscopic PBHs that might seed early structure formation or produce X-ray backgrounds \cite{Tanaka:2015sba,Yang:2020zcu,Hektor:2018qqw,Mena:2019nhm,Villanueva-Domingo:2021cgh}. These studies and others highlight the potential for future high-redshift 21cm observations to be highly constraining of exotic energy injection sources during the Dark Ages and the epoch of Cosmic Dawn. We expect LED PBHs may similarly have a strong impact on future 21cm observables.

A bound on PBH evaporation in the galaxy was recently placed based on measurements of the ISM temperature in the Leo T dwarf galaxy. \cite{Kim2021}. These require careful accounting of heating and cooling effects in the ISM -- based on the results of \cite{Kim2021}, which are stronger than the INTEGRAL constraints of \cite{Laha:2020ivk} between $1$ and $3 \times 10^{17}$ g, they could lead to improved limits in a small portion of the parameters space for LED BHs. 

Finally, if the compactified extra dimensions have a toroidal geometry, the production and subsequent decay of Kaluza-Klein (KK) modes during reheating sets constraints such that any reheating temperature that would result in PBHs forming would already be severely constrained \cite{Hannestad:2001nq}. However, constraints based on the production and decay of KK modes are highly dependent on the compactification geometry, the decay products and the existence of additional branes \cite{Hannestad:2001nq}. Conversely, the PBH results in this work are only sensitive to the precise compactification geometry when $r_h \sim R$ (or alternatively stated as $\Mbh \sim M_{\rm 4D}$) and the results for all other values of $\Mbh$ are insensitive to such details. This makes observational constraints based on KK mode production and PBH production complementary to each other. 

\section{Conclusions}
\label{sec:conclusions}
In this article we have derived the full cosmological evolution of PBHs in the presence of LEDs including their production, accretion and evaporation history. We then derived bounds on the existence of those low mass PBHs using BBN, CMB, isotropic photon flux, and galactic centre X-rays. In doing so, we have also recomputed or updated the constraints on 4D PBHs from each of those sources. The constraints on heavier PBHs from gravitational lensing, mergers, and accretion rely on physics at scales larger than the size of the LEDs and therefore will be unchanged from previous analysis.

The abundance and mass of the PBHs for a given number of extra dimensions depend on $M_\star$ and $T_{\rm RH}$. We have set conservative constraints on the $M_\star$-$T_{\rm RH}$ parameter space by ensuring that the PBHs are not overabundant. Stricter constraints could be set on the properties of the extra dimensions by ensuring the produced PBHs are not ruled out by the astrophysical constraints derived here. To do so would require recomputing the astrophysical constraints over a full range of $M_\star$ values and has been left to future work.

We have also found that in the case of two LEDs, the PBHs produced in the early Universe would survive until today and could, with the appropriate reheating temperature, comprise the entirety of the dark matter abundance observed today. In the cases of $n>2$, PBHs would still be created in the early Universe however they would be light enough such that they would have evaporated before today. In those cases the PBHs could still have interesting cosmological impacts even if they are not a dark matter candidate.

In addition to their prospect as dark matter candidates, black holes can produce all gravity-coupled degrees of freedom as they evaporate, as long as the BH temperature is high enough, and the particle mass is kinematically accessible. In the case of BHs produced at colliders, this provides a potential window into the dark sector \cite{Song:2019lxb}. PBHs produced in the early Universe could also evaporate to yield the relic abundance of dark matter \cite{Bernal:2020bjf,Gondolo:2020uqv,Bernal:2021bbv,Barman:2021ost,Sandick:2021gew,Cheek:2021odj,Cheek:2021cfe}; this behaviour would change in the presence of extra dimensions. 

The possibility of large extra dimensions opens a new direction in the search for primordial black holes, including the alluring possibility of producing PBH dark matter without relying on large or non-Gaussian primordial fluctuations. The distinctive evaporation rate and spectra of these BHs mean that any positive detection would point directly at the existence of higher spatial dimensions and provide tantalizing clues about the origin of the Planck scale, bringing together two of the deepest mysteries of the cosmos: dark matter, and the unification of gravity with particle physics. 

\acknowledgements
We thank George Johnson, Thomas Siegert, Patrick St\"ocker, and Marco Ajello for helpful correspondence.  AF is supported by a McLaughlin Fellowship. KJM is supported by the National Science Foundation under Grant No.~2108931, a Goodnight Early Career Innovator Award, and a Visiting Fellowship at the Perimeter Institute. NS thanks the UK Science and Technology Facilities Council (STFC) for funding this work through
support for the Quantum Sensors for the Hidden Sector (QSHS) collaboration under grants ST/T006102/1,
ST/T006242/1, ST/T006145/1, ST/T006277/1, ST/T006625/1, ST/T006811/1, ST/T006102/1 and ST/T006099/1. ACV is supported by the Arthur B.~McDonald Canadian Astroparticle Physics Research Institute and NSERC, with equipment funded by the Canada Foundation for Innovation and the Province of Ontario, and housed at the Queen's Centre for Advanced Computing. Research at Perimeter Institute is supported by the Government of Canada through the Department of Innovation, Science, and Economic Development, and by the Province of Ontario.

\appendix
\section{Detailed solutions of BH mass spectrum}
\label{sec:massspecapp}
\textbf{\textit{Approximate solutions assuming radiation domination and non-evaporating BHs.}} We start from an approximate yet more intuitive approach to solve for the BH mass spectrum.
 Since BH accretion appears to be instantaneous and the asymptotic mass $M_{as}$ is independent of the initial BH mass, we can assume all BHs created at a temperature $T_i$ (and a time $t_i$) obtain the mass $M_{as}$. Neglecting BH evaporation during the accretion process, their mass spectrum follows
\begin{equation}
    \dfrac{1}{a^3}\dfrac{d}{dt}(a^3h_M)=\Gamma_M\delta\left(\Mbh-M_{as}(T_i)\right)\,,
    \label{eq:dndM}
\end{equation}
where $h_M=dn_\bullet/dM$ and $\Gamma_M=\int dM d\Gamma/dM$ with $d\Gamma/dM$ given in Eq.~\eqref{eq:BHprodrate}. The Dirac delta function on the right hand side of the equation indicates that BH of a specific mass $\Mbh$ after accretion can only be produced at a temperature $T_i$ which satisfies the condition $\Mbh=M_{as}(T_i)$. Since the temperature is also a function of time $t$, Eq.~\eqref{eq:dndM} translates into
\begin{equation}
    \dfrac{\partial h_M}{\partial t}+3Hh_M=\dfrac{\Gamma_M}{|dM_{as}/dt|}\delta(t-t_i)=\Gamma_M\left|\frac{dM_{as}}{dT_i}\frac{dT_i}{dt}\right|^{-1}\delta(t-t_i)\,,
    \label{eq:dndM2}
\end{equation}
where $dM_{as}/dT$ can be derived from Eq.~\eqref{eq:asympmass}. Integrating Eq.~\eqref{eq:dndM2} over an infinitesimal time step around $t_i$, we find
\begin{equation}
    h_M(M_{as})=\Gamma_M\left|\frac{dM_{as}}{dT_i}\frac{dT_i}{dt}\right|^{-1}.
    \label{eq:dndM3}
\end{equation}
If radiation dominates, $dT/dt$ is given in Eq.~\eqref{eq:dTdt} and
\begin{equation}
    h_M(M_{as}(T_i))=\Gamma_M \left(\sqrt{\dfrac{16\pi^3}{45}g_\star}\dfrac{n+1}{n-1}\dfrac{M_{as}}{M_{pl}}T_i^2\right)^{-1}\,.
    \label{eq:hM_rd}
\end{equation}
After $t_i$
the right hand side of Eq.~\eqref{eq:dndM2} is vanishes and the mass spectrum drops as $a^{-3}\propto T^3$ in a radiation dominated universe.

\textbf{\textit{Exact solutions.}}
Now we turn to a more rigorous treatment without assuming instantaneous accretion. At time $t_i$, the number of microscopic BHs produced is $h_{t}\equiv \frac{d n_\bullet}{d t}|_{t=t_i}$. 
Since the accreted BH mass is fairly insensitive to the initial masses, we assume all BHs are created at a minimum mass required for efficient accretion $M_i=M_{i,\min}(T_i)$, defined in Eq.~\eqref{eq:Mimin}, and they evolve collectively afterwards. We use $M(t_i,t)$ to denote the mass of BHs that evolve from $t_i$ to $t$, and $h_t({t_i,t})$ to show the evolution of the BH mass spectrum. The latter is described by
\begin{equation}
    \dfrac{1}{a^3}\dfrac{d}{dt}(a^3h_t)=\Gamma_M\delta(t-t_i)\,.
    \label{eq:dndt}
\end{equation}
 Eq.~\eqref{eq:dndt} can be further split into two equations, one for BH production at $t_i$
\begin{equation}
    h_t(t_i,t=t_i)=\Gamma_M(t_i)\,,
    \label{eq:htinitial}
\end{equation}
and the other for the redshift of the spectrum at $t>t_i$
\begin{equation}
    \dfrac{\partial h_t}{\partial t}+3Hh_t=0\,.
\end{equation}
The evolution of BH mass follows Eq.~\eqref{eq:dMdtcombine}, which reads
\begin{equation}
    \dfrac{\partial M(t_i,t)}{\partial t}=\left(-\alpha+\beta\dfrac{T^4}{T_H^4}\right)T_H^2\,.
\end{equation}
At time $t$, the BH energy density is given by
\begin{equation}
    \rho_{\rm BH}(t)=\int_{t_{\rm RH}}^{t}h_t(t_i,t)M(t_i,t)dt_i\,,
\end{equation}
where $t_{\rm RH}$ is the time of reheating, and the radiation density evolves as
\begin{equation}
    \dfrac{d\rho_r}{dt}+4H\rho_r=-\int _{t_{\rm RH}}^{t}h_t(t_i,t)\dfrac{\partial M(t_i,t)}{\partial t}dt_i-\Gamma_M M(t_i,t_i)\,,
\end{equation}
where the loss or gain of radiation is caused by the change in BH mass. The second term on the right-hand-side of the equation can usually be neglected since the accretion energy loss is supposed to be much more efficient than Planckian mass BH productions. Combining BH and radiation, the expansion of the Universe is governed by the Friedmann equation
\begin{equation}
    H^2=\dfrac{8\pi}{3M_{pl}^2}(\rho_r+\rhobh)\,,
    \label{eq:Friedmann}
\end{equation}
where $\rho_r$ is given in Eq.~\eqref{eq:rhor}. Eqs.~\eqref{eq:htinitial} to~\eqref{eq:Friedmann} provide a complete set of integro-differential equations to solve for the BH mass $M(t_i,t)$ and mass spectrum $h_t(t_i,t)$. 

Instead of solving the above equations directly, we can reduce the number of equations by switching to the temperature basis, where we find
\begin{equation}
    h_T(T_i,T=T_i)=\Gamma_M(T_i)\left(\Bigr\rvert\dfrac{dT}{dt}\Bigr\rvert_{T=T_i}\right)^{-1}\,,
    \label{eq:hTinitial}
\end{equation}
\begin{equation}
    \dfrac{\partial h_T}{\partial T}\dfrac{dT}{dt}+3Hh_T=0\,,
\end{equation}
\begin{equation}
    \dfrac{\partial M(T_i,T)}{\partial T}=\left(-\alpha+\beta\dfrac{T^4}{T_H^4}\right)T_H^2\left(\dfrac{dT}{dt}\right)^{-1}\,,
    \label{eq:partialMpartialT}
\end{equation}
\begin{equation}
    \rho_{\bullet}(T)=\int^{T_{\rm RH}}_{T}h_T(T_i,T)M(T_i,T)dT_i\,,
    \label{eq:rhoBHatT}
\end{equation}
\begin{equation}
    \dfrac{d\rho_r}{dT}\dfrac{dT}{dt}+4H\rho_r=-\int ^{T_{\rm RH}}_{T}h_T(T_i,T)\dfrac{\partial M(T_i,T)}{\partial t}dT_i\,,
    \label{eq:drhodT}
\end{equation}
where $h_{T}(T_i,T)\equiv \frac{d n}{d T}|_{T=T_i}$ and Eq~\eqref{eq:Friedmann} stays the same. We can substitute Eqs.~\eqref{eq:rhor} and~\eqref{eq:rhoBHatT} into Eq.~\eqref{eq:drhodT} to find $dT/dt$
\begin{equation}
    \dfrac{dT}{dt}=-T\left(\dfrac{1}{M_{pl}}\sqrt{\dfrac{8\pi}{3}(\rho_r+\rho_{\rm BH})}+\dfrac{1}{4\rho_r}\int ^{T_{\rm RH}}_{T}h_T(T_i,T)\left(-\alpha+\beta\dfrac{T^4}{T_H^4}\right)T_H(M(T_i,T))^2dT_i\right)\,.
    \label{eq:dTdtlong}
\end{equation}
Since the variation of $g_\star$ is mild during accretion, we set $dg_\star/dt=0$. Eq.~\eqref{eq:dTdtlong} can further be plugged into Eqs.~(\ref{eq:hTinitial}--\ref{eq:partialMpartialT}) to obtain the equations for $h_T$ and $M$, where Eq.~\eqref{eq:htinitial} and $M(T_i,T=T_i)=M_{i,\min}(T_i)$ serve as initial conditions. However, this formalism may not work if BHs dominate the energy density of the Universe after accretion and reheat the plasma significantly when they decay. It is crucial to solving the equations on the time basis to keep track of BH evolution in this scenario.

In the above integro-differential equations, $T$ and $t$ always appear in the differentials while $T_i$ and $t_i$ appear in integrals. After we obtain $h_T$ and $M$, we can map it to the mass spectrum using the relation
\begin{equation}
    h_M(M,T)\equiv \dfrac{dn}{dM}\Bigr\rvert_{M=M(T_i,T)}=h_T(T_i,T)\left(\dfrac{\partial M(T_i,T)}{\partial T_i}\right)^{-1}\,.
\end{equation}

\begin{figure}
    \centering
    \includegraphics[width=0.48\textwidth]{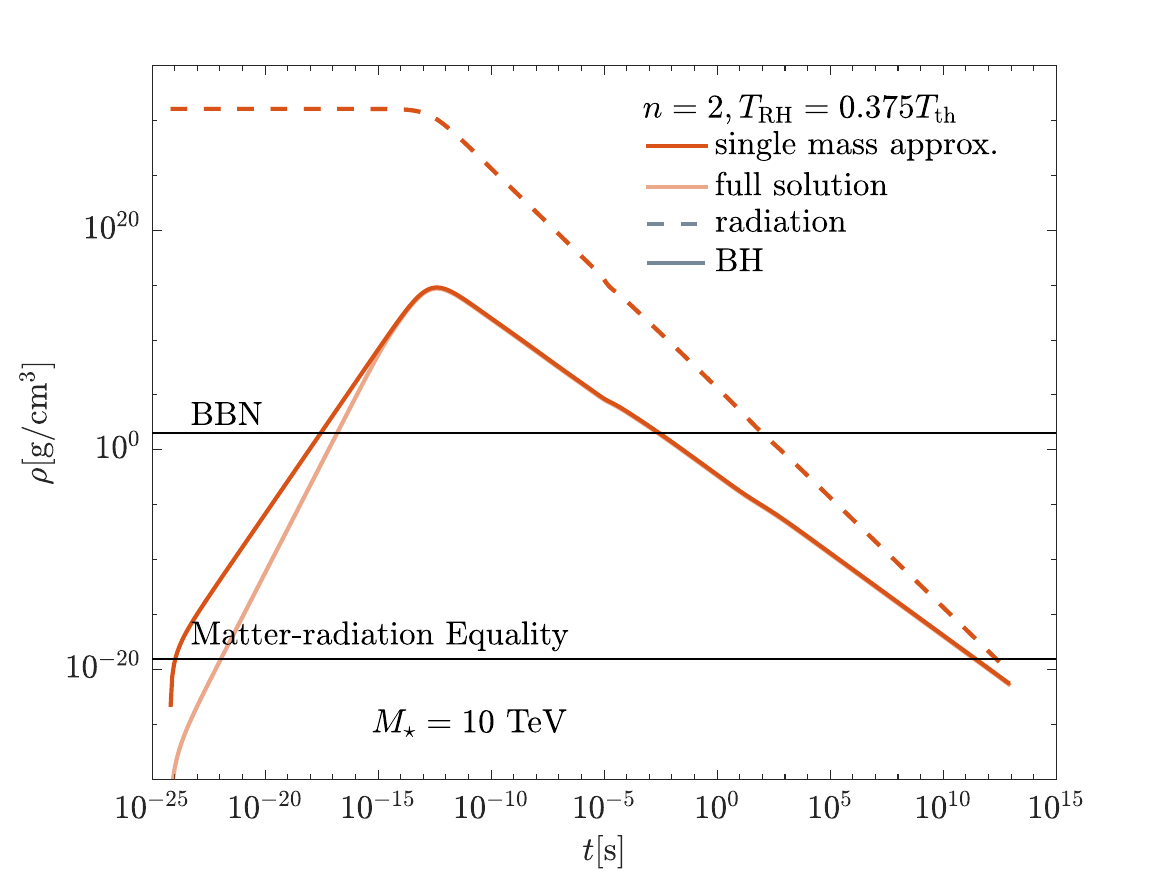}
    \includegraphics[width=0.48\textwidth]{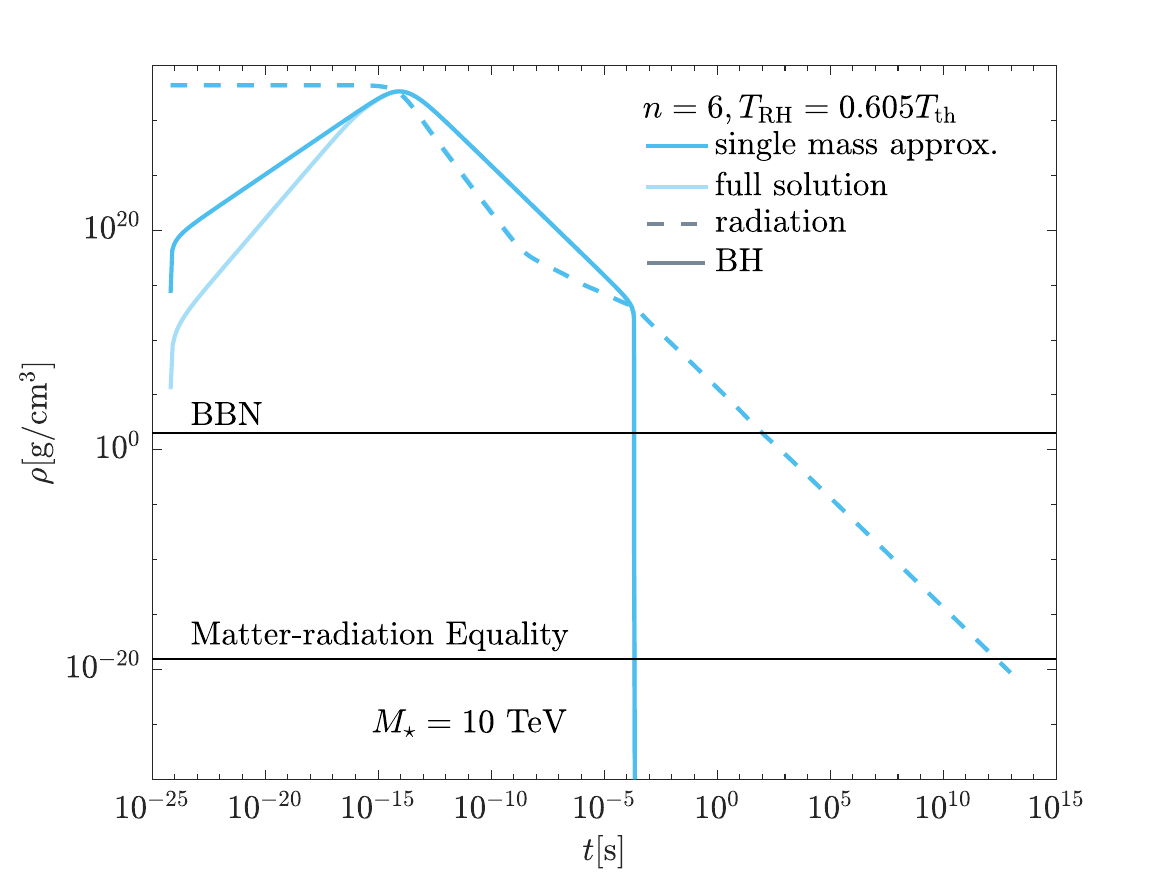}    
    \caption{Evolution of BH and radiation energy density. Left: The lighter orange lines describe the solutions to the full integro-differential equations, and the darker orange lines represent the solutions of single mass approximation at production (see text in Sec.~\ref{sec:BHevo}). Solid and dashed lines depict the energy density of BHs and radiation respectively. Horizontal lines show the expected radiation density of the plasma when BBN and matter-radiation equality take place in standard cosmology. We assume the fundamental scale $M_\star=10$~TeV and $n=2$ extra dimensions with the reheating temperature $T_{\rm RH}=0.375T_{\rm th}=1.09$~TeV. Right: Same as the left panel but for 6 extra dimensions and $T_{\rm RH}=0.605T_{\rm th}=3.77$~TeV. }
    \label{fig:rhooft_exact}
\end{figure}

\begin{figure}
    \centering
    \includegraphics[width=0.6\textwidth]{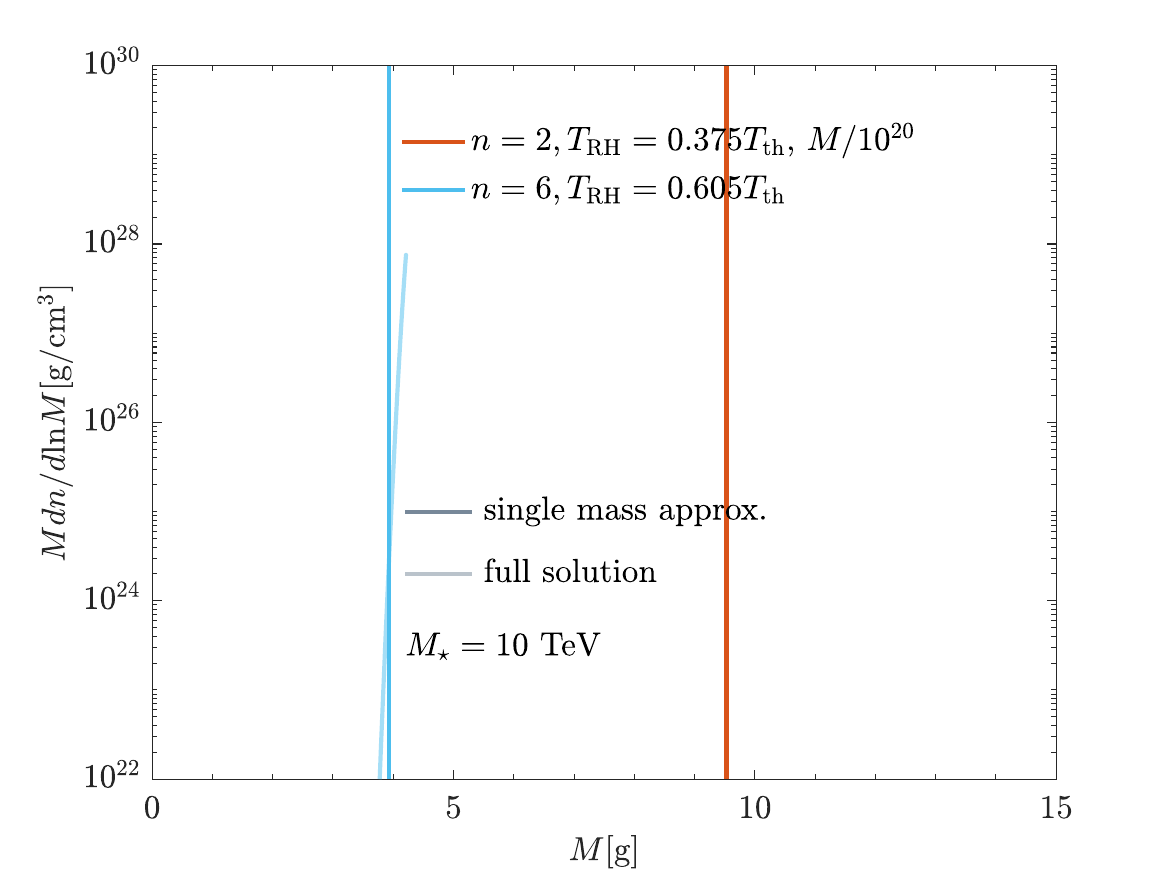}
    \caption{BH mass spectrum at $t=10^{-10}$~s. The orange and blue lines show the BH mass distribution for $n=2$, $T_{\rm RH}=0.375T_{\rm th}=1.09$~TeV and $n=6$, $T_{\rm RH}=0.605T_{\rm th}=3.77$~TeV, respectively. The lighter coloured lines depict the solutions to the full integro-differential equations, and the  darker vertical lines are obtained from single mass approximation. BH mass in $n=2$ is rescaled to fit into the plot range of the figure. The Planck scale $M_\star=10$~TeV is assumed.}
    \label{fig:hM_exact}
\end{figure}

We solve the full integro-differential equations on time basis, and show the evolution of BH energy density in Fig.~\ref{fig:rhooft_exact}, as well as the mass spectrum at $10^{-10}$~s in Fig.~\ref{fig:hM_exact}. We choose two typical scenarios. In the first scenario, $n=2$, $T_{\rm RH}=1.09$~TeV, the BH energy density remains subdominant until the plasma temperature drops below 0.75~eV. In the other scenario, BHs dominate the energy budget of the Universe at about $10^{-15}$s, and then evaporate away before BBN. In both cases, we find good agreement with the energy density evolution obtained from the single mass approximation described in Sec.~\ref{sec:BHevo}. The mass distribution spreads more in the full solution. However, the peak mass of the distribution agrees with single mass approximation, up to a close to 1 factor.

\section{Secondary particle production  from pion and muon decay}
\label{sec:pimudecay}
We use \texttt{Hazma}~\cite{Coogan:2019qpu} code to compute the secondary gamma from $\pi^0$ decay $\pi^0\rightarrow \gamma\gamma$, and from radiative muon and charged pion decay through the processes $\mu^-\rightarrow e^-\bar{\nu}_e\nu_\mu\gamma$ and $\pi^-\rightarrow l^-\bar{\nu}_l\gamma$ where $l=\mu,\ e$. For the emission of electrons from muon decay, the radiative process is subdominant and we consider the tree-level differential decay spectrum in the rest frame of a muon
\begin{equation}
    \dfrac{d\Gamma_\mu}{dE_e}=\dfrac{G_F}{12\pi^3}\sqrt{E_e^2-m_e^2}\left(E_e(m_\mu^2+m_e^2-2m_\mu E_e)+2(E_e m_\mu-m_e^2)(m_\mu-E_e)\right)\,.
\end{equation}
Given muon energy $E'_\mu$ in the lab frame, the energy and momentum of electron in the lab frame is related to their muon rest frame values via the Lorentz boost
\begin{align}
    E_e&=\gamma_\mu E'_e(1+\beta_\mu\sqrt{1-m_e^2/E_e^{'2}}\cos\theta')\,,\label{eq:Eboost}\\
    p_\parallel&=\gamma_\mu E'_e(\beta_\mu +\sqrt{1-m_e^2/E_e^{'2}}\cos\theta')=\sqrt{E_e^2-m_e^2}\cos\theta\,,\label{eq:pbost}
\end{align}
where $\gamma_\mu=E'_\mu/m_\mu$, $\beta_\mu=\sqrt{1-1/\gamma_\mu^2}$, and $\theta^{(')}$ is the angle between the rest (lab) frame electron momentum and muon momentum. The decay spectrum is also boosted via a Jacobian
\begin{equation}
    \dfrac{d\Gamma_\mu^2}{dE'_ed\cos\theta'}=\begin{vmatrix}
    \frac{dE_e}{dE'_e}&\frac{dE_e}{d\cos\theta'}\\
    \frac{d\cos\theta}{dE'_e}&\frac{d\cos\theta}{d\cos\theta'}
    \end{vmatrix}\dfrac{d\Gamma_\mu^2}{dE_ed\cos\theta}=\dfrac{1}{2}J_{\rm lab}\dfrac{d\Gamma_\mu}{dE_e}\,.
\end{equation}
The last equality holds as the rest frame spectrum is independent of $\cos\theta$. The Jacobian can be evaluated using Eqs.~\eqref{eq:Eboost} and~\eqref{eq:pbost}. Explicitly.
\begin{equation}
    J_{\rm lab}=\dfrac{\beta_e\gamma_e}{\sqrt{\gamma_\mu^2\gamma_e^2(1+\beta_\mu\beta_e\cos\theta')^2-1}}\,,
\end{equation}
where $\beta_e$ and $\gamma_e$ are defined accordingly with $E'_e$. The lab frame electron spectrum is then obtained by integrating over the angular distribution,
\begin{equation}
    \dfrac{d\Gamma_\mu}{dE'_e}=\int d\cos\theta' J_{\rm lab}\dfrac{d\Gamma_\mu}{dE_e}(E'_e,\cos\theta')\,.
    \label{eq:mudecaylab}
\end{equation}
The normalized differential decay spectrum in Eq.~\eqref{eq:flux_sec} is therefore
\begin{equation}
    \dfrac{df_\mu}{dE'_e}=\dfrac{1}{N_\mu}\dfrac{d\Gamma_\mu}{dE_e}(E'_e,\cos\theta')\,,
    \label{eq:mutoe}
\end{equation}
and the normalization
\begin{equation}
    N_\mu=\dfrac{G_F}{24\pi^3}m_e^5\left(3z{\rm arccosh}(z)-(z^2+2)\sqrt{z^2-1}\right)\,,
\end{equation}
where $z\equiv m_\mu/m_e$. In the rest frame, assuming neutrinos are massless with energy $E_{1,2}$ and momentum $\vec{p}_{1,2}$, we have
\begin{equation}
    E_1=m_\mu-E_e-E_2=|\vec{p}_e+\vec{p}_2|\geq |p_e-E_2|\,.
\end{equation}
This sets kinematically limits on the electron energy in the rest frame
\begin{equation}
    m_e\leq E_e\leq \dfrac{m_\mu^2+m_e^2}{2m_\mu}\simeq \dfrac{m_\mu}{2}\,.
    \label{eq:Eelimitrest}
\end{equation}
After boost we find the cutoff of electron energy at $\cos\theta=\pm 1$, i.e.,
\begin{align}
    E'_{e,\min}&=\gamma_\mu m_e\,,\\
    E'_{e,\max}&\simeq \dfrac{\gamma_\mu m_\mu}{2}(1+\beta_\mu\sqrt{1-4/z^2)}\,.
\end{align}
Imposing the condition Eq.~\eqref{eq:Eelimitrest} on the right hand side of Eq.~\eqref{eq:Eboost} we can also find the limits of the angular integral in Eq.~\eqref{eq:mudecaylab},
\begin{align}
    (\cos\theta')_{\min}&=\max\left\{\dfrac{1}{\beta_\mu\beta_e}\left(\dfrac{m_e}{\gamma_\mu E'_e-1}\right),-1\right\}\,,\\
    (\cos\theta')_{\max}&\simeq \min\left\{\dfrac{1}{\beta_\mu\beta_e}\left(\dfrac{m_\mu}{2\gamma_\mu E'_e-1}\right),1\right\}\,.
\end{align}

Next we consider charged pion decay. Since $\pi^-\rightarrow \mu^-\bar{\nu}_\mu$ dominates electron production, we will only consider this decay channel. The kinematics in two-body decay is rather simplified and in the pion rest frame muon obtains a single energy
\begin{equation}
    E_\mu^{\rm CM}=\dfrac{m_\pi^2+m_\mu^2}{2m_\pi}\,,
\end{equation}
and the normalized decay spectrum
\begin{equation}
    \dfrac{df_\pi^2}{dE_\mu d\cos\theta}=\dfrac{1}{2}\delta (E_\mu-E_{\mu,\rm CM})
\end{equation}
Boost this into the lab frame and integrate over $\cos\theta'$ we find
\begin{equation}
    \dfrac{df_\pi}{dE'_\mu}=\dfrac{1}{2\beta_\pi\gamma_\pi\sqrt{E_{\mu,\rm CM}^2-m_\mu^2}}\,,
\end{equation}
where $\beta_\pi$ and $\gamma_\pi$ are similarly defined as before. The limits of muon energy in the lab frame are reached at
\begin{equation}
    E'_{\mu,\min/\max}=\gamma_\pi(E_{\mu,\rm CM}\mp \beta_\pi\sqrt{E^2_{\mu,\rm CM}-m_\mu^2})\,.
\end{equation}
The electron spectrum from $\pi^{\pm}$ decay can be attained directly after integrating over the intermediate muon energy,
\begin{equation}
    \dfrac{df_\pi}{dE'_e}=\int dE'_\mu \dfrac{df_\pi}{dE'_\mu}\dfrac{df_\mu}{dE'_e}\,,
\end{equation}
with $df_\mu/dE'_e$ given in Eq.~\eqref{eq:mutoe}. 

\begin{figure}
    \centering
    \includegraphics[width=0.6\textwidth]{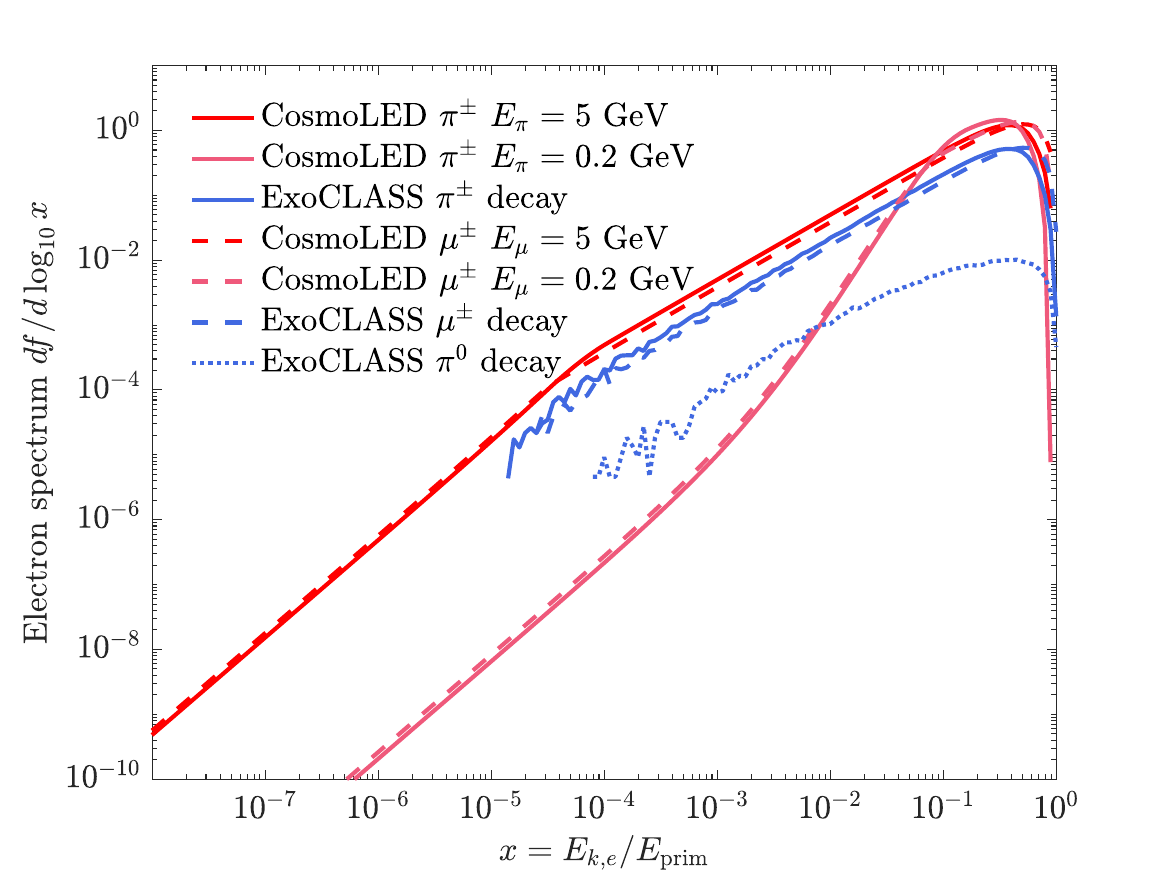}
    \caption{Secondary electron spectrum from pion and muon decay. The solid, dashed and dotted lines show the decay spectrum from $\pi^\pm$, $\mu^\pm$ and $\pi^0$ respectively. The red and pink lines are obtained from CosmoLED (this work) at the primary particle energy $E=5$~GeV and 0.2~GeV. The blue lines depict the secondary electron spectra computed with \texttt{ExoCLASS}, which are independent of primary particle energy.}
    \label{fig:efrommupi}
\end{figure}

The secondary electrons from muon and pion decay are compared with \texttt{ExoCLASS}~\cite{Stocker:2018avm} spectra in Fig.~\ref{fig:efrommupi}. For direct comparison we define $x\equiv E_{k,e}/E_{\rm prim}$, the ratio between the kinetic energy of electron and the energy of primary particles. We do not include $e^{\pm}$ from $\pi^0$ decay, which is considered to be subdominant. The \texttt{ExoCLASS} spectra computed from \texttt{PYTHIA v8.219}~\cite{Sjostrand:2014zea} are independent of energy. We show the spectra at $E_{\rm prim}=5$~GeV, and 0.2~GeV. The high energy spectra are close to that of \texttt{ExoCLASS}, but the difference is more pronounced as the primary particle energy is close to their mass. 

For secondary photons, \texttt{ExoCLASS} does not consider the contribution from muon and charmed pion decay. The \texttt{ExoCLASS} secondary photon spectrum $\pi^0$ decay agrees with that in \texttt{Hazma} at high $\pi^0$ energies.

\section{Derivation of Photon Flux Change from the Universe Expansion} \label{sec:RSFluxChangeDerivation}
In this appendix we derive the rate of change of a differential flux of photons due to the expansion of the Universe as expressed in Eq.~\eqref{eq:fluxChangeRS}. This is done by studying the change in flux over a redshift step of size, $dz$, and taking the limit of $dz\rightarrow 0$.

A flux of photons evolving over a differential redshift step will change due to the Universe expanding in two ways: the number density decreases proportionally to the volume of the Universe and photons lose energy. Due to the change in photon energy, the differential flux changes from $\frac{d\Phi}{dE}\rightarrow\frac{d\Phi}{dE'}$ where
\begin{equation}
    E' = \frac{1+z+dz}{1+z} E = (1 + \frac{dz}{1+z})E \equiv E + \delta E.
\end{equation}
The change in flux over a differential redshift step is therefore
\begin{equation} \label{eq:fluxChangeRS}
\delta\frac{d\Phi}{dE}(E,z) \equiv \frac{d\Phi}{dE'}(E, z + dz) - \frac{d\Phi}{dE}(E, z) = \bigg(\frac{1+z+dz }{1+z}\bigg)^3 \frac{d\Phi}{dE'}(E-\delta E, z) - \frac{d\Phi}{dE}(E, z)
\end{equation}
where changing the flux $\frac{d\Phi}{dE'}$ from redshift $z+dz$ to redshift $z$ requires accounting for the changing in volume and the fact that photons ending at energy $E$ must have originated at energy $E-\delta E$.

Ignoring any terms that are higher than first order in $dz$ Equation \eqref{eq:fluxChangeRS} becomes
\begin{equation}
\delta\frac{d\Phi}{dE}(E,z) = (1 + 3\frac{dz}{1+z}) \frac{dE}{dE'}\frac{d\Phi}{dE}(E-\delta E, z) - \frac{d\Phi}{dE}(E, z).
\end{equation}
Using
\begin{equation}
    \frac{dE'}{dE} = 1 + \frac{dz}{1+z},
\end{equation}
\begin{equation}
    \frac{dE}{dE'} = 1 - \frac{dz}{1+z},
\end{equation}
and
\begin{equation}
    \frac{d\Phi}{dE}(E-\delta E, z) = \frac{d\Phi}{dE}(E, z) - \delta E \frac{d^2\Phi}{dE^2}(E, z)
\end{equation}
the change in flux becomes
\begin{equation}
\delta\frac{d\Phi}{dE}(E,z) = \bigg[1 + 3\frac{dz}{1+z}\bigg] \bigg[1 - \frac{dz}{1+z}\bigg]\bigg[\frac{d\Phi}{dE}(E, z) - \delta E \frac{d^2\Phi}{dE^2}(E, z)\bigg] - \frac{d\Phi}{dE}(E, z).
\end{equation}
Again removing terms higher order in $dz$,
\begin{equation}
\delta\frac{d\Phi}{dE}(E,z) =
2 \frac{dz}{1+z} \frac{d\Phi}{dE}(E, z) - \delta E \frac{d^2\Phi}{dE^2}(E, z).
\end{equation}
and therefore
\begin{equation} \label{eq:fluxChangeExpansion}
    \frac{d\Phi_\textrm{exp}}{dEdz}(E,z) = \frac{2}{1+z}\frac{d\Phi}{dE}(E,z) - \frac{E}{1+z}\frac{d^2\Phi}{dE^2}(E,z)
\end{equation}

This can be related to the cosmological continuity equation by integrating Eq.~\eqref{eq:fluxChangeExpansion} over all energies.
\begin{equation}
    \frac{d\Phi_\textrm{exp}}{dz}(z) = \int_0^\infty dE \bigg[\frac{2}{1+z}\frac{d\Phi}{dE}(E,z) - \frac{E}{1+z}\frac{d^2\Phi}{dE^2}(E,z)\bigg].
\end{equation}
By integrating the second term by parts this leads to
\begin{equation}
    \frac{d\Phi_\textrm{exp}}{dz}(z) =  \frac{2}{1+z}\Phi(z) - \frac{1}{1+z}\bigg(\bigg[E\frac{d\Phi}{dE}(E,z)\bigg]_{E=0}^\infty - \int_0^\infty dE \frac{d\Phi}{dE}(E,z)\bigg).
\end{equation}
The boundary term goes to zero for all physical spectra so that the number density continuity equation is recovered
\begin{equation}
    \frac{d\Phi_\textrm{exp}}{dz}(z) =  \frac{3}{1+z}\Phi(z).
\end{equation}
\section{Numerical Evaluation of EBL Flux} \label{sec:numericEBL}
\subsection{Discrete Differential Equation}

Sec.~\ref{sec:EBL} describes how the EBL X-ray and gamma-ray spectrum instantaneously changes as a function of energy and redshift. This results in a integro-differential equation that cannot be simply integrated to determine the resulting EBL spectrum today. In this appendix we describe the numerical methods used to solve that system.

The EBL contribution was calculated by tracking the evolution of the photon spectrum over discretized redshift steps starting at $z=1100$. The photon flux at the $i^\textrm{th}$ redshift step, $z_i$, is given by
\begin{equation} \label{eq:eblStepSpecChange}
    \frac{d\Phi_{\gamma,\textrm{EBL}}}{dE_i}(E_i,z_i) =  \frac{V_{i-1}}{V_i}\frac{dE_{i-1}}{dE_i}\frac{d\Phi_{\gamma,\textrm{EBL}}}{dE_{i-1}}(E_i,z_{i-1}) e^{-\tau(E_i,z_{i-1},z_i)} + \frac{d\Phi_{\gamma,\textrm{comp}}}{dE_idz_i}(E_i,z_i)  \Delta z + \frac{d\Phi_{\gamma,\textrm{inj}}}{dE_idz_i}(E_i,z_i)  \Delta z 
\end{equation}
where $\Phi_{\gamma,\textrm{EBL}}$ is the extragalactic isotropic photon flux, $E_{i-1}$ and $E_{i}$ are the photon energies at redshifts $z_{i-1}$ and $z_i$ respectively, $V_i$ is the Universe volume at redshift $z_i$, $\tau$ is the absorption probability of a photon with energy $E_i$ travelling between redshift $z_{i-1}$ and $z_i$, and $\Delta z = z_i - z_{i-1}$. 

The second and third term of Eq.~\eqref{eq:eblStepSpecChange} which describe the change in flux due to Compton scattering and photon injection are determined from Eqs.~\eqref{eq:ComptonNetChange} and \eqref{eq:eblInjSpecChange} respectively. Calculating the change in flux due to Compton scattering in this way requires performing an integral for each energy bin in the discretized spectrum. That is computationally slow so often approximations are used to simplify this step. A more in-depth discussion about Compton scattering can be found in the next subsection.

The first term in Eq.~\eqref{eq:eblStepSpecChange} accounts for the change of flux due to the expansion of the Universe and the absorption of photons. While the instantaneous changes in flux due to these processes are described by Eqs.~\eqref{eq:photFluxChangeRS} and \eqref{eq:photFluxChangeABS} separately, it is convenient to combine them into one term that accounts for the total effect.

Evolving the EBL spectrum with the total effect of the Universe expanding between two redshifts also has the advantage of not needing to calculate derivatives as in Eq.~\eqref{eq:photFluxChangeRS}. This is done by directly taking into account the two effects that the expansion of the Universe has on the photon flux. Firstly, the increasing volume decreases the number density of photons. This is accounted for in the $\frac{V_{i-1}}{V_i}$ factor that contributes
\begin{equation}
    \frac{V_{i-1}}{V_i} = \bigg( \frac{1+z_i}{1+z_{i-1}}\bigg)^3,
\end{equation}
Secondly, the expansion causes photons to lose energy via redshifting so that
\begin{equation} \label{eq:redshiftEloss}
    E_i = \frac{1+z_i}{1+z_{i-1}} E_{i-1}.
\end{equation}

As discussed in the next subsection, Compton scattering can sometimes be approximated as causing a fractional energy loss rate for all photons which would be treated as an additional term to Eq.~\eqref{eq:redshiftEloss}. When the fractional energy loss approximation is not used so the only difference between $E_i$ and $E_{i-1}$ in Eq.~\eqref{eq:redshiftEloss} comes from adiabatic expansion, the first term of Eq.~\eqref{eq:eblStepSpecChange} can be written explicitly so that
\begin{equation} \label{eq:eblStepSpecChange2}
    \frac{d\Phi_{\gamma,\textrm{EBL}}}{dE_i}(E_i,z_i) =  \bigg( \frac{1+z_i}{1+z_{i-1}}\bigg)^2 \frac{d\Phi_{\gamma,\textrm{EBL}}}{dE_{i-1}}\bigg( \frac{1+z_i}{1+z_{i-1}} E_{i-1},z_{i-1}\bigg) e^{-\tau} + \frac{d\Phi_{\gamma,\textrm{comp}}}{dE_idz_i}(E_i,z_i)  \Delta z + \frac{d\Phi_{\gamma,\textrm{inj}}}{dE_idz_i}(E_i,z_i)  \Delta z .
\end{equation}

The exponent $\tau$ in the first term of Eq.~\eqref{eq:photonFluxTotalChange} comes from integrating the instantaneous change due to absorption as described in Eq.~\eqref{eq:photFluxChangeABS}. This exponential suppression accounts for the absorption probability over the time step due to photoionization of neutral gas, pair production from atoms and ions, photon-photon scattering, and pair production off the CMB. Depending on the treatment of Compton scattering, it may also be included in the $\tau$ calculation. Assuming that the discretized redshift steps are sufficiently small, $\tau$ can be calculated using 
\begin{equation} \label{eq:iblAttenuation}
    \tau(E,z_{i-1},z_i) \approx \Delta z \frac{d\tau}{dz}(E,z_i),
\end{equation}
where $\frac{d\tau}{dz}$ is determined as in Ref.~\cite{zdziarski1989absorption}. 

With these numerical methods, Eq.~\eqref{eq:eblStepSpecChange2} can be used to evolve the EBL spectrum and determine the expected observed flux today. The two computational bottlenecks in this method are the integrals required in solving the upscattered photon flux from ICS and the change in photons flux from Compton scattering. As discussed in Sec.~\ref{sec:extragalactic}, accounting for photons from ICS does not improve the constraints set from the isotropic X-ray and gamma ray spectrum so ICS can be safely ignored to improve computational speed. On the other hand, the treatment of Compton scattering can have an impact on the strength of the constraints so understanding which approximations can be used requires further discussion.

\subsection{Compton Scattering Approximation}

The instantaneous rate of change to the EBL flux due to Compton scattering is fully described by Eq.~\eqref{eq:ComptonNetChange}. However, when using this method for incorporating the effect of Compton scattering into the discretized evolution of the X-ray background as done in Eq.~\eqref{eq:eblStepSpecChange} there are two potential issues that need to be addressed.

One potential issue is that by assuming the total change in flux due to Compton scattering is equal to $ \frac{d\Phi_{\gamma,\textrm{comp}}}{dEdz} \Delta z $  as done in Eq.~\eqref{eq:eblStepSpecChange} there is an implicit assumption that during a redshift step photons either do not scatter or scatter once. It does not allow for multiple Compton scatters of a single photon within a single step. This is valid as long as the redshift steps are sufficiently short. The maximum scattering rate is for low energy photons in the Thomson limit where $\sigma_c \approx \sigma_T$. Therefore, the condition that must be true for this treatment of Compton scattering to be valid is
\begin{equation}
    \sigma_T n_e(z) \Delta t \ll 1
\end{equation}
where $\Delta t$ is the absolute time of the redshift step.

\begin{figure*}[hbt]
	\centering	\includegraphics[width=0.6\textwidth]{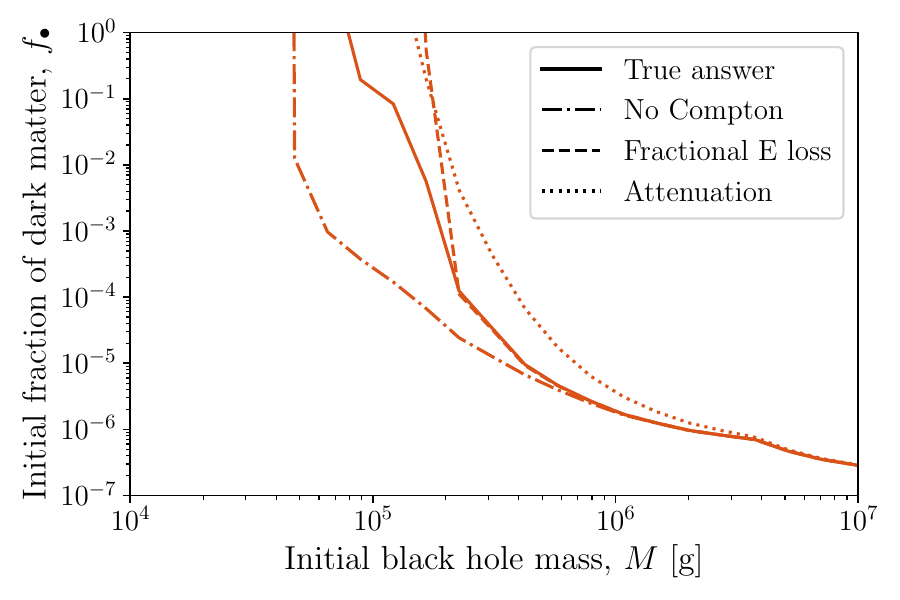}
	\caption{A comparison of EBL constraints for PBHs with $n=2$ and $M_\star=10$~TeV which have fully evaporated before today using different approximations for Compton scattering.}
	\label{fig:comptonApprox}
\end{figure*}

The other issue with this treatment of Compton scattering is that using Eq.~\eqref{eq:ComptonNetChange} to determine the effect of Compton scattering during a redshift step requires computing an integral to determine the change for each energy bin. This can be computationally intensive. Therefore, there are different approximations that can be used depending on the regime of interest and the accuracy needed. They are:
\begin{itemize}
    \item {\it Attenuation} - A simple approximation is to ignore the downscattered photons and assume that all photons that Compton scatter are fully absorbed. This would be implemented by treating Compton scattering as an additional component of $\frac{d\tau}{dz}(E,z)$ in Eq.~\eqref{eq:iblAttenuation} where the Compton component is given by
    \begin{equation}
        \frac{d\tau}{dt}(E,z)\bigg|_\textrm{compton} = n_e(z) \sigma_c(E) .
    \end{equation}
    
    This is generally a conservative and computationally simple approximation to make. This approximation is able to do a good job of estimating how much the flux of high energy photons is attenuated but it breaks down with low energy photons because while they may scatter frequently, they only lose a small fraction of their energy on each scatter. Additionally, if the calculation needs to accurately calculate the shape of the low energy flux this approximation cannot be used. By ignoring the downscattered photons, the predicted flux of low energy photons will be too small. 
    
    \item {\it Fractional Energy Loss} - The opposite limit of attenuation is where all photons scatter however they only lose a small fraction of their energy on each scatter. That is true in the case of photons with $E \ll m_e$. With the additional assumption that all photons of a given energy lose energy at the same rate which again is valid in the limit of each photon scattering many times, Compton scattering can be included as an additional form of energy loss similar to redshifting. Eq.~\eqref{eq:redshiftEloss}, which describes how the photon energy changes of a redshift step becomes
    \begin{equation}
     E_i = \frac{1+z_i}{1+z_{i-1}} E_{i-1} - (z_{i-1} - z_{i})\frac{dE_\textrm{Compton}}{dz}(E_{i-1},z_{i-1})
    \end{equation}
    with $\frac{dE}{dz}$ determined as in Ref.~\cite{zdziarski1989absorption}. This does make determining the derivative $\frac{dE_{i-1}}{dE_i}$ in Eq.~\eqref{eq:eblStepSpecChange} more challenging. Therefore, when using this approximation Compton scattering and redshifting were treated sequentially. The photon spectrum was first changed accounting for redshifting and then the effect of Compton scattering was accounted for. Instead of calculating $\frac{dE_{i-1}}{dE_i}$ directly, we integrated the differential flux, $\frac{d\Phi_\gamma}{dE}$, to determine the total flux $\Phi_\gamma(E)$, and then took the derivative with respect to the shifted energy bins $E'$.
    While Ref.~\cite{zdziarski1989absorption} provides an expression for $\frac{dE}{dz}$ for all energies, the assumptions underlying this approximation are not valid for high energy photons or when only some photons scatter during a single step. The constraints found using this approximation do match the complete calculation more closely than the {\it attenuation} approximation however due to the assumptions breaking down some accuracy is sacrificed in comparison to using Eq.~\eqref{eq:ComptonNetChange}.
    
    \item The last approximation is to use Eq.~\eqref{eq:ComptonNetChange} to determine the proper Compton scattering effect only for black holes that have fully evaporated before today. The Universe is transparent to Compton scattering for photons originating at $z < 100$ and if the black holes still exist today, the signal will be dominated by photons produced recently. This is the approximation that was used to produce the final constraints in this work. For black holes evaporated before today we perform the full computationally intensive calculation and for black holes that are still around we use the {\it fractional energy loss} approximation.
\end{itemize}
A comparison of the effect the different Compton scattering approximations have on the constraints on PBH abundance with $n = 2$ can be seen in Fig. \ref{fig:comptonApprox}. For more massive PBHs that finish evaporating at later times Compton scattering stops being important and all approximations converge. For $n   > 2$ the pattern is similar except the effect of Compton scattering is less and therefore the differences between the various approximations are less important.

\bibliography{LEDBH.bib}
\end{document}